\documentclass[usenatbib]{mnrasb}
\pdfoutput=1
\usepackage{graphicx}

\usepackage{amsmath}
\usepackage{amssymb}
\usepackage{listings}
\usepackage{dsfont}
\usepackage{courier}

\newcommand{\nv}{\hat{\boldsymbol{\theta}}}
\newcommand{\wtj}[6]{\left(\begin{array}{ccc} #1 & #2 & #3\\#4 & #5 & #6\end{array} \right)}
\newcommand{\nmt}{{\tt NaMaster}}
\newcommand{\va}{{\mathbf{a}}}
\newcommand{\vb}{{\mathbf{b}}}

\newcommand{\mI}{{\mathds{1}}}

\title[A unified pseudo-$C_\ell$ framework]{A unified pseudo-$C_\ell$ framework}
  \author[D. Alonso et al.]{David Alonso$^{1,2}$\thanks{david.alonso@physics.ox.ac.uk}, Javier Sanchez$^{3}$, An\v{z}e Slosar$^{4}$ \newauthor{(The LSST Dark Energy Science Collaboration)}\\
  $^{1}$ School of Physics and Astronomy, Cardiff University, The Parade, Cardiff, CF24 3AA, United Kingdom\\
  $^{2}$ Department of Physics, University of Oxford, Denys Wilkinson Building, Keble Road, Oxford OX1 3RH, United Kingdom\\
  $^{3}$ Department of Physics and Astronomy, University of California, Irvine, CA 92697, USA\\
  $^{4}$ Brookhaven National Laboratory, Physics Department, Upton, NY 11973, USA}

\begin{document}
  \date{\today}
  \pagerange{1--27} \pubyear{2018}
  \maketitle

\begin{abstract}
  The pseudo-$C_\ell$ is an algorithm for estimating the angular power and cross-power spectra that is very fast and in realistic cases also nearly optimal. The algorithm can be extended to deal with contaminant deprojection and $E/B$ purification, and can therefore be applied in a wide variety of scenarios of interest for current and future cosmological observations. This paper presents \nmt, a public, validated, accurate and easy-to-use software package that, for the first time, provides a unified framework to compute angular cross-power spectra of any pair of spin-0 or spin-2 fields, contaminated by an arbitrary number of linear systematics and requiring $B$- or $E$-mode purification, both on the sphere or in the flat-sky approximation. We describe the mathematical background of the estimator, including all the features above, and its software implementation in \nmt. We construct a validation suite that aims to resemble the types of observations that next-generation large-scale structure and ground-based CMB experiments will face, and use it to show that the code is able to recover the input power spectra in the most complex scenarios with no detectable bias. \nmt~can be found at \url{https://github.com/LSSTDESC/NaMaster}, and is provided with comprehensive documentation and a number of code examples.
\end{abstract}


  \section{Introduction}\label{sec:intro}
    Two-point functions have proven to be the most useful summary statistic for cosmology in terms of data compression, due to the near-Gaussian nature of the modes of the cosmic density field that can be easily analyzed to extract cosmological information~\citep{1994ApJ...430L..85G,1995PhRvL..74.4369B,1997PhRvD..55.5895T,2001PhRvD..64f3001T}. Since the only true observables in any astronomical dataset are the intensity and polarization as a function of frequency $\nu$ and sky position $\nv$, angular 2-point correlators play a particularly central role in most cosmological analyses, both in the form of configuration-space two-point correlation functions $\xi(\theta)$ and as Fourier-space\footnote{In this paper we use the term ``Fourier'' to describe both the flat space Fourier expansion as well as harmonic-space expansion on the sphere. We will also use the term ``power spectrum'' to describe both the auto power spectrum of a field and cross-power spectra between two fields. } power spectra $C_\ell$ \citep{2012MNRAS.427.1891A,2011PhRvD..84f3505B}.  Although correlation functions are often easier to estimate from complex observations~\citep{1993ApJ...412...64L}, a Fourier space analysis is advantageous for many other reasons. The most important is that Fourier space is a natural space for working with statistics of translation-invariant fields. Therefore, Fourier space analysis has a clearer separation between linear and non-linear scales, as well as Gaussian and non-Gaussian modes~\citep{2008PhRvD..77j3013H}. Moreover, individual band-power measurements are typically significantly less correlated compared to configuration-space analysis which makes covariance matrix estimation easier. This has motivated a large body of literature aimed at designing optimal and efficient power spectrum estimators and used extensively in Cosmic Microwave Background (CMB) two-point measurements.
    
    While maximum-likelihood~\citep{1998PhRvD..57.2117B,2003PhRvD..67b3001W} or minimum-variance and quadratic estimators~\citep{1997PhRvD..55.5895T,2000MNRAS.317L..23H,2001PhRvD..64f3001T,2018arXiv180702484V} can recover the power spectrum with virtually no loss of information (see also the approaches of e.g. \cite{2004ApJS..155..227E,2008MNRAS.389.1284T,2016MNRAS.455.4452A}, based on Bayesian sampling methods), their computational implementation is prohibitively expensive for high resolution data, since these methods scale as $\ell_{\rm max}^6$ with the maximum multipole $\ell_{\rm max}$ (or $\ell_{\rm max}^4$ for optimal algorithms such as that of \cite{2003PhRvD..67b3001W}). For this reason, pseudo-$C_\ell$ algorithms \citep{1973ApJ...185..413P,2001PhRvD..64h3003W,2001ApJ...561L..11S,2001PhRvD..64h3003W,2002ApJ...567....2H,2002MNRAS.336.1304H,2004MNRAS.350..914C,2005MNRAS.358..833T} have become a popular alternative that preserves the $\ell_{\rm max}^3$ scaling of spherical harmonic transforms (and faster in the flat-sky approximation). This method can be used on fields with arbitrary spin \citep{2003ApJS..148..161K} in both curved and flat skies, has been extended to deal with contaminant deprojection in scalar fields \citep{2017MNRAS.465.1847E}, and can be optimized to measure the CMB $E/B$-mode power spectrum \citep{2002PhRvD..65b3505L,2003PhRvD..67b3501B,2006PhRvD..74h3002S,2009PhRvD..79l3515G,2011PhRvD..83h3003B}. Due to the advantages of this method, a number of useful, publicly available codes currently exist to carry out some of these calculations \footnote{See, for instance: \href{http://www2.iap.fr/users/hivon/software/PolSpice/}{PolSpice}, \href{https://gitlab.in2p3.fr/tristram/Xpol}{Xpol}, \href{https://gitlab.in2p3.fr/tristram/Xpure}{Xpure} and \href{https://github.com/thibautlouis/hyperQube}{hyperQube}.}. Unfortunately, no public, validated and maintained code exists simultaneously supporting all of the capabilities described above in a consistent manner. This paper presents a public software package, \nmt, that unifies all of these features providing a general framework to estimate pseudo-$C_\ell$ angular power spectra in a wide variety of scenarios. The paper also presents a few novel results that were derived as part of the development of \nmt. These include extending the contaminant deprojection formalism to spin-$2$ fields, combining it with $E/B$ purification and providing a consistent derivation of the same functionality in the flat-sky approximation.
    
    The paper is structured as follows: Section \ref{sec:maths} introduces the pseudo-$C_\ell$ estimator in the curved sky, including mode deprojection and $E/B$ purification. This is then generalized in the flat-sky approximation. In Section \ref{sec:structure} we briefly describe the implementation of these methods on \nmt, as well as its structure and usage. Section \ref{sec:validation} presents the stress tests used to validate the code. We envisage this code to be useful in analysis of both the Cosmic Microwave Background data and in the tomographic large-scale structure from photometric data, so this section illustrates not just its validity but also applicability in typical user-case scenarios in both fields. We conclude in Section \ref{sec:discussion}. The appendices provide further details about the mathematical background used throughout the paper, and present some additional validation tests.
    
  \section{The pseudo-\texorpdfstring{$C_\ell$}{Cl} estimator}\label{sec:maths}
    Here we introduce the pseudo-$C_\ell$ estimator and its relevant extensions. Many of the results presented here can be found in the literature (e.g. \cite{2002ApJ...567....2H,2003ApJS..148..161K,2006PhRvD..74h3002S,2009PhRvD..79l3515G,2013A&A...554A.112R,2017MNRAS.465.1847E} and references therein), although some others are new as far as we are aware. In what follows we will use boldface symbols for vector fields (e.g. ${\bf a}$) and sans-serif symbols for matrices (e.g. ${\sf C}$).
    
    \subsection{Spherical harmonic transforms and pseudo-\texorpdfstring{$C_\ell$}{Cl}s}\label{ssec:maths.pcl101}
      Let ${\bf a}(\nv)$ be a spin-$s_a$ quantity defined on the sphere, where $\nv$ is the unit vector pointing in a particular direction $(\theta,\varphi)$. We define its spherical harmonic coefficients through a spherical harmonic transform (SHT) as:
      \begin{align}\label{eq:sht1}
        {\bf a}_{\ell m}&\equiv\mathcal{S}\left[{\bf a}(\nv)\right]^{s_a}_{\ell m}\equiv\int d\nv\,{\sf Y}^{s_a\dag}_{\ell m}(\nv)\,{\bf a}(\nv), \\\label{eq:sht2}
        {\bf a}(\nv)&=\mathcal{S}^{-1}\left[{\bf a}_{\ell m}\right]^{s_a}_{\nv}\equiv\sum_{\ell m}\,{\sf Y}^{s_a}_{\ell m}(\nv)\,{\bf a}_{\ell m}.
      \end{align}
      Note that, in general we will use vector notation such that, for a complex spin-$s_a$ field $a$, we form the vector ${\bf a}\equiv({\rm Re}(a),{\rm Im}(a))$. The harmonic coefficients above are decomposed in a similar manner into $E$ and $B$ modes: ${\bf a}_{\ell m}\equiv(a^E_{\ell m},a^B_{\ell m})$ (e.g. see Appendix \ref{app:shts} and \cite{1997PhRvD..55.1830Z}). The matrices ${\sf Y}^s_{\ell m}$ are defined in terms of the spin-weighed spherical harmonics, and are described in detail in Appendix \ref{app:shts}. Some technical details regarding band-limits and the practicalities of SHTs in a discretized sphere are discussed in \ref{app:shts.blim}. To simplify the notation, we will also often abbreviate the pair $(\ell,m)$ as ${\bf l}$. 
      
      For two isotropic fields ${\bf a}$ and ${\bf b}$, the power spectrum is given by their covariance matrix:
      \begin{equation}\label{eq:cldef}
        \left\langle{\bf a}_{\bf l}{\bf b}_{{\bf l}'}^\dag\right\rangle\equiv{\sf C}^{ab}_\ell\delta_{\ell\ell'}\delta_{mm'}.
      \end{equation}
      Note that for a general non-zero spin, vectors $\va$ and $\vb$ have two-components, each of which is a complex number. Nevertheless the rotational invariance requires that elements of the $2\times2$ matrix ${\sf C}_\ell$ be real.

      \subsubsection{Power-spectrum estimation}
        The problem we are attempting to solve is how to estimate ${\sf C}^{ab}_\ell$ from a single realization of ${\bf a}$ and ${\bf b}$ measured on a cut sky. In general, the input data for measuring the power spectrum come pixelized on a sphere. Instead of dealing with an infinitely sampled field, we have a finite number of measurements
        \begin{equation}
          \va_i = \va_i^{\rm true} + {\bf n}_i,
        \end{equation}
        where $\va_i^{\rm true}$ is is the true underlying field whose power spectrum we would like to estimate and ${\bf n}_i$ is the noise vector, which in general can be non-white and non-homogeneous. We will denote the pixel covariance of the noise component as $\tilde{\sf N}$\footnote{Note that the matrices $\tilde{\sf C}$ and $\tilde{\sf N}$ used in this section are the data and noise pixel-pixel covariance matrices, and therefore are different from the power spectrum matrices, ${\sf C}_\ell$ and ${\sf N}_\ell$, used in the rest of the paper (e.g. Eq. \ref{eq:cldef}).}. Note that for the sake of power-spectrum estimation the there is no conceptual difference between inhomogeneous noise and a survey mask: masked areas can be simply though of as areas of infinite noise. Conversely, when we talk about ``multiplying by the mask'' we really mean a more general process of dividing by the expected variance field.

        The underlying power spectrum in Eq.~\ref{eq:cldef} means that there is a non-trivial true underlying covariance matrix $\tilde{\sf C}^{\rm true} = \left<\va^{\rm true} \va^{{\rm true}\dagger}\right>$, which in addition to the noise component leads to the actual data covariance $\tilde{\sf C} = \left<\va \va^\dagger\right>=\tilde{\sf C}^{\rm true} + \tilde{\sf N}$.

        In the standard incarnation of the optimal quadratic estimator~\citep{1998PhRvD..57.2117B}, the measurement is produced on inverse covariance weighted data, namely
        \begin{equation}\label{eq:fisher_bpw}
          {\sf F_{ij}} B_j = \frac{1}{2}\va^\dagger \tilde{\sf C}^{-1}\,{\sf P}_i\,\tilde{\sf C}^{-1} \va -
          \frac{1}{2}\rm{Tr}\left[\tilde{\sf C}^{-1}{\sf P}_i\tilde{\sf C}^{-1}\,{\sf N} \right],
        \end{equation}
        where ${\sf F}$ is the Fisher matrix, the covariance matrix is modeled as a linear sum over band-power parameters $B_i$, i.e.
        \begin{equation}\label{eq:bpws_qmv}
          \tilde{\sf C} = \sum_i {\sf P}_i\,B_i + {\sf N},
        \end{equation}
        (which defines the response matrices ${\sf P}_i$) and the last term in Eq. \ref{eq:fisher_bpw} corresponds to a bias of the convolved estimates due to the presence of noise. Thus, this optimal estimator corresponds to i) inverse-variance weighting the data, ii) calculating the spectrum of the weighted data using the response matrices and iii) deconvolving these estimates using a mode-coupling matrix which in this particular case happens to be the same as Fisher matrix. The main computational challenge when implementing this estimator usually lies in storing the covariance matrix and its inverse (or computing it on the fly), particularly in order to estimate the Fisher matrix. This motivates the pseudo-$C_\ell$ estimator, which replaces these computationally very intensive steps with numerically more efficient ones at the expense of optimality.

        The main improvement is gained by replacing the inverse covariance by its diagonal, assuming that the data is uncorrelated between pixels. The response matrices are also replaced by simple spherical harmonic transforms and multiplication of the corresponding harmonic coefficients, followed by an average over bandpowers. Both of these steps reduce to exact optimal estimator in the limit of uncorrelated data. The resulting estimates are still biased and require a mode-decoupling matrix, which can be calculated as we show in Section \ref{sssec:maths.pcl101.mcoup}. 

        The pseudo-$C_\ell$ estimates will be optimal when these steps are close to what an optimal quadratic estimator would do. First, this requires that $\tilde{\sf C}$ be close to diagonal, which is true when either the noise is large and uncorrelated, or when the underlying power spectrum is close to white. Multiplication by the mask mixes modes, and replacement of multiplication by the response matrix with band-power averaging, makes sense only when the mask is ``well-behaved'', i.e. compact with no high-frequency structure in it. If these conditions are not met, the pseudo-$C_\ell$ estimator will still be unbiased (by construction), but it will get progressively less optimal. In practical surveys, however, the loss of optimality rarely exceeds 10-20\% and is typically smaller~\citep{2013MNRAS.435.1857L}.

        In what follows we will work with two-fields $\va$ and $\vb$ and spell out in detail how to calculate the cross-power spectrum. The auto-power spectrum case is the same (with $\vb=\va$), but one must additionally subtract the noise-contribution to the estimate. So far, the discussion has been completely general, but from now onwards, we will focus on the power estimation of auto, and cross-power spectra for the spin-0 and spin-2 fields only, as these are the most commonly used in cosmology.

      \subsubsection{Mode coupling}\label{sssec:maths.pcl101.mcoup}
        Let $v(\nv)$ be a sky mask or weights map for ${\bf a}$, and let us define ${\bf a}^v\equiv v(\nv){\bf a}(\nv)$. As discussed above we start by considering a na\"ive estimator for ${\sf C}^{ab}_\ell$ from the masked fields:
        \begin{equation}\label{eq:pcldef}
          {\rm PCL}_\ell\left({\bf a}^v,{\bf b}^w\right)\equiv\frac{1}{2\ell+1}\sum_{m=-\ell}^\ell{\bf a}^v_{\bf l}{\bf b}^{w\dag}_{\bf l}.
        \end{equation}
        The incomplete sky coverage couples different $\ell$ modes, and makes this estimator biased. The pseudo-$C_\ell$ method is based on computing an analytical prediction for this bias and correcting for it. This is straightforward to do using the results in Appendix \ref{app:shts}. The harmonic coefficients of a masked field are given by
        \begin{equation}
          {\bf a}^v_{\bf l}=\sum_{{\bf l}_1{\bf l}_2}{\sf D}^{s_a}_{{\bf l}{\bf l}_1{\bf l}_2}{\bf a}_{{\bf l}_1}v_{{\bf l}_2},
        \end{equation}
        where ${\sf D}$ is defined in Eq. \ref{eq:dmat}. Using the statistical isotropy of the unmasked field (Eq. \ref{eq:cldef}) and the orthogonality relation Eq. \ref{eq:3jorth}, we obtain:
        \begin{equation}\label{eq:modecoup}
          {\rm vec}\left[\left\langle{\rm PCL}_\ell\left({\bf a}^v_{\bf l},{\bf b}^w_{\bf l}\right)\right\rangle\right]=\sum_{\ell'}{\sf M}^{s_as_b}_{\ell\ell'}\cdot{\rm vec}\left[{\sf C}^{ab}_{\ell'}\right],
        \end{equation}
        where we have defined the vectorization operator \citep{2008PhRvD..77j3013H}
        \begin{align}
          &{\rm vec}\left[\left(\begin{array}{cc}
                          C^{EE} & C^{EB}\\
                          C^{BE} & C^{BB}
                         \end{array}\right)\right]\equiv
          \left(
          \begin{array}{c}
            C^{EE} \\
            C^{EB} \\
            C^{BE} \\
            C^{BB} \\
          \end{array}
          \right),\\
          &{\rm vec}\left[\left(\begin{array}{c}
                                 C^{TE}\\C^{TB}
                                \end{array}\right)\right]\equiv
            \left(\begin{array}{c}
                    C^{TE}\\C^{TB}
            \end{array}\right),
            \hspace{6pt}
            {\rm vec}\left[C^{TT}\right]={\rm C}^{TT},
        \end{align}
        and $T$ here stands for any spin-$0$ field.
      
        The mode-coupling matrix ${\sf M}_{\ell\ell'}$ in Eq. \ref{eq:modecoup} can be computed in terms of the harmonic coefficients of the two masks and, for spin-0 and spin-2 fields, is given by \citep{2002ApJ...567....2H,2003ApJS..148..161K}:
        \begin{align}\label{eq:m00}
          &{\sf M}^{00}_{\ell\ell'}=\frac{2\ell'+1}{4\pi}\sum_{\ell''}P^{vw}_{\ell''}\wtj{\ell}{\ell'}{\ell''}{0}{0}{0}^2\,\mI,\\\label{eq:m02}
          &{\sf M}^{02}_{\ell\ell'}=M^{0+}_{\ell\ell'}\,\mI,\\\nonumber
          &M^{0+}_{\ell\ell'}=\frac{2\ell'+1}{4\pi}\sum_{\ell''}P^{vw}_{\ell''}\wtj{\ell}{\ell'}{\ell''}{0}{0}{0}\wtj{\ell}{\ell'}{\ell''}{2}{-2}{0},\\\label{eq:m22}
          &{\sf M}^{22}_{\ell\ell'}=\left(
                        \begin{array}{cccc}
                          M^{+}_{\ell\ell'} & 0 & 0 & M^{-}_{\ell\ell'} \\
                          0 & M^{+}_{\ell\ell'} & -M^{-}_{\ell\ell'} & 0 \\
                          0 & -M^{-}_{\ell\ell'} & M^{+}_{\ell\ell'} & 0 \\
                          M^{-}_{\ell\ell'} & 0 & 0 & M^{+}_{\ell\ell'}
                        \end{array}\right),\\\nonumber
          &M^{\pm}_{\ell\ell'}=\frac{2\ell'+1}{4\pi}\sum_{\ell''}P^{vw}_{\ell''}\wtj{\ell}{\ell'}{\ell''}{2}{-2}{0}^2\frac{1\pm(-1)^{\ell+\ell'+\ell''}}{2},
        \end{align}
        where $\mI$ is the unit matrix, and $P^{vw}_\ell\equiv(2\ell+1)\,{\rm PCL}_\ell(v,w)$.
        \begin{figure*}
          \centering
          \includegraphics[width=0.9\textwidth]{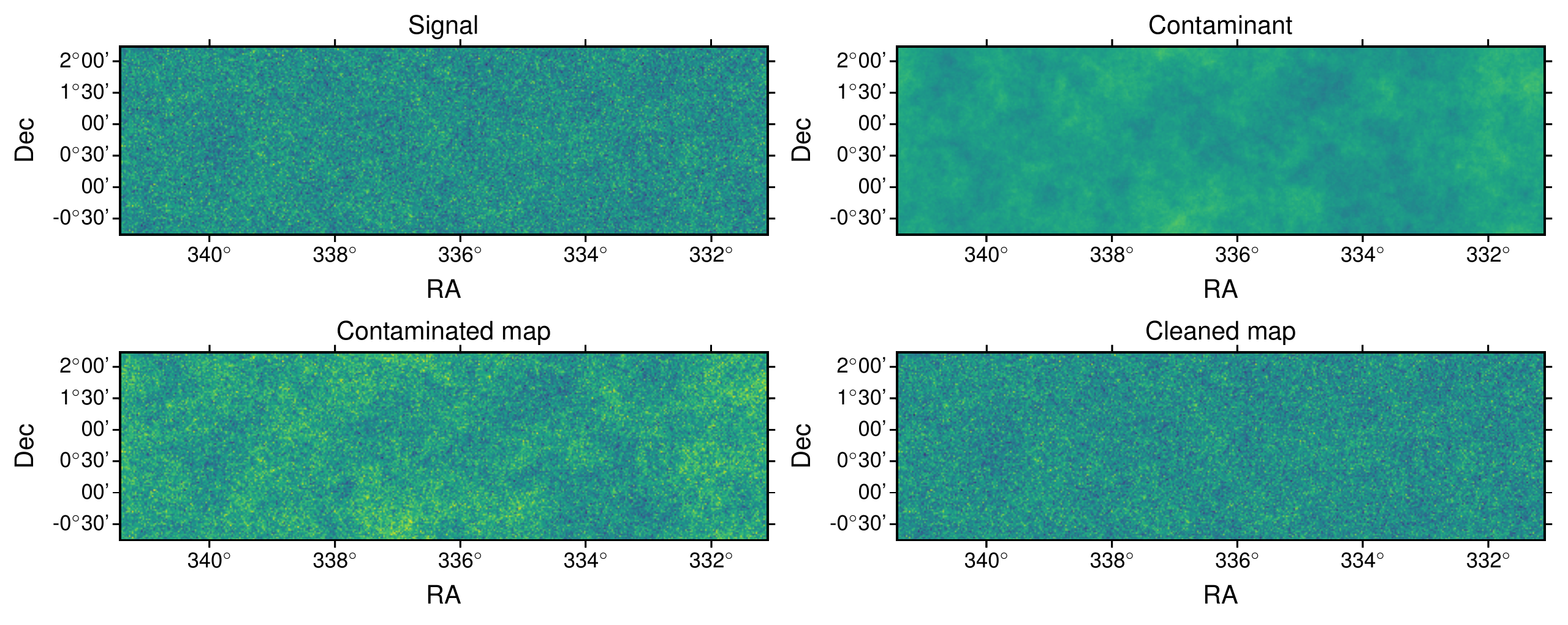}
          \caption{Depiction of the contaminant deprojection process. {\sl Top left:} underlying true signal map (galaxy overdensity used in Section \ref{sssec:validation.suite.lss}). {\sl Top right:} sky contaminant (dust contamination used in Section \ref{sssec:validation.suite.lss}). {\sl Bottom left:} contaminated map obtained by adding the previous two maps. {\sl Bottom right:} cleaned map obtained through Eq. \ref{eq:clean_f}. Since only one contaminant is considered in this case, the cleaning is almost perfect. Residual contamination at the power spectrum level must still be subtracted in order to avoid a biased estimate.}
          \label{fig:lss_cont}
        \end{figure*}
      
      \subsubsection{Bandpowers}\label{sssec:maths.pcl101.bpws}
        Given the loss of information implicit in masking the originally curved-sky field, it is in general not possible to invert the mode-coupling matrix in Eq. \ref{eq:modecoup} directly. One solution to this problem is to convolve the theory prediction with the mode-coupling matrix at the likelihood level. A more usual approach is instead to bin the coupled pseudo-$C_\ell$ into bandpowers. A bandpower $q$ is defined as a set of $N_q$ multipoles $\vec{\ell}_q\equiv(\ell_q^1,...,\ell_q^{N_q})$ and a set of weights defined at those multipoles $w^\ell_q$, and normalized such that $\sum_{\ell\in\vec{\ell}_q}w_q^\ell=1$. The $q$-th bandpower for the coupled pseudo-$C_\ell$ is then defined as
        \begin{align}\label{eq:bpws}
          {\rm vec}\left[\tilde{\sf B}^{ab}_q\right]&\equiv\sum_{\ell\in\vec{\ell}_q}w_q^\ell\,{\rm vec}\left[{\rm PCL}_{\ell}({\bf a}^v_{\bf l},{\bf b}^w_{\bf l})\right]\\\nonumber
          \left\langle{\rm vec}\left[\tilde{\sf B}^{ab}_q\right]\right\rangle&=\sum_{\ell\in\vec{\ell}_q}w_q^\ell\sum_{\ell'}{\sf M}^{s_as_b}_{\ell\ell'}\,{\rm vec}\left[{\sf C}^{ab}_{\ell'}\right],
        \end{align}
        where in the second line we have taken the expectation value of the first one. One then proceeds by assuming that the true power spectrum is a step-wise function, taking constant values over the multipoles corresponding to each bandpower: ${\sf C}^{ab}_\ell\equiv\sum_q{\sf B}^{ab}_q\Theta(\ell\in\vec{\ell}_q)$ (where $\Theta$ is a binary step function). Inserting this in the equation above, it is then possible to find an unbiased estimator for ${\sf B}^{ab}_q$ in terms of the binned pseudo-bandpowers $\tilde{\sf B}^{ab}_q$:
        \begin{equation}\label{eq:decoup}
          {\rm vec}\left[{\sf B}^{ab}_q\right]=\sum_{q'}\left({\cal M}^{s_as_b}\right)^{-1}_{qq'}\,{\rm vec}\left[\tilde{\sf B}^{ab}_{q'}\right],
        \end{equation}
        where the binned coupling matrix ${\cal M}^{s_as_b}$ is
        \begin{equation}
          {\cal M}^{s_as_b}_{qq'} \equiv \sum_{\ell\in\vec{\ell}_q}\sum_{\ell'\in\vec{\ell}_{q'}}w_q^\ell\,{\sf M}^{s_as_b}_{\ell\ell'}.
        \end{equation}
      
        Since, in general, the true power spectrum is not a step-wise function, the theoretical prediction must be corrected for the bandpower binning before making any comparison with the data:
        \begin{equation}\label{eq:theory_correct}
          {\rm vec}\left[{\sf B}^{ab,{\rm th}}_q\right]=\sum_{\ell}{\cal F}^{s_as_b}_{q\ell}\cdot{\rm vec}\left[{\sf C}^{ab,{\rm th}}_\ell\right],
        \end{equation}
        where the filter ${\cal F}^{s_as_b}_{q\ell}$ is given by
        \begin{equation}\label{eq:bpresponse}
          {\cal F}^{s_as_b}_{q\ell}=\sum_{q'}({\cal M}^{s_as_b})^{-1}_{qq'}\sum_{\ell'\in\vec{\ell}_{q'}}w_{q'}^{\ell'}{\sf M}^{s_as_b}_{\ell'\ell}.
        \end{equation}
        The action of the filter ${\cal F}^{s_as_b}$ is therefore given by a sequence of three operations:
        \begin{enumerate}
          \item Coupling the different multipoles (Eq. \ref{eq:modecoup}).
          \item Binning into bandpowers (Eq. \ref{eq:bpws}).
          \item Decoupling the bandpowers (Eq. \ref{eq:decoup}).
        \end{enumerate}
        Depending on the structure of the mode-coupling matrix, ${\cal F}^{s_as_b}$ can be significantly different from a na\"ive binning operator.
        
    \subsection{Contaminant deprojection}\label{ssec:maths.deproj}
      In a typical data-analysis case, we deal with contaminants that pollute the measured signal. Usually, we know something about these contaminants. As an example, in case of CMB analysis, we might have a map of a given foreground at a different frequency which can be marginalized out\footnote{It is worth noting that the formalism described here assumes that a given contaminant template has no significant contribution from the signal itself. This might not be the case, for example, if one blindly took a high-frequency sky map to serve as a template of dust, since that may still contain a significant CMB contribution. Care must therefore be exercised when defining the contaminant templates to marginalize over.}. But even when we do not, we often have a tracer map of a given effect, for example, the map of dust might tell us something about systematic errors related to imperfect correction to dust reddening, or a map of co-added point spread function (PSF) fluctuations could be used to remove the possible correlations between measured ellipticities. Even if the dust map is non-linearly and perhaps stochastically related to the amount of reddening, as long as that relation is local in space, we get a linear contamination on large scales, very much like complex astrophysics of galaxy formation nevertheless reduces to linear biasing on very large scales (see Appendix \ref{app:lincont}). Thus, we generically deal with a linear model of the kind:
      \begin{equation}\label{eq:aobs}
        {\bf a}^{\rm obs}(\nv)={\bf a}^v+\sum_{i=1}^{N_a}\alpha_i{\bf f}^i(\nv),
      \end{equation}
      where the observed map ${\bf a}^{\rm obs}$ is composed of a masked signal map that receives contributions from $N_a$ known contaminants ${\bf f}^i$ and where the contamination coefficients $\alpha_i$ are unknown (note that we have implicitly applied the same mask to the contaminant templates ${\bf f}^i$). The deprojection of these contaminants from the data can be done and propagated into the estimated power spectrum in a natural way in the case of quadratic estimators \citep{1992ApJ...398..169R,2004PhRvD..69l3003S,2016MNRAS.456.2095E}, and the method can be extended to the pseudo-$C_\ell$ algorithm \citep{2017MNRAS.465.1847E}. The process is illustrated in Figure \ref{fig:lss_cont}. A best-fit value for the coefficients $\alpha_i$ can be found assuming \emph{uncorrelated weights} across the map\footnote{The assumption of uncorrelated weights is the key difference between pseudo-$C_\ell$ and optimal quadratic estimators \citep{2013MNRAS.435.1857L}. Note that, since the templates are already multiplied by the mask, the weights are not equal across the map, but correspond to the mask values in general.}
      \begin{align}\label{eq:deproj_coeff}
        &\hat{\alpha}_i=F_{ij}\int d\nv\,{\bf f}^{j\dag}(\nv)\,{\bf a}^{\rm obs}(\nv),\\\nonumber
        &({\sf F}^{-1})_{ij}\equiv\int d\nv\, {\bf f}^{i\dag}(\nv)\,{\bf f}^j(\nv),
      \end{align}
      where there is an implicit summation over the repeated index $j$ in Eq. \ref{eq:deproj_coeff} (which we will omit from now on). The clean map is then given by
      \begin{align}
        {\bf a}^c(\nv)
        &\equiv{\bf a}^{\rm obs}(\nv)-{\bf f}^i(\nv)F_{ij}\int d\nv'\,{\bf f}^{j\dag}(\nv')\,{\bf a}^{\rm obs}(\nv')
      \end{align}
      which, when substituting Eq. \ref{eq:aobs}, yields
      \begin{align}\label{eq:clean}
        {\bf a}^c(\nv)&={\bf a}^v(\nv)-{\bf f}^i(\nv)F_{ij}\int d\nv'\,{\bf f}^{j\dag}(\nv')\,{\bf a}^v(\nv').
      \end{align}      
      The harmonic coefficients of ${\bf a}^c$ are therefore given by
      \begin{equation}\label{eq:clean_f}
        {\bf a}^c_{\bf l}={\bf a}^v_{\bf l}-{\bf f}^i_{\bf l}\,F_{ij}\sum_{{\bf l}'}{\bf f}^{j\dag}_{{\bf l}'}{\bf a}^v_{{\bf l}'}
      \end{equation}
      
      Although in general we expect the second term in Eq. \ref{eq:clean} to be very small, since the signal and contaminant maps are uncorrelated, due to this residual statistical contamination, a direct computation of the pseudo-$C_\ell$ of the cleaned map would yield a biased estimator beyond the mode coupling induced by the mask $v$. This bias can however be estimated analytically and corrected for. This was done in \cite{2017MNRAS.465.1847E} for scalar quantities, and we generalize their result here to fields of arbitrary spin.
      
      Considering a second map ${\bf b}$ with mask $w$, $N_b$ contaminants ${\bf g}^i$ and correlation matrix $G_{ij}$, the mode-coupled pseudo-$C_\ell$ of ${\bf a}^c$ and ${\bf b}^c$ is given by:
      \begin{align}\nonumber
        {\rm PCL}_\ell({\bf a}^c,{\bf b}^c)=&{\rm PCL}_\ell({\bf a}^v,{\bf b}^w)\\\nonumber
        &-\frac{G_{ij}}{2\ell+1}\sum_m\sum_{{\bf l}'}{\bf a}^v_{\bf l}{\bf b}^{w\dag}_{{\bf l}'}{\bf g}^j_{{\bf l}'}{\bf g}^{i\dag}_{\bf l}\\\nonumber
        &-\frac{F_{ij}}{2\ell+1}\sum_m\sum_{{\bf l}'}{\bf f}^i_{\bf l}{\bf f}^{j\dag}_{{\bf l}'}{\bf a}^v_{{\bf l}'}{\bf b}^{w\dag}_{\bf l}\\\label{eq:dpbias0}
        &+\frac{F_{ij}G_{rs}}{2\ell+1}\sum_m\sum_{{\bf l}',{\bf l}''}{\bf f}^i_{\bf l}{\bf f}^{j\dag}_{{\bf l}'}{\bf a}^v_{{\bf l}'}{\bf b}^{w\dag}_{{\bf l}''}{\bf g}^r_{{\bf l}''}{\bf g}^{s\dag}_{\bf l}.
      \end{align}
      Taking the ensemble average of the equation above, we find that the first term is the usual pseudo-$C_\ell$ estimator. The combination of the last three terms is what we will refer to as the \emph{deprojection bias}, and can be computed as\footnote{These equations can be easily derived by expanding ${\bf a}^v$ and ${\bf b}^v$ in terms of the spherical harmonics of ${\bf a}$, ${\bf b}$, $v$ and $w$, making use of Eq. \ref{eq:cldef} and repeatedly employing the definitions of the spherical harmonic transform (Eqs. \ref{eq:sht1} and \ref{eq:sht2}).}:
      \begin{align}\nonumber
        \Delta {\sf C}^{ab}_\ell=&-G_{ij}\,{\rm PCL}_\ell\left(\tilde{\bf g}^j_{\bf l},{\bf g}^i_{\bf l}\right)-F_{ij}\,{\rm PCL}_\ell\left({\bf f}^i_{\bf l},\tilde{\bf f}^j_{\bf l}\right)\\\label{eq:dpbias}
        &+F_{ij}G_{rs}\left[\int d\nv\,{\bf f}^{j\dag}(\nv)\tilde{\bf g}^r(\nv)\right]{\rm PCL}_\ell\left({\bf f}^{i},{\bf g}^s\right),
      \end{align}
      where
      \begin{align}\label{eq:dp.g.aux}
        \tilde{\bf g}^i_{\bf l}&\equiv{\cal S}\left[v(\nv){\cal S}^{-1}\left[{\sf C}^{ab}_{\ell_1}{\cal S}\left[w\,{\bf g}^i\right]^{s_b}_{{\bf l}_1}\right]^{s_a}_{\nv}\right]^{s_a}_{\bf l},\\\label{eq:dp.f.aux}
        \tilde{\bf f}^i_{\bf l}&\equiv{\cal S}\left[w(\nv){\cal S}^{-1}\left[{\sf C}^{ab\dag}_{\ell_1}{\cal S}\left[v\,{\bf f}^i\right]^{s_a}_{{\bf l}_1}\right]^{s_b}_{\nv}\right]^{s_b}_{\bf l}.
      \end{align}
      
      The procedure to obtain an unbiased estimator of the power spectrum in the presence of contaminants can therefore be summarized as follows:
      \begin{enumerate}
        \item Produce clean versions of both maps (i.e. estimate the best-fit coefficients in Eq. \ref{eq:deproj_coeff} and subtract the contamination as in Eq. \ref{eq:clean}).
        \item Estimate the pseudo-$C_\ell$ of the clean maps (Eq. \ref{eq:pcldef}).
        \item Estimate the deprojection bias in Eq. \ref{eq:dpbias} and subtract it from the pseudo-$C_\ell$ above.
        \item Use the methods described in Sections \ref{sssec:maths.pcl101.mcoup} and  \ref{sssec:maths.pcl101.bpws} to account for the mode-coupling matrix.        
      \end{enumerate}
      
      Two further points must be addressed before moving on. First, the computation of the deprojection bias requires an estimate of the true power spectrum ${\sf C}^{ab}_\ell$ (see Eqs. \ref{eq:dp.g.aux} and \ref{eq:dp.f.aux}). In practice, this can be done through an iterative approach, starting from an initial guess of ${\sf C}^{ab}_\ell$ and replacing it by the estimated power spectrum in the previous iteration. Alternatively, the ensemble average of Eq. \ref{eq:dpbias0} can be seen as a convolution of ${\sf C}^{ab}_\ell$, which could be treated by computing the convolution matrix as a modified version of the standard mode-coupling matrix that accounts for mode deprojection. Estimating this convolution matrix in this case becomes significantly more computationally intensive, and therefore an iterative approach is preferred. 
      
      Secondly, estimating the deprojection coefficients (Eq. \ref{eq:deproj_coeff}) involves inverting the correlation matrix of all the contaminant templates. This will not be in general possible (e.g. if a set of templates are linearly related). The problem can however be solved self-consistently by using the Moore-Penrose pseudo-inverse \citep{2017MNRAS.465.1847E}. This is trivial for a symmetric and positive-definite matrix such as ${\sf F}$, and corresponds to setting the inverse of all of its zero eigenvalues (or all those below a given threshold) to zero\footnote{The choice of eigenvalue threshold in \nmt~is accessible to the user.}.
      
    \subsection{\texorpdfstring{$E/B$}{E/B} purification}\label{ssec:maths.pureb}
      As described in \cite{2002PhRvD..65b3505L,2003PhRvD..67b3501B,2006PhRvD..74h3002S,2010PhRvD..82b3001Z,2010A&A...519A.104K,2011PhRvD..83h3003B}, $B$-mode purification refers to the map-level removal of the contamination from $E$-modes in the $B$-mode component of a given map caused by an incomplete sky coverage and vice-versa\footnote{Note that this is a different effect from the leakage between different polarized and unpolarized channels in CMB observations due to instrumental effects (asymmetric beams, pair differencing, polarization angle calibration etc.)}. The procedure is particularly useful in situations in which the $E$-mode component of the signal is significantly larger than the $B$ modes, as is the case in for the CMB. In this case, removing the leakage at the power spectrum level (i.e. the standard pseudo-$C_\ell$ approach) produces a sub-optimal estimator in which the variance in the $B$-mode power spectrum is dominated by the variance of the leaked $E$-modes. This section presents the purification method and the associated modifications to the pseudo-$C_\ell$ algorithm as described in the references above. In this section we describe the algorithm in terms of purifying $B$-modes from $E$-mode contamination, but we note that the reverse, i.e. purifying $E$-modes from $B$-mode contamination is conceptually exactly the same. Our discussion is also specific for spin-2 fields.
      
      We define a field ${\bf f}$ to be a $B$ mode if $({\bf D}^{E}_s)^\dag{\bf f}=0$, where ${\bf D}^{E}_s$ is the differential operator defined in Eq.~\ref{eq:ederiv}. At the same time, and under the definition of the dot product:
      \begin{equation}
        ({\bf f},{\bf g})\equiv\int d\nv\,{\bf f}^\dag(\nv){\bf g}(\nv),
      \end{equation}
      we define a \emph{pure} $B$ mode as a field that is orthogonal to all $E$ modes.
      
      Since ${\bf D}^{E\dag}_s{\bf D}^B_s=0$, one can always generate a $B$ mode by applying ${\bf D}^{B}_s$ to a scalar field. It is then possible to show that $E$ and $B$ modes thus defined are orthogonal in the full sky:
      \begin{equation}
        ({\bf D}^E_s\varphi,{\bf D}^B_s\psi)=\int d\nv\,({\bf D}^E_s\varphi)^\dag{\bf D}^B_s\psi=0,
      \end{equation}
      where $\varphi$ and $\psi$ are two arbitrary scalar fields. This can be done by integrating by parts and noting that the celestial sphere has no boundaries. On a cut sky, however, and for $s=2$, this is only true if the fields satisfy Neumann and Dirichlet boundary conditions simultaneously (i.e. vanishing value and first derivative on the boundary of the cut sky region). 
  
      Let $w(\nv)$ be a sky window function defining the sky region to be analyzed (and the weight to be applied in each pixel). The standard pseudo $B$-mode of a spin-2 field ${\bf P}$ is then given by
      \begin{align}\nonumber
        \tilde{B}_{\bf l}&\equiv\int d\nv\,w(\nv)\left(_s{\bf Y}^{B}_{\bf l}(\nv)\right)^\dag{\bf P}\\
                         &=\int d\nv\,w(\nv)({\bf D}^B_sY_{\bf l})^\dag{\bf P}(\nv),
      \end{align}
      Since ${\bf D}^B_sY_{\ell m}$ is a $B$-mode, in the absence of $w$ this expression would correspond to a projection that filters out all the $E$-modes from ${\bf P}$. However, $w(\nv){\bf D}^B_sY_{\ell m}$ is not a $B$-mode, and therefore $\tilde{B}_{\ell m}$ receives contributions from ambiguous $E$ modes (which then propagate into the variance of the pseudo-$C_\ell$ estimator of the power spectrum). The idea behind $B$-mode purification is to move $w$ to the right of ${\bf D}^B_s$, defining the pure $B$ component:
      \begin{equation}
        B^p_{\bf l}=\int d\nv\left({\bf D}^B_s(w\,Y_{\bf l})\right)^\dag\,{\bf P}(\nv).
      \end{equation}
      Since ${\bf D}^B_s(wY_{\bf l})$ is a $B$-mode quantity, $B^p_{\bf l}$ should receive contributions only from $B$-modes.
  
      Expanding ${\bf D}^B_2(wY_{\bf l})$, we can write $B^p_{\bf l}$ as:
      \begin{equation}\label{eq:pureb}
        B^p_{\bf l}=\left(\tilde{P}_2\right)^B_{\bf l}+2\frac{\beta_{\ell,2}}{\beta_{\ell,1}}\left(\tilde{P}_1\right)^B_{\bf l}+\beta_{\ell,2}\left(\tilde{P}_0\right)^B_{\bf l},
      \end{equation}
      where $(a)^B_{\bf l}$ stands for the $B$-mode component of field ${\bf a}$, and we have defined the fields $\tilde{P}_n=(\eth^{2-n}w)^*(Q+iU)$, where $Q$ and $U$ are the real and imaginary parts of the field $P$ (see Appendix \ref{app:shts} for the definitions of $\eth$ and $\beta_{\ell,s}$).
     
      The $B(E)$-purification of a given field is simply achieved by applying Eq. \ref{eq:pureb} to the $B(E)$-mode component of the field. Note that doing so requires the computation of the first and second-order derivatives of the weights map $w$, and therefore purification methods require masks for which this quantities are well defined. This usually involves tapering the mask boundaries to avoid sharp edges. A thorough discussion of these methods can be found in \cite{2009PhRvD..79l3515G}. Note that the derivatives of $w$ can be computed analytically in harmonic space:
      \begin{align}
        \eth^nw={\cal S}^{-1}\left[\left(-\frac{w_{\bf l}}{\beta_{\ell,n}},0\right)\right],
      \end{align}
      
      At this stage, the pseudo-$C_\ell$ method proceeds as usual, applying Eq. \ref{eq:pcldef} to the purified field. The purification process however requires a slight modification of the analytical form of the mode-coupling matrix. Specifically, if either $E$ or $B$ modes of a spin-2 field have been purified Eqs. \ref{eq:m02} and \ref{eq:m22} must be modified by exchanging one factor of
      \begin{equation}\nonumber
        \wtj{\ell}{\ell'}{\ell''}{2}{-2}{0}
      \end{equation}
      for
      \begin{align}\nonumber
        &\wtj{\ell}{\ell'}{\ell''}{2}{-2}{0}+2\frac{\beta_{\ell,2}}{\beta_{\ell,1}\beta_{\ell'',1}}\wtj{\ell}{\ell'}{\ell''}{1}{-2}{1}\\
        &\,\,\,+\frac{\beta_{\ell,2}}{\beta_{\ell'',2}}\wtj{\ell}{\ell'}{\ell''}{0}{-2}{2}.
      \end{align}
      per purified field.
      
      It is worth noting that, as described in \cite{2009PhRvD..79l3515G}, the performance of $E/B$ purification can be optimized by an appropriate choice of window function. The corresponding optimal spin-0, spin-1 and spin-2 window functions can be estimated using a preconditioned conjugate-gradient method. The resulting spin window functions are no longer related by simple covariant derivatives. The performance of this method, compared with simple apodization methods, will depend on the noise properties of the experiment and the footprint geometry. This optimal method is currently not available in \nmt, and will be implemented in future releases of the code.
      
      Finally, it is also possible to combine purification and contaminant deprojection. Since $E/B$ purifications involves destroying part of the signal (e.g. projecting out all leaked $E$ modes from a $B$-mode component), it is more optimal to use the non-pure maps to compute the contamination coefficients ($\alpha_i$ in Eq. \ref{eq:deproj_coeff}). Therefore we will assume here that purification takes place between the steps (i) and (ii) described at the end of Section \ref{ssec:maths.deproj}, and not before step (i). In this case, although no additional modifications are needed in the mode-coupling matrix, beyond those we have just described, the expression for the deprojection bias (Eq. \ref{eq:dpbias}) must be slightly altered. In particular, although the fields $\tilde{\bf g}^i$ and $\tilde{\bf f}^i$ in Equations \ref{eq:dp.g.aux} and \ref{eq:dp.f.aux} are still computed from the non-pure versions of ${\bf f}^i$ and ${\bf g}^i$, all the maps entering the three instances of ${\rm PCL}_\ell$ in Eq. \ref{eq:dpbias} must be consistently $E/B$-purified before computing those pseudo-$C_\ell$s.
        
      \subsection{Beams and noise}\label{ssec:maths.beam_noise}
        In any practical scenario, the observed sky map will contain a mixture of signal ${\bf s}$ (i.e. the component whose power spectrum we are really interested in) and noise ${\bf n}$ (i.e. a stochastic contaminant that we do not have a template for). Furthermore, the signal component will often be smoothed on the smallest scales by an instrumental beam (although this latter effect is less relevant for galaxy surveys). We discuss how to deal with both of these complications here.
        
        \subsubsection{Beam deconvolution}\label{sssec:maths.beam_noise.beam}
          Including the effect of a spherically symmetric instrumental beam amounts to a simple redefinition of the mode-coupling matrix
          \begin{equation}
            {\sf M}^{s_as_b}_{\ell_1\ell_2}\rightarrow{\sf M}^{s_as_b}_{\ell_1\ell_2}W_{\ell_2}^{ab},
          \end{equation}
          where $W^{ab}_\ell$ is the product of the harmonic transform of the beams for maps $a$ and $b$. Note that since any continuous tangent vector field on the sphere must, at some location, be zero (the so-called hairy ball theorem), it is impossible to have an anisotropic beam that is uniform over the entire sphere.  However, it is possible that the beam is both anisotropic and spatially-varying. In this case, the pseudo-$C_\ell$ power spetrum estimation will produce biased estimates that will be suppressed by an \emph{effective} beam. Correcting for the beam in such cases can also be done analytically, but we leave this for future work.  In practice, if required, the
          suppression factor can be found by applying the PCL estimator to a set of mock input maps with the same underlying signal but  with and without application of the beam smoothing (to both data maps and contaminant maps) and taking the ratio.

        \subsubsection{Noise bias}
         In auto-correlation, the two-point correlation of the noise component ${\bf n}$ will contaminate our estimate of the power spectrum of ${\bf s}$, and we must correct for this noise bias. Depending on the statistical properties of ${\bf n}$, different strategies can be used to do this:
         \paragraph*{Homogeneous noise.}
           If the noise is homogeneous, and can be described by a known noise power spectrum ${\sf N}_\ell$, the procedure is simple: first, convolve the known noise power spectrum with the pseudo-$C_\ell$ mode-coupling matrix. Then, estimate the contribution from contaminant deprojection by using ${\sf N}_\ell$ in lieu of ${\sf C}^{ab}_\ell$ in Eqs. \ref{eq:dpbias}, \ref{eq:dp.g.aux} and \ref{eq:dp.f.aux}. The total noise bias is then the sum of both contributions, corrected for the pseudo-$C_\ell$ convolution using Eq. \ref{eq:theory_correct}.
          
         \paragraph*{Uncorrelated noise.}
           If the noise is inhomogeneous but uncorrelated (i.e. $\langle {\bf n}(\nv)\,{\bf n}^\dag(\nv')\rangle=\mI\,\sigma_n^2(\nv)\delta(\nv,\nv')$\footnote{Here $\delta(\nv,\nv')$ is the Dirac $\delta$-function on the sphere, and $\sigma_n^2$ is the local noise variance in one steradian.}), it is also possible to find an analytical estimate of the noise bias.
           \begin{itemize}
             \item It is not difficult to prove that the pseudo-$C_\ell$ of the noise contribution is given by
                   \begin{align}
                     {\rm PCL}_\ell({\bf n},{\bf n})&=\mI\,\int \frac{d\nv}{4\pi}v^2(\nv)\sigma_n^2(\nv)\\
                                                    &=\mI A^2\,\int \frac{d\nv}{4\pi}\sigma_n^{-2}(\nv),
                   \end{align}
                   where, in the second line, we have assumed that we use an inverse-variance noise weighting scheme, in which the weights map is $v(\nv)=A\,\sigma_n^{-2}(\nv)$.
             \item The additional bias from contaminant deprojection can also be computed analytically, starting from Eq. \ref{eq:dpbias0}, as:
                   \begin{align}\nonumber
                     \Delta{\sf C}^{nn}_\ell=&-2F_{ij}{\rm PCL}_\ell\left({\bf f}^i,v^2\sigma_n^2{\bf f}^j\right)\\
                     &+F_{ij}F_{rs}\left[\int d\nv\, v^2\sigma_n^2\,{\bf f}^{j\dag}{\bf f}^r\right]\,{\rm PCL}_\ell\left({\bf f}^i,{\bf f}^s\right).
                   \end{align}
           \end{itemize}

        \paragraph*{General case.} In general, if the noise is both non-white and inhomogeneous, the total noise bias can be estimated by averaging the result of applying the pseudo-$C_\ell$ estimator to a large number of noise realizations with the same noise properties as the data, assuming that those properties are sufficiently well characterised to produce those simulations. If that is not the case, a common approach to avoiding the noise bias altogether is to use only cross-correlations of data splits with independent noise contributions.

    \subsection{Flat-sky pseudo-\texorpdfstring{$C_\ell$}{Cl}s}\label{ssec:maths.flat}
      Using the curved-sky expression presented above and in Appendix \ref{app:shts} for the analysis of small sky patches, where the curvature of the sphere can be neglected, is numerically inefficient for two main reasons:
      \begin{enumerate}
        \item Curved-sky data formats usually store information about the full sphere, even if the data being stored is limited to a small sky patch. This can lead to an inefficient use of memory and storage space, because the standard numerical implementations of spherical transforms operate on full-sky data.
        \item Spherical harmonic transforms are notoriously slower than discrete Fourier transforms (DFTs), both due to the scaling of both algorithms (${\cal O}(N_{\rm pix}^{3/2})$ for SHTs vs. ${\cal O}(N_{\rm pix}\log(N_{\rm pix}))$ for DFTs) and because by SHTs involve operations over the full sphere.
        \item Fourier transforms over a flat patch offer exact quadrature, which makes them inherently more numerically stable\footnote{Note that it is also possible to achieve exact quadrature with certain spherical pixelization schemes (e.g. \cite{2011IJMPD..20.1053D}).}.
      \end{enumerate}
      For this reason, \nmt\, also supports the computation of power spectra from flat-sky maps. Appendix \ref{app:flat} describes the dictionary between spin-$s$ fields and their Fourier/harmonic coefficients defined in flat and curved skies, and we present the flat-sky versions of the pseudo-$C_\ell$ methods described in the previous sections here.
      
      \subsubsection{Standard pseudo-$C_\ell$}\label{sssec:maths.flat.pcl101}
        Using the definitions presented in Appendix \ref{app:flat}, it is easy to show that the Fourier coefficients of a masked field are given by \citep{2013MNRAS.435.2040L}
        \begin{equation}\label{eq:pcl_flat_masked}
          {\bf a}^v_{\bf l}\equiv{\cal D}(v\,{\bf a})^{s_a}_{\bf l}=\sum_{\bf k}\frac{\Delta {\bf k}^2}{2\pi}{\sf R}(s_a(\varphi_{\bf l}-\varphi_{\bf k}))\,{\bf a}_{\bf k}\,v_{{\bf l}-{\bf k}}.
        \end{equation}
        Here, ${\cal D}$ stands for a standard 2D discrete Fourier transform, and ${\sf R}$ is a rotation matrix in 2 dimensions.  Note that, in the context of flat-sky fields, the vector ${\bf l}$ is a 2D wave vector, and must not be confused with the abbreviation ${\bf l}\equiv(\ell,m)$ used in previous sections. The quantity ${\bf a}_{\bf l}$ is therefore the Fourier transform of the real-space flat-sky field ${\bf a}({\bf x})$, where ${\bf x}$ is the angular displacement with respect to a given reference point. Using Eq. \ref{eq:pcl_flat_masked} it is then straightforward to estimate the covariance of two masked fields at the same wavenumber:
        \begin{equation}\label{eq:flat_coup}
          \left\langle{\bf a}^v_{\bf l}{\bf b}^{w\dag}_{\bf l}\right\rangle=\sum_{\bf k}\frac{\Delta{\bf k}^2}{(2\pi)^2}{\sf R}(s_a\Delta\varphi){\sf C}^{ab}_{\bf k}{\sf R}^\dag(s_b\Delta\varphi)\,v_{{\bf l}-{\bf k}}w^*_{{\bf l}-{\bf k}},
        \end{equation}
        where $\Delta\varphi\equiv\varphi_{\bf l}-\varphi_{\bf k}$.
    
        At this stage it is natural to connect directly with the final bandpowers since, unlike in the curved-sky case, there is no natural minimal binning ($\Delta\ell=1$) for the un-masked power spectrum. In this case we will define a bandpower ${\sf B}_q$, indexed by an integer $q$ as the average of the covariance above over a set of ${\bf l}$ values $S_q$:
        \begin{equation}
          {\rm vec}\left[{\sf B}^{ab}_q\right]\equiv\frac{\Delta{\bf k}^2}{N_q}\sum_{{\bf l}\in S_q}{\rm vec}\left[{\bf a}^v_{\bf l}{\bf b}^{w\dag}_{\bf l}\right],
        \end{equation}
        where $N_q$ is the number of Fourier-space pixels covered by $S_q$. Typically, $S_q$ will be the set of pixels within an annulus defined by an interval $\left(|{\bf k}|_{\rm min},|{\bf k}|_{\rm max}\right)$.  Taking the expectation value of the equation above, we can relate this estimator to the true power spectrum as:
        \begin{equation}\label{eq:predflat}
          \langle{\rm vec}\left[{\sf B}^{ab}_q\right]\rangle=\sum_{{\bf l}\in S_b}N_q^{-1}\sum_{\bf k}{\sf M}^{s_a\,s_b}_{{\bf l}{\bf k}}\cdot\,{\rm vec}\left[\hat{\sf C}^{ab}_{\bf k}\right],
        \end{equation}
        where the un-binned mode-coupling matrix is:
        \begin{align}\label{eq:m00_flat}
          {\sf M}^{00}_{{\bf l}\,{\bf k}}&\equiv\frac{(2\pi)^2}{L_x^2L_y^2}v_{{\bf l}-{\bf k}}w^*_{{\bf l}-{\bf k}},\\\label{eq:m02_flat}
          {\sf M}^{02}_{{\bf l}\,{\bf k}}&\equiv\frac{(2\pi)^2}{L_x^2L_y^2}v_{{\bf l}-{\bf k}}w^*_{{\bf l}-{\bf k}}
          \left(
          \begin{array}{cc}
           {\rm c} & -{\rm s} \\
           {\rm s} &  {\rm c}
          \end{array}\right)\\\label{eq:m22_flat}
          {\sf M}^{22}_{{\bf l}\,{\bf k}}&\equiv\frac{(2\pi)^2}{L_x^2L_y^2}v_{{\bf l}-{\bf k}}w^*_{{\bf l}-{\bf k}}\left(
          \begin{array}{cccc}
           {\rm c}^2 & -{\rm c\,}{\rm s} & -{\rm c}\,{\rm s} & {\rm s}^2 \\
           {\rm c}\,{\rm s} & {\rm c}^2 & -{\rm s}^2 & -{\rm c}\,{\rm s} \\
           {\rm c}\,{\rm s} & -{\rm s}^2 & {\rm c}^2 & -{\rm c}\,{\rm s} \\
           {\rm s}^2 &  {\rm c}\,{\rm s} &  {\rm c}\,{\rm s} & {\rm c}^2
          \end{array}\right),
        \end{align}
        where $\rm{c} \equiv \cos2\Delta\varphi$ and $\rm{s}\equiv \sin2\Delta\varphi$. The mode-coupling matrix for the bandpowers is therefore given by:
        \begin{equation}\label{eq:modcoup_flat}
          \mathcal{M}^{s_as_b}_{qq'}=\sum_{{\bf l}\in S_q}N_q^{-1}\sum_{{\bf k}\in S_{q'}}{\sf M}^{s_as_b}_{{\bf l}\,{\bf k}}.
        \end{equation}
        
        At this point it is important to note that, with this procedure, computing the coupling matrix becomes an ${\cal O}(N_{\rm pix}^2)$ problem (or at best ${\cal O}(\ell_{\rm max}^2N_{\rm pix})$), which is worse than the ${\cal O}(\ell_{\rm max}^3$) pseudo-$C_\ell$ algorithm we described in the curved sky case. The key in that case is the orthogonality relation Eq. \ref{eq:3jorth}, which significantly simplifies the expressions \ref{eq:m00}-\ref{eq:m22}. In a flat sky, and in the presence of a discrete and finite Cartesian grid, it is not possible however to average over azimuthal Fourier degrees of freedom at constant $l$ in order to obtain a numerically stable version of the mask power spectrum for arbitrary weight maps $v$ and $w$. This would not be a problem in the continuum limit ($(\Delta x,\Delta y)\rightarrow0$), where similar simplifying relations can be found to vastly improve the computational efficiency of the method. Taking the continuum approximation is unfortunately not accurate enough in practice, as described in \cite{2016arXiv161204664A}, and a fully discrete approach is usually necessary. This slower performance is normally compensated by the smaller pixel numbers and faster DFTs that can be used in the flat sky, so that the flat-sky approximation is still preferable over a curved-sky treatment for small sky patches. Appendix \ref{app:flat.pcl} discusses these issues in detail, providing the continuum-limit expressions mentioned above.
   
      \subsubsection{Contaminant cleaning}\label{sssec:maths.flat.deproj}
        Using the same notation as in Section \ref{ssec:maths.deproj}, the contaminant-cleaned version of ${\bf a}$ is simply given by(c.f. Eqs. \ref{eq:clean} and \ref{eq:clean_f}):
        \begin{align}
          &{\bf a}^c({\bf x})={\bf a}^v({\bf x})-{\bf f}^i({\bf x})F_{ij}\sum_{\bf x}\Delta{\bf x}^2\,{\bf f}^{j\dag}({\bf x}'){\bf a}^v({\bf x}'),\\
          &{\bf a}^c_{\bf l}={\bf a}^v_{\bf l}-{\bf f}^i_{\bf l}F_{ij}\sum_{\bf k}\Delta{\bf k}^2\,{\bf f}^{j\dag}_{\bf k}{\bf a}^v_{\bf k},
        \end{align}
        where $({\sf F}^{-1})_{ij}=\sum_{\bf x}\Delta{\bf x}^2{\bf f}^{i\dag}({\bf x}){\bf f}^j({\bf x})$.
    
        The deprojection bias also takes a similar form (c.f. Eq. \ref{eq:dpbias}):
        \begin{align}\nonumber
          \Delta{\sf C}^{ab}_{\bf l}=&-G_{ij}\,\tilde{\bf g}^j_{\bf l}\,{\bf g}^{i\dag}_{\bf l}-F_{ij}\,{\bf f}^i_{\bf l}\,\tilde{\bf f}^{j\dag}_{\bf l}\\\label{eq:dpbias_flat}
          &+F_{ij}G_{rs}\left[\sum_{\bf x}\Delta{\bf x}^2{\bf f}^{j\dag}({\bf x})\tilde{\bf g}^r({\bf x})\right]\,{\bf f}^i_{\bf l}{\bf g}^{s\dag}_{\bf l},
        \end{align}
        where
        \begin{align}\label{eq:dp.g.aux_flat}
          &\tilde{\bf g}^i_{\bf l}\equiv{\cal D}\left[v({\bf x}){\cal D}^{-1}\left[{\sf C}^{ab}_{{\bf l}_1}{\cal D}\left[w{\bf g}^i\right]^{s_b}_{{\bf l}_1}\right]^{s_a}_{\bf x}\right]^{s_a}_{\bf l},\\\label{eq:dp.f.aux_flat}
          &\tilde{\bf f}^i_{\bf l}\equiv{\cal D}\left[w({\bf x}){\cal D}^{-1}\left[{\sf C}^{ab\dag}_{{\bf l}_1}{\cal D}\left[v{\bf f}^i\right]^{s_a}_{{\bf l}_1}\right]^{s_b}_{\bf x}\right]^{s_b}_{\bf l}.
        \end{align}

      \subsubsection{$E$ and $B$ purification}\label{sssec:maths.flat.pureb}
        The logic behind $E/B$ purification is the same in flat and curved skies, and we will not repeat it here. The analogue of a pure $B$-mode (Eq. \ref{eq:pureb}) in the flat-sky approximation is:
        \begin{equation}
          B^p_{\bf l}=\left(\tilde{P}_2\right)^B_{\bf l}+2l^{-1}\left(\tilde{P}_1\right)^B_{\bf l}+l^{-2}\left(\tilde{P}_0\right)^B_{\bf l},
        \end{equation}
        (and a similar relation for the pure $E$ component), where $P_n=(\eth^{2-n}w)^*(Q+iU)$, and $w$ is the sky mask. The derivatives of $w$ can be taken by using the following relation:
        \begin{equation}
          \eth^nw={\cal D}^{-1}\left[\left(-l^n\,w_{\bf l},0\right)\right],\\
        \end{equation}
    
        Note that, in the flat sky, a mathematically (but not computationally) simpler relation for the pure component can be found, given by:
        \begin{equation}
          B^p_{\bf l}=\int\frac{d{\bf k}^2}{2\pi}B_{\bf k}w_{{\bf l}-{\bf k}}\frac{k^2}{\ell^2},
        \end{equation}
        and similarly for $E$ modes. Comparing with Eq. \ref{eq:flat_coup} (where  ${\sf R}$ is defined in Eq. \ref{eq:flat_sht}), it is easy to see that the key to work out the expressions for the pure-$E$ and $B$ coupling matrices is simply to replace all factors of ${\rm c}$ and ${\rm s}$ with $k^2/\ell^2$ and 0 respectively in Equations \ref{eq:m02_flat} and \ref{eq:m22_flat}. Again, under the assumption that purification takes place after contaminant deprojection, the only modification to the expression for the deprojection bias is to make sure that all functions of ${\bf l}$ in Eq. \ref{eq:dpbias_flat} are consistently purified (but not any of the other fields appearing there or in Eqs. \ref{eq:dp.g.aux_flat} and \ref{eq:dp.f.aux_flat}).

  \section{Code structure}\label{sec:structure}
    All of the methods described above have been implemented in a software package called \nmt\footnote{The code source is hosted in \url{https://github.com/LSSTDESC/NaMaster}, and its documentation can be found at \url{http://namaster.readthedocs.io/}.}. We briefly describe the code structure here.
    
    The code is written in C and wrapped into {\tt python} to facilitate its use and its combination with other software libraries for astronomy. \nmt~itself is based on a number of these libraries:
    \begin{itemize}
      \item {\tt HEALPix} \citep{2005ApJ...622..759G} is currently the only pixelization scheme supported for curved-sky calculations. Through {\tt HEALPix}, \nmt~also makes use of {\tt cfitsio}~\citep{1999ASPC..172..487P}.
      \item The basic curved-sky operation carried out by \nmt~is the spherical harmonic transform. For this, the code makes use of {\tt libsharp}~\citep{2013A&A...554A.112R}.
      \item In flat sky, SHTs are replaced by 2D DFTs, for which \nmt~uses the {\tt FFTW} library~\citep{FFTW05}.
      \item The GNU Scientific Library \cite{gslbook} is also used for some numerical calculations.
    \end{itemize}
    
    As described in the previous sections, some of the operations (e.g. SHTs, deprojection, purification) take place in each field individually, while others depend on the correlation of a pair of fields (e.g. PCL, mode-coupling matrix, bias deprojection). It is therefore inefficient to carry out the per-field operations on every field every time a power spectrum is computed, especially if a large number of fields are being cross-correlated. For this reason, at the \texttt{python} level, \nmt~is structured around two main classes that incorporate the two different types of operations we just described. Each of these classes are associated to a counterpart structure in C:
    
    \paragraph*{NmtField.} This class stores all the necessary information about one individual observed spin-0 or spin-2 field. A {\tt NmtField} object is defined by a sky mask or weights map (e.g. $v$ in Eq. \ref{eq:aobs}), 1 or 2 sky maps (depending on the spin), corresponding to an observation of the field on the sky (${\bf a}^{\rm obs}$ in Eq. \ref{eq:aobs}), a set of contaminant templates (${\bf f}_i$ in Eq. \ref{eq:aobs}), and the choice to purify or not the $E$ or $B$ mode component of the field. Once initialized, this class carries out the following operations:
    \begin{itemize}
      \item SHT or DFT of the field (Eqs. \ref{eq:sht1} or \ref{eq:dft}).
      \item Contaminant deprojection (Eq. \ref{eq:clean}).
      \item $E$-mode and/or $B$-mode purification (Eq. \ref{eq:pureb}).
    \end{itemize}
    
    \paragraph*{NmtWorkspace.} This class stores the information necessary to compute an unbiased estimate of the power spectrum of two {\tt NmtField}s. The main objective of these objects is to compute mode-coupling matrices. {\tt NmtWorkspace}s are equipped with read/write methods to avoid redoing these calculations when estimating the power spectra of several fields with a common set of masks (e.g. for a large number of tomographic bin or simulations).
    
    \nmt~also includes other convenience classes to handle bandpowers, covariance matrices etc. as well as routines to carry out useful operations (e.g. mask apodization, Gaussian simulations) and to wrap up commonly used sequences of operations (e.g. pseudo-$C_\ell$ computation followed by binning and deconvolution).
    
    The workflow for a typical pseudo-$C_\ell$ run would be:
    \begin{lstlisting}
from pymaster import *

...

f2 = NmtField(mask, [map_q,map_u],
            templates = [[dust_q,dust_u]],
            purify_b = True, beam=beam_ell)
f0 = NmtField(mask, [map_t])

b = NmtBin(nside, nlb = 10)

wsp = NmtWorkspace()
wsp.compute_coupling_matrix(f0, f2, b)

cl_bias = deprojection_bias(f0, f2, cl_theory)

cl_coupled = compute_coupled_cell(f0, f2)

cl_decoupled = wsp.decouple_cell(cl_coupled, cl_bias)
    \end{lstlisting}
    Briefly:
    \begin{enumerate}
      \item In line 1 we import \nmt's python module.
      \item All preliminary I/O operations (reading maps, masks etc.) take place implicitly in line 3.
      \item In line 5 we create a field with spin 2. The field is defined in terms of a mask and maps of the $(Q,U)$ Stokes parameters. We also provide a set of contaminants to deproject, an instrumental beam and request for the field's $B$-modes to be purified. Deprojection and purification will take place at this stage in that order, as explained in Section \ref{ssec:maths.pureb}. The field's SHT is also estimated upon initialization.
      \item We define a second field in line 8. This one is a simpler spin-0 field with no contaminants or beam.
      \item We define the output bandpower structure in line 10. In this case we will use bins of 10 multipoles, although more general schemes are supported.
      \item In lines 12 and 13 we use an {\tt NmtWorkspace} object to compute the mode-coupling matrix.
      \item The deprojection bias is computed in line 15 using a best-guess for the true underlying power spectrum.
      \item The pseudo-$C_\ell$ is computed in line 17.
      \item The deconvolved bandpowers are finally estimated using the workspace in line 19. Note that the mode-coupling matrix must have been precomputed in line 13 for this to be successful.
    \end{enumerate}
    
    Further details regarding the different features of \nmt~can be found in the online documentation \url{http://namaster.readthedocs.io/}.

  \section{Code performance and validation} \label{sec:validation}
    The core of our code-validation suite is based on two science examples. These examples are not meant to be realistic mocks of actual data, however, they do contain the dynamic range and noise-levels of a typical experimental set-up that we expect in the next decade, while providing us with an exact, known true power spectrum to compare against. This section presents these two examples and the results of this validation in terms of accuracy and computational performance. Results will be presented for both curved-sky and flat-sky realizations. Note that, in all cases, we use the true signal power spectra as input to compute the bias from contaminant deprojection (see Eqs. \ref{eq:dp.g.aux} and \ref{eq:dp.f.aux}). This allows us to verify that the code works as expected, and to isolate any residual bias associated with software bugs, rather than an imperfect guess of the true power spectrum.

    \subsection{Validation suite} \label{ssec:validation.suite}
      We base our validation suite on two science cases of relevance for the most relevant next-generation cosmology experiments. We describe these here.
      
      \subsubsection{Galaxy clustering and weak lensing} \label{sssec:validation.suite.lss}
        \begin{figure}
          \centering
          \includegraphics[width=0.99\columnwidth]{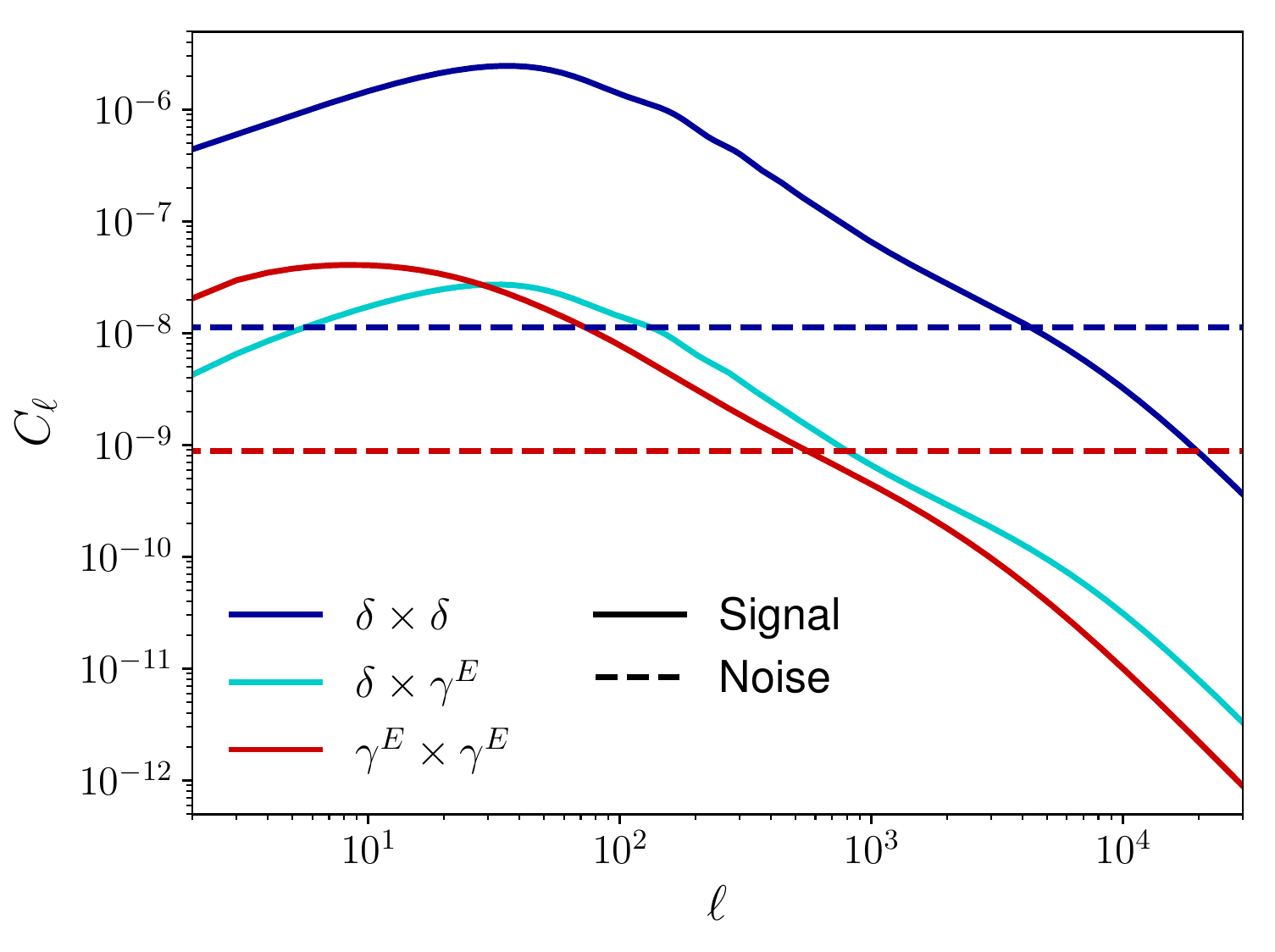}
          \caption{Signal and noise power spectra used in the LSS validation set. Dark blue, cyan and red curves show the spectra associated with the $\delta$-$\delta$, $\delta$-$\gamma_E$ and $\gamma_E$-$\gamma_E$ correlations respectively, where $\gamma_E$ is the lensing $E$ mode. All other cross-correlations are zero. The dashed lines show the noise power spectra, associated with shot noise (red) and intrinsic shape scatter (blue).} \label{fig:lss_cls}
        \end{figure}
        \begin{figure*}
          \centering
          \includegraphics[width=0.99\textwidth]{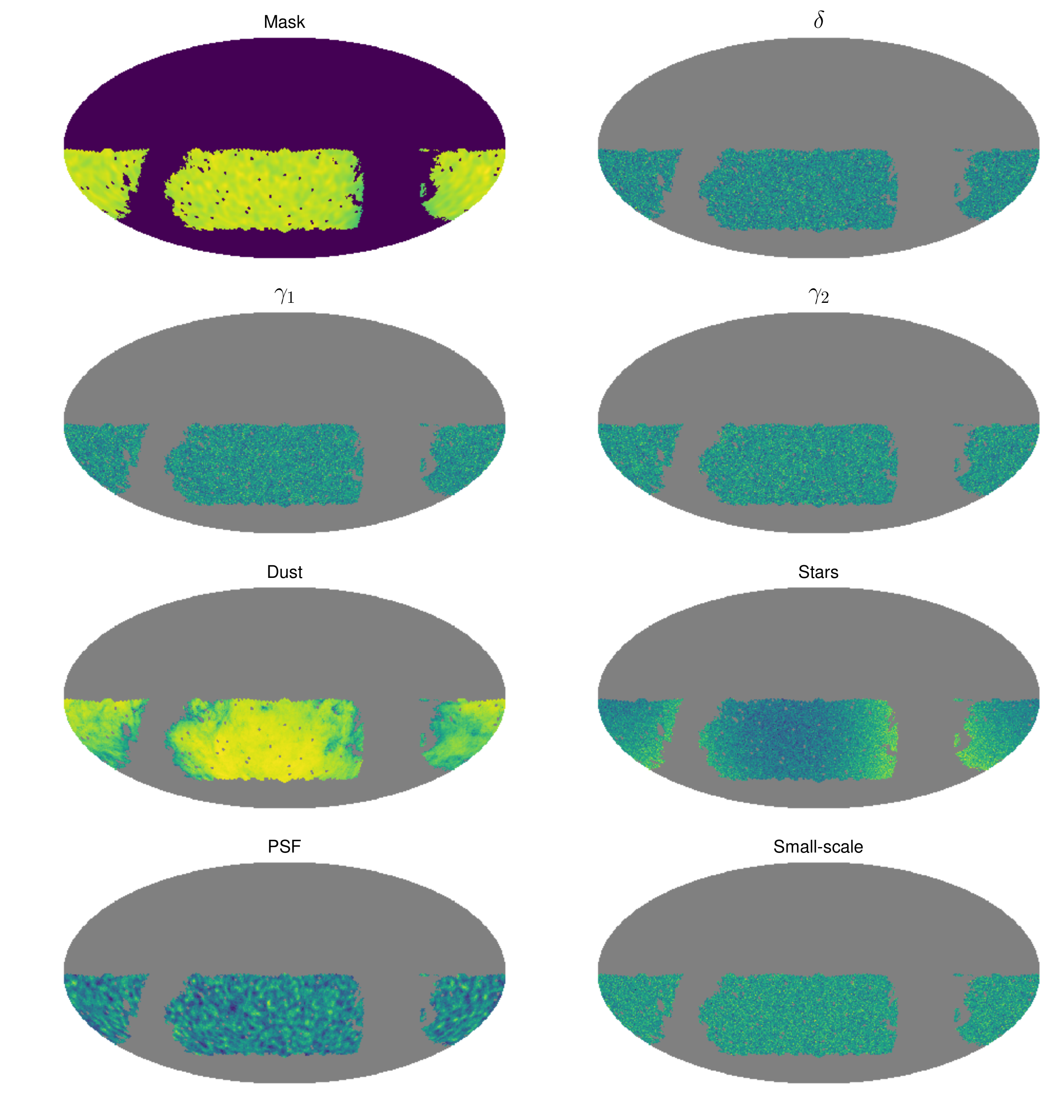}
          \caption{Inputs for the large-scale structure, curved-sky validation set. From top to bottom and left to right: sky mask (1), Gaussian realization of the galaxy overdensity $\delta$ (2) and cosmic shear $\gamma_i$ (3,4), dust contaminant for $\delta$ (5), star contaminant for $\delta$ (6), PSF contaminant for $\gamma$ (7, $Q$ component) and small-scale contaminant for $\gamma$ (8, $Q$ component).}
          \label{fig:lss_mocks_sph}
        \end{figure*}
        \begin{figure*}
          \centering
          \includegraphics[width=0.9\textwidth]{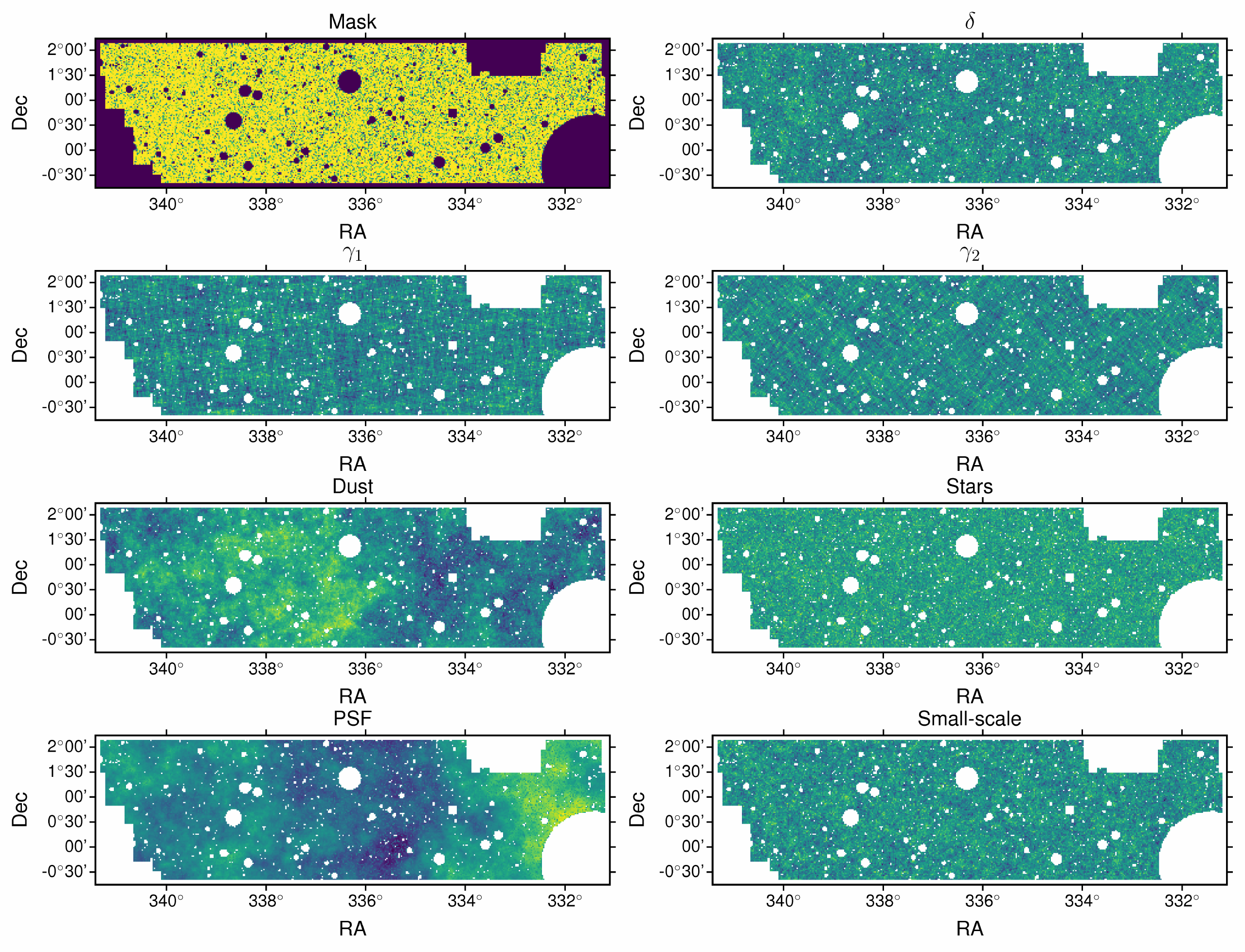}
          \caption{Same as Fig. \ref{fig:lss_mocks_sph} for the large-scale structure, flat-sky validation set.}
          \label{fig:lss_mocks_flat}
        \end{figure*}
        \begin{figure}
          \centering
          \includegraphics[width=0.99\columnwidth]{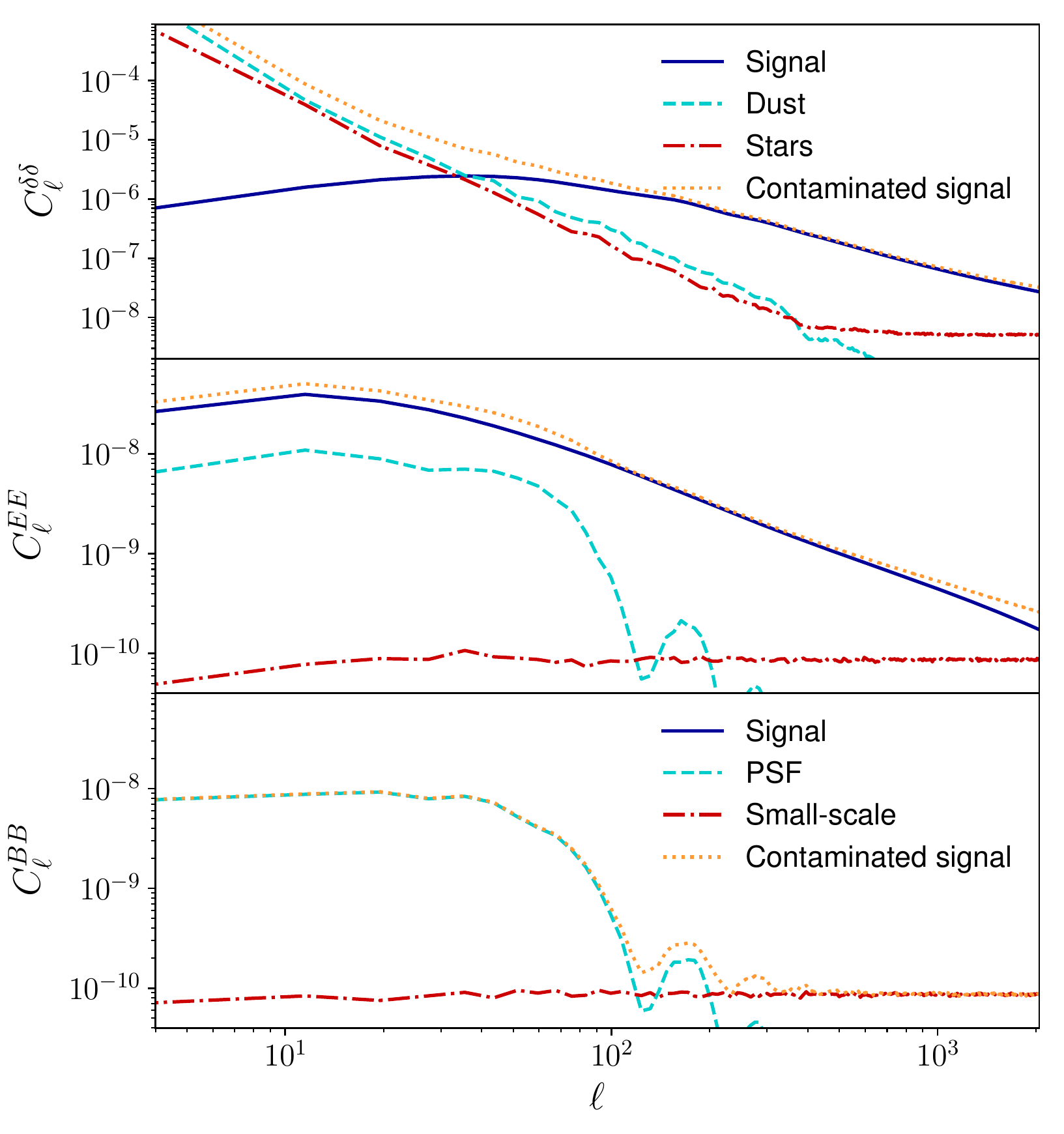}
          \caption{{\sl Top:} signal power spectrum (solid dark blue) for the auto-correlation of the galaxy overdensity, as well as the contaminants associated to dust (dashed cyan) and stars (dot-dashed red). {\sl Middle:} $E$-mode lensing power spectrum (solid dark blue) and the spectra of the PSF and small-scale contaminants used in our validation set (dashed cyan and dot-dashed red respectively). {\sl Bottom:} same as middle panel for the $B$-mode power spectrum. Note that in this case the input signal power spectrum is $0$. In all cases the orange dotted line shows the combination of signal and contaminants.} \label{fig:lss_cont_cls}
        \end{figure}
        The first validation suite exemplifies the use of pseudo-$C_\ell$ algorithms in the analysis of large-scale structure data. Current and next-generation imaging surveys such as DES~\citep{2016MNRAS.460.1270D},
        KiDS~\citep{2017A&A...604A.134D}, HSC~\citep{2018PASJ...70S...4A}, LSST~\citep{2008arXiv0805.2366I} and Euclid~\citep{2011arXiv1110.3193L} will constrain the growth and geometry of structure through cosmic time by using the overdensity of galaxies as a proxy for the matter inhomogeneities (a spin-0 field) and their correlated shape distortions (a spin-2 field at first order), mostly caused by weak lensing. These measurements will mostly follow a tomographic approach, using the angular auto- and cross-power spectra between these observables across different bins of photometric redshift \citep{2017arXiv170609359K,2018MNRAS.474.4894J,2017arXiv170801530D}.
        
        Large-scale structure data are often characterized by inhomogeneous sky coverages and complex mask structures, associated with the presence of bright stars and observational artifacts. They are also affected by many sky contaminants, associated to Galactic sources (dust absorption, star density) and observing conditions (seeing, airmass etc.) as discussed in \citet{2016ApJS..226...24L}. The typically flatter power spectra for both galaxy clustering and lensing (e.g. when compared with the damping tail of CMB temperature fluctuations), makes the pseudo-$C_\ell$ approach ideal for this kind of analysis \citep{2013MNRAS.435.1857L,2016ApJS..226...24L}. This science case therefore allows us to validate several aspects of the code simultaneously:
        \begin{itemize}
          \item The unbiased reconstruction of the 6 possible correlations between spin-0 and spin-2 fields.
          \item The inclusion of inhomogeneous noise and the correction of the associated noise bias.
          \item The effects of deprojection for several contaminant templates in different-spin fields.
          \item The code's ability to handle complex masks and high-resolution pixelization.
        \end{itemize}

        For this validation suite we therefore use a mock dataset mimicking a typical observation of a LSST-like survey in a particular redshift bin. Slightly different set-ups are used for the curved-sky and flat-sky validation suites. The characteristics of these data are as follows:
        \begin{itemize}
          \item {\bf Signal:} we generate maps of the galaxy overdensity $\delta$ and of the shear field $(\gamma_1,\gamma_2)$ for a redshift bin centred at $z\simeq1$ with a width $\Delta z\sim0.1$ and Gaussian photo-$z$ tails with width $\sigma_z=0.06$. For $\delta_g$ we assume unit bias $b(z)=1$ and no contributions from redshift-space distortions or magnification. For $\gamma$, we assume no contribution from intrinsic alignments. The maps are generated as Gaussian realizations of the corresponding angular power spectra computed with the Core Cosmology Library\footnote{\url{https://github.com/LSSTDESC/CCL}} for cosmological parameters compatible with the 2015 Planck measurements~\citet{2016A&A...594A..13P}. These power spectra are shown in Fig. \ref{fig:lss_cls}. Sample realizations of $\delta$ and $\gamma$ can be seen in Figures \ref{fig:lss_mocks_sph} and \ref{fig:lss_mocks_flat} for the curved-sky and flat-sky cases respectively.
          \item {\bf Mask:} for the curved-sky data, we use a mask built as a combination of the sky coverage output by the LSST {\tt OpSim} database~\citep{2014SPIE.9150E..15D} and a more conservative galactic cut based on the reddening map of \citet{1998wfsc.conf..297S}. To add an extra layer of complexity, we drill 100 additional 1-degree holes on the resulting combined mask. We also explore the improvement stemming from the down-weighting of higher-noise in an inverse-variance way in the presence of inhomogeneous noise (see description below), which is the reason why the default version of this mask is not binary, but traces the inverse of the noise variance (see top left panel of Fig. \ref{fig:lss_mocks_sph}). For the flat-sky data, we use a high-resolution mask constructed from the bright-object mask used for the first public data release from the HSC collaboration~\citep{2018PASJ...70S...4A}. The chosen footprint corresponds to the VVDS field, covers approximately 20 deg$^2$ and contains structure on a wide range of scales. As can be seen in the top-left panel of Figure~\ref{fig:lss_mocks_flat}, the mask is not binary, and its value at each pixel correspond to the pixel's fractional masked area.
          \item {\bf Contaminants:} we contaminate the signal maps described above with several different types of residuals. In all cases we add contaminants linearly, to avoid deviating from the model described in Section \ref{ssec:maths.deproj}. The contaminant amplitudes are chosen so that any residual contamination in the relevant range of multipoles will be detected statistically by the validation suite. Maps of these contaminants are shown in the 4 bottom panels of Figure~\ref{fig:lss_mocks_sph} and Figure~\ref{fig:lss_mocks_flat}, and their impact on the angular power spectrum is presented in Fig. \ref{fig:lss_cont_cls} for the curved-sky case.
          \begin{itemize}
            \item {\sl Dust.} For the curved-sky suite, we include a linear contaminant for $\delta_g$ proportional to the fluctuations around the mean of the dust reddening map of~\citet{1998wfsc.conf..297S}. This contamination affects mostly the largest scales, and simulates the effect of an imperfect correction of the dust absorption on the galaxy number density. The amplitude of the dust fluctuations is chosen so that their power spectrum within the unmasked region be $\sim20\%$ of the galaxy power spectrum at $\ell\sim120$. For the flat-sky suite, we generate a flat-sky Gaussian realization of a field with a power spectrum $C_\ell\propto\ell^{-2.4}$, corresponding to the large-$\ell$ behaviour of the curved-sky dust map. The amplitude was chosen such that the contaminant power spectrum be $\sim10\%$ of the signal at $\ell\sim400$.
            \item {\sl Stars.} We add a second contaminant to $\delta_g$ associated with star contamination. For the curved-sky realizations this contaminant is proportional to the fluctuation around the mean of a star density map generated using the LSST {\tt CatSim} catalog~\citep{2014SPIE.9150E..14C}. We modify this contaminant slightly by adding a white-noise component that dominates the small-scale spectrum beyond $\ell\simeq400$. We do this in order to simulate the noise-like distribution of stars on small scales as well as to include a contaminant source that, unlike dust, can surpass the signal on small scales. The amplitude of this contaminant is chosen such that its power spectrum is about 3\% of the signal at $\ell\sim400$.
            \item {\sl PSF fluctuations.} We add a large-scale contaminant for weak lensing in the form of fluctuations in the image point-spread function (PSF) with a characteristic scale corresponding to the telescope field of view. We generate this contaminant as a spin-2 field with equal $E$ and $B$-mode amplitude and a power spectrum
            \begin{equation}
              C^{EE}_\ell=C^{BB}_\ell\propto\left[2\frac{J_1(\ell\theta_{\rm FoV})}{\ell\theta_{\rm FoV}}\right]^2,
            \end{equation}
            where $J_1(x)$ is the order-1 Bessel function, and $2J_1(\ell\theta)/\ell\theta$ is the Fourier transform of a circular aperture with radius $\theta$. For this we assume $\theta_{\rm FoV}=1.75^\circ$, and therefore we try to mimic the signature of PSF fluctuations that vary between different pointings with a 3.5$^\circ$ diameter. The amplitude of the power spectrum is chosen so that the contaminant amounts to $30\%$ of the signal power spectrum at $\ell\sim50$. We use the same prescription for both the curved-sky and the flat-sky cases. 
            \item {\sl Small-scale contamination.} Finally we include an additional small-scale contaminant to weak lensing with a flat power spectrum, unit $E$-$B$ ratio, and an amplitude fixed so the contaminant's power spectrum is $20\%$ of the signal at $\ell=500$. Although we do not identify this contaminant with any specific source, it could arise from e.g. the impact of stars or blending on shape measurement.
          \end{itemize}
          \item {\bf Noise:} for the curved-sky case we generate inhomogeneous noise realizations. To do this, we first generate maps with Gaussian, white and homogeneous noise with a pixel variance given by $1/n_{\rm pix}$ for $\delta$ and $\sigma_\gamma^2/n_{\rm pix}$ for each shear component, where $n_{\rm pix}$ is the average number of sources in a pixel's area, and $\sigma_\gamma=0.28$ is the intrinsic ellipticity scatter per component. These noise realizations are then locally rescaled by a factor $f(\nv)=f_{\rm star}(\nv)\,f_{\rm depth}(\nv)$ made up of two components:
          \begin{itemize}
            \item $f_{\rm star}(\nv)$ is a function of the local star density modeled with the same star map used to generate the galaxy clustering contaminant described above. The function interpolates between $f_{\rm star}=1$ for low star densities and $f_{\rm star}=1.5$ for high star densities. The specific functional form used is
            \begin{equation}
              f_{\rm star}=\frac{r_{\rm star}-4}{2r_{\rm star}-4},
            \end{equation}
            where $r_{\rm star}=N_{\rm star}/{\rm max}(N_{\rm star})$ and $N_{\rm star}$ is the star density map.
            \item $f_{\rm depth}(\nv)$ is a map of stochastic depth variations generated as a Gaussian random field with mean 1 and fluctuations with a typical amplitude of 3-4\% and a characteristic scale of $\theta_{\rm FoV}=1.75^\circ$.
          \end{itemize}
          The logic behind these simulated noise maps is that both the presence of stars and fluctuations in the survey depth will modulate the number density of observed objects, and with it the noise in both $\delta$ and $\gamma$. A more optimal estimate of the power spectrum could therefore be achieved by using an inverse-variance weighting scheme in which the mask is inversely proportional to $f(\nv)$.
          
          For simplicity, the flat-sky simulations use only the homogeneous part of these noise realizations.

          \item {\bf Pixelization:} the main suite of curved-sky simulations used to validate the code use \texttt{HEALPix} resolution parameter $N_{\rm side}=1024$, corresponding to a pixel size of $\sim3.4'$. This resolution allows us to validate the code on a large number of realizations without facing strong computational challenges on the main scales of interest ($\ell\lesssim2000$).  As noted in Appendix~\ref{app:full_sky_nside}, however, the code has also been validated with a smaller suite of high-resolution simulations with $N_{\rm side}=4096$, and we have been able to run it successfully up to a maximum $N_{\rm side}=8192$ ($\theta_{\rm pix}\sim0.4'$). The flat-sky simulations use square pixels of $0.6$ arcmin a side.
        \end{itemize}
        
        We should note that, unlike the simulated maps used here, the true galaxy distribution is markedly non-Gaussian. However, none of the expressions presented in Section \ref{sec:maths} for the pseudo-$C_\ell$ method assume Gaussianity, and therefore the method will deliver unbiased estimates of the harmonic-space two-point function for non-Gaussian fields too.
        
      \subsubsection{CMB $B$-modes from the ground} \label{sssec:validation.suite.cmb}
        \begin{figure}
          \centering
          \includegraphics[width=0.99\columnwidth]{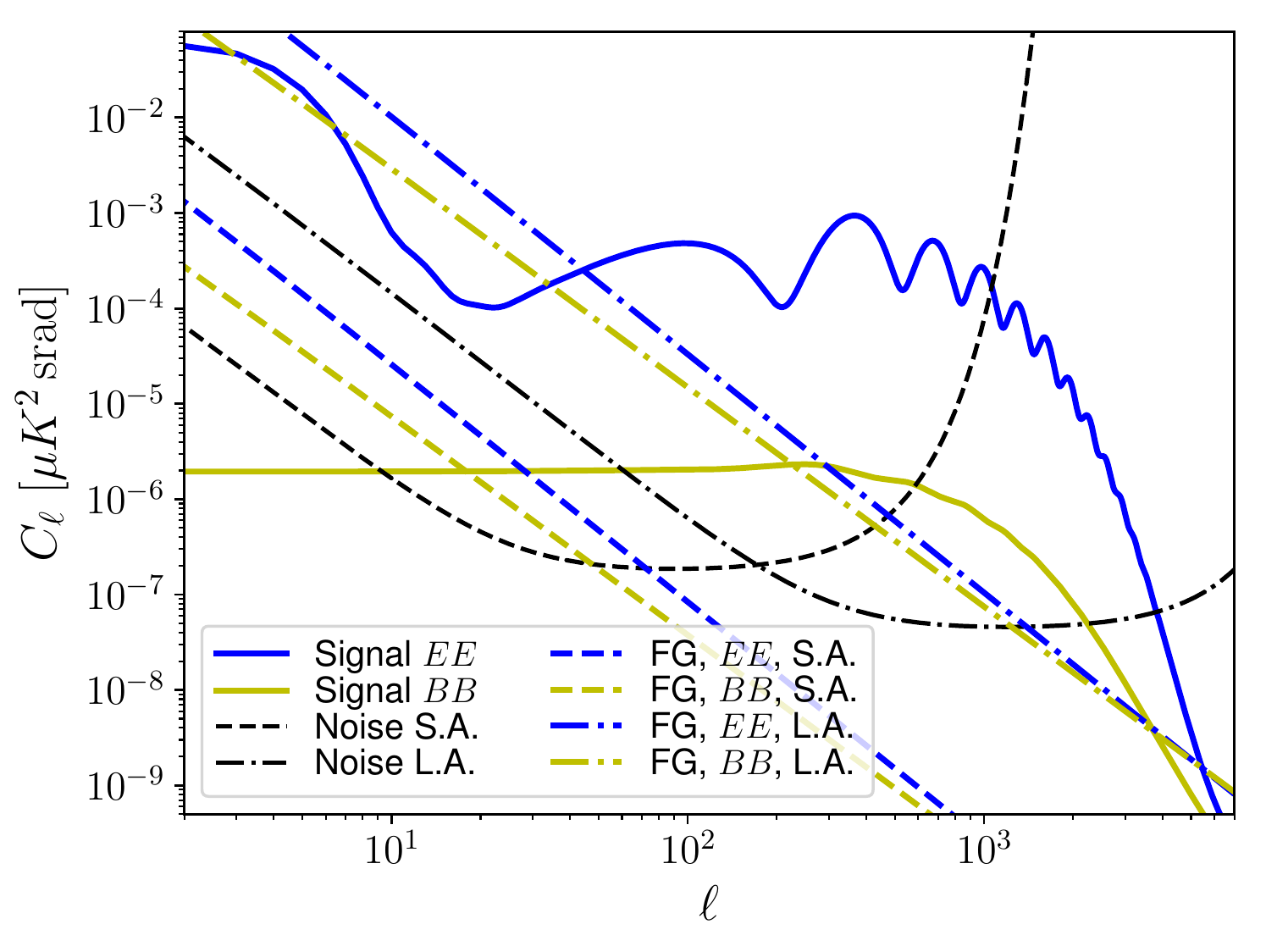}
          \caption{$E$-mode and $B$-mode power spectra (blue and yellow respectively) for the cosmological CMB signal (solid lines), for the curved-sky foreground contaminants (dashed lines) and for the flat-sky foreground contaminants (dot-dashed lines). The noise power spectra used for the full- and flat-sky simulations are shown as black dashed and dot-dashed lines respectively. These noise curves are intended to depict two different types of experiments: a small-aperture telescope (S.A., dashed lines) targetting primordial tensor modes, and a large-aperture telescope (L.A., dot-dashed lines) targetting the small-scale lensing $B$-modes.} \label{fig:cmb_cls}
        \end{figure}
        \begin{figure*}
          \centering
          \includegraphics[width=0.99\textwidth]{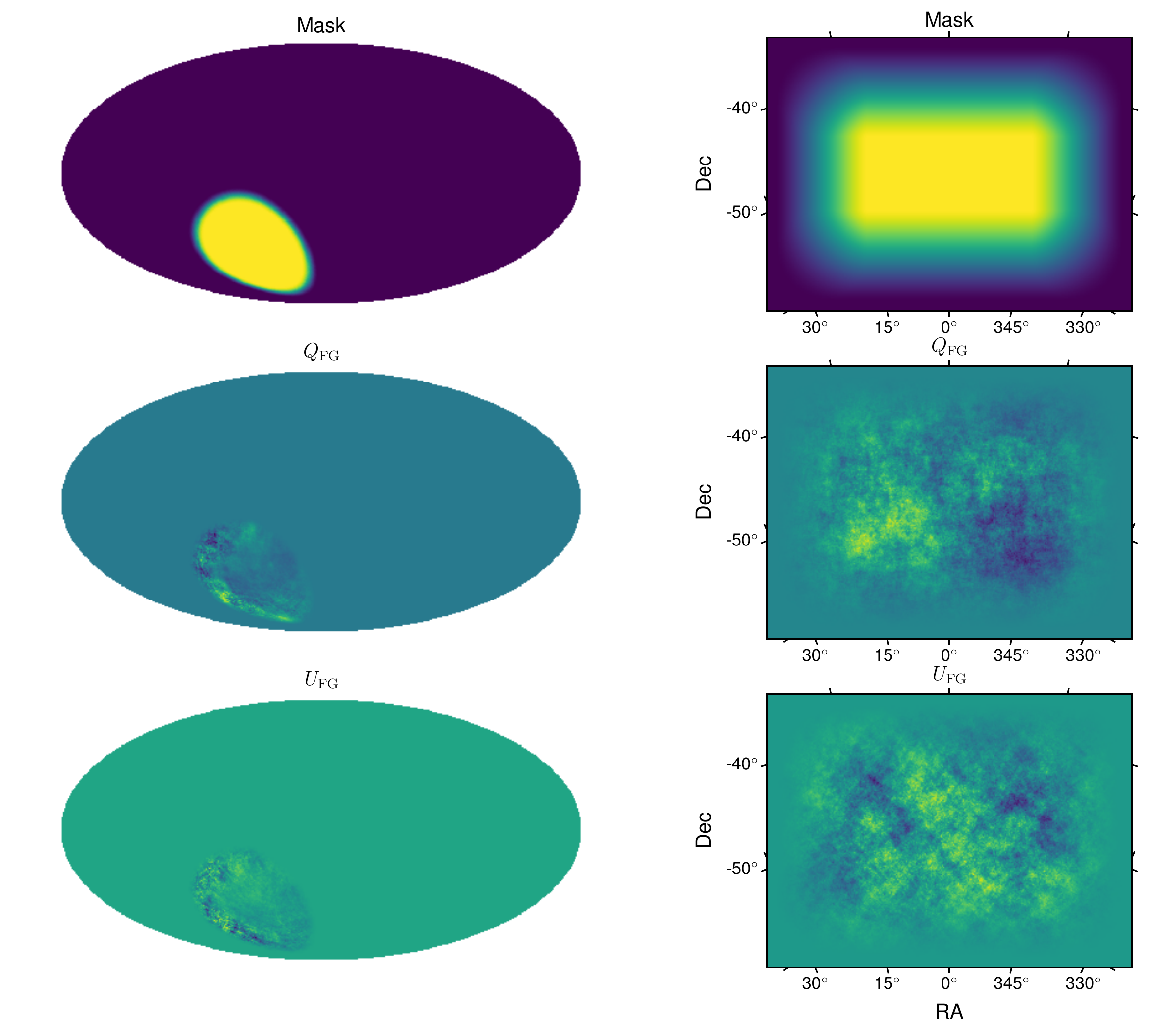}
          \caption{{\sl Left panels:} mask (top), and $Q$ and $U$ components (middle, bottom) of the foreground contaminant used in the CMB curved-sky validation set. {\sl Right panels:} same as left panels for the flat-sky validation set. All maps have units of $\mu K_{\rm CMB}$.}
          \label{fig:cmb_mocks}
        \end{figure*}
        Our second example corresponds to the measurement of $B$-modes in the polarized CMB emission, closely associated with the presence of primordial gravitational waves. This is one of the main science cases targeted by current ground-based Stage-3 experiments such as BICEP/Keck~\citep{2016PhRvL.116c1302B}, Advanced ACTPol~\citep{2016SPIE.9910E..14D,2017JCAP...06..031L} or the Simons Array~\citep{2016JLTP..184..805S,2017ApJ...848..121P}, as well as next-generation facilities such as the Simons Observatory~\citep{2018arXiv180807445T} or CMB Stage-4~\citep{2016arXiv161002743A}. Potential future space missions, such as LiteBIRD~\citep{2018JLTP..tmp..124S} will be able to make low-resolution, curved-sky maps of the polarized CMB to put constraints on the primordial $B$-mode signal from its reionization bump, at $\ell<30$, and will likely make use of pixel-based likelihoods and optimal quadratic methods. Ground-based experiments, on the other hand, will be systematics-limited at low $\ell$, and will be able to make higher-resolution maps, for which pseudo-$C_\ell$ methods are better suited.
        
        The measurement of CMB $B$ modes is affected by two challenges: the presence of foreground residuals left in the data by any component separation method, and the much larger amplitude of the $E$-mode signal, which can dominate the uncertainties in the $B$-mode power spectrum due to $E$-to-$B$ leakage. This science case therefore allows us to validate the code when both contaminant deprojection and $B$-mode purification are required, as well as in the presence of a non-negligible instrumental beam.
        
        Our first validation suite is therefore inspired on a Southern-Hemisphere, high-sensitivity polarization experiment with the following characteristics:
        \begin{itemize}
          \item {\bf Signal:} we generate $Q$ and $U$ maps of the CMB as Gaussian realization of a lensed power spectrum corresponding to the best-fit parameters from Planck~\citep{2016A&A...594A..13P}. For simplicity we assume no delensing and zero primordial $B$-modes, so that the $B$-mode power spectrum is entirely due to the effects of CMB lensing. The signal power spectra are shown in Fig. \ref{fig:cmb_cls}.
          \item {\bf Mask:} for the curved-sky case we assume a survey targeting a 4000 deg$^2$ footprint. The corresponding mask is analytically apodized using the ``$C^2$'' apodization method of \cite{2009PhRvD..79l3515G} with an apodization scale of 10 degrees. This apodization method is based on smoothing the transition at the mask edge with a sinusoidal function that interpolates between 0 and 1 in a differentiable manner. The observed area encompasses one of the lowest-foreground regions observable from the Southern Hemisphere. For the flat-sky case we assume a smaller sky fraction of 500 deg$^2$ with a similar apodization. The assumed masks are shown in the top panels of Figure~\ref{fig:cmb_mocks}.
          \item {\bf Noise:} for the curved-sky and flat-sky case we assume white noise levels of 1 and 0.5$\mu K\,{\rm arcmin}$ in intensity respectively. We also include a $1/f$-like component associated with atmospheric noise with a characteristic scale $\ell_{\rm knee}$ and tilt $\alpha_{\rm knee}$ such that the noise power spectrum can be written as
          \begin{equation}
            N_\ell=\sigma_N^2\left[1+\left(\frac{\ell_{\rm knee}}{\ell}\right)^{\alpha_{\rm knee}}\right],
          \end{equation}
          where $\sigma_N^2$ is the white-noise variance in one steradian. We use $\alpha_{\rm knee}=2.4$ and $\ell_{\rm knee}=10$ and 300 for the curved-sky and flat-sky cases respectively. The beam-deconvolved noise power spectra in both cases are shown in Fig. \ref{fig:cmb_cls}.
          \item {\bf Beam:} we assume a Gaussian beam with full-width at half-maximum of $\theta_{\rm FWHM}=20$ and 1.4 arcmin for the curved-sky and flat-sky cases respectively. This resolution corresponds to what could be achieved at 150 GHz with a small aperture experiment (0.4m, S.A.) and a large aperture experiment (6m, L.A.) in either case. The choice of $\ell_{\rm knee}$ described above for the noise curves is also consistent with the atmospheric noise levels achievable with those types of telescopes \citep{2018arXiv180807445T}. As shown in Fig. \ref{fig:cmb_cls}, the combination of beam size and $\ell_{\rm knee}$ makes the two experiments sensitive to the $B$-mode power spectrum in different ranges of scales, corresponding to the primordial $B$-mode signal around the recombination bump (S.A.) and to the high-ell lensing signal (L.A.).
          \item {\bf Contaminants:} we assume a dust-like residual for both the curved-sky an flat-sky cases. For the curved-sky set-up, we use maps of the dust emission at 150 GHz generated by {\tt PySM}~\citep{2017MNRAS.469.2821T} and scaled so that the amplitude of the residual power spectrum is about 10\% of the lensing $B$-mode signal at $\ell=50$. For the flat-sky signal we generate a Gaussian realization of a dust-like component with power spectra
          \begin{equation}
            C_\ell^{EE}\propto\ell^{-2.5},\hspace{12pt}
            C_\ell^{BB}\propto\ell^{-2.3},
          \end{equation}
          where the spectral tilts were estimated from the high-$\ell$ behaviour of the {\tt PySM} maps in the low-foreground footprint. The amplitudes of these power spectra were fixed by requiring that the $B$-mode power be about 20\% of the lensing $B$-mode power spectrum at $\ell=550$ and so that $C_\ell^{EE}/C_\ell^{BB}\simeq2$, as found in \citet{2016A&A...594A..13P}. The foreground residual maps we use are shown in the bottom panels of Figure~\ref{fig:cmb_mocks}, and their power spectra are compared with the signal and noise spectra in Fig. \ref{fig:cmb_cls}.
          \item {\bf Pixelization:} the curved-sky case, targetting the $B$-mode recombination bump at $\ell\sim100$, uses a HEALPix resolution parameter $N_{\rm side}=256$, corresponding to $14$-arcmin pixels and sufficient to measure power spectra to multipoles $\ell\sim500$. The flat-sky case, targetting the lensing $B$-modes, uses square 2-arcmin pixels. These correspond to scales $\ell\lesssim5000$, well within the signal-dominated regime given the noise level and beam sizes quoted above.
        \end{itemize}

    \subsection{Curved-sky large-scale structure validation}\label{ssec:validation.lss_full}
      \begin{figure*}
        \centering
        \includegraphics[width=0.99\columnwidth]{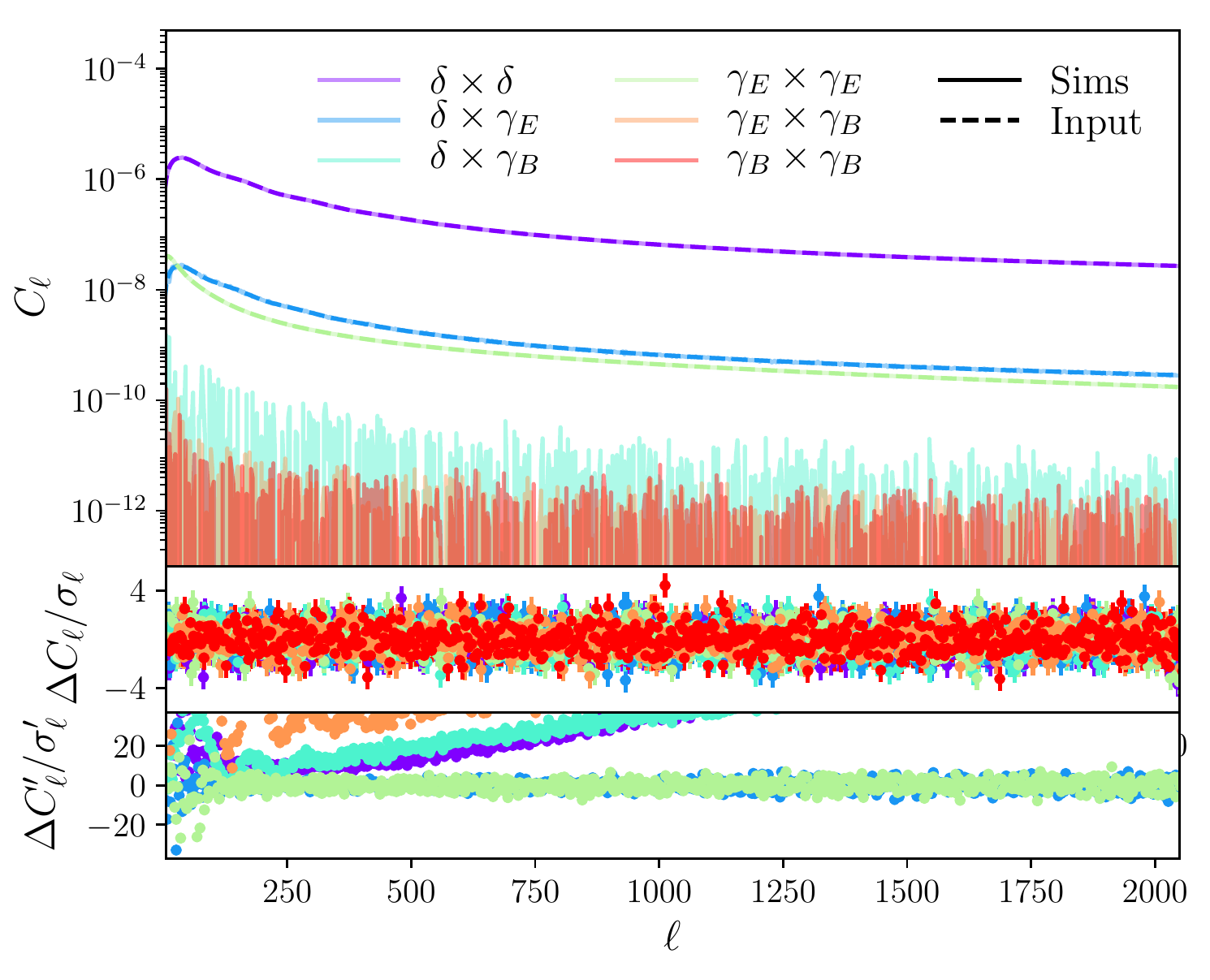}
        \includegraphics[width=0.99\columnwidth]{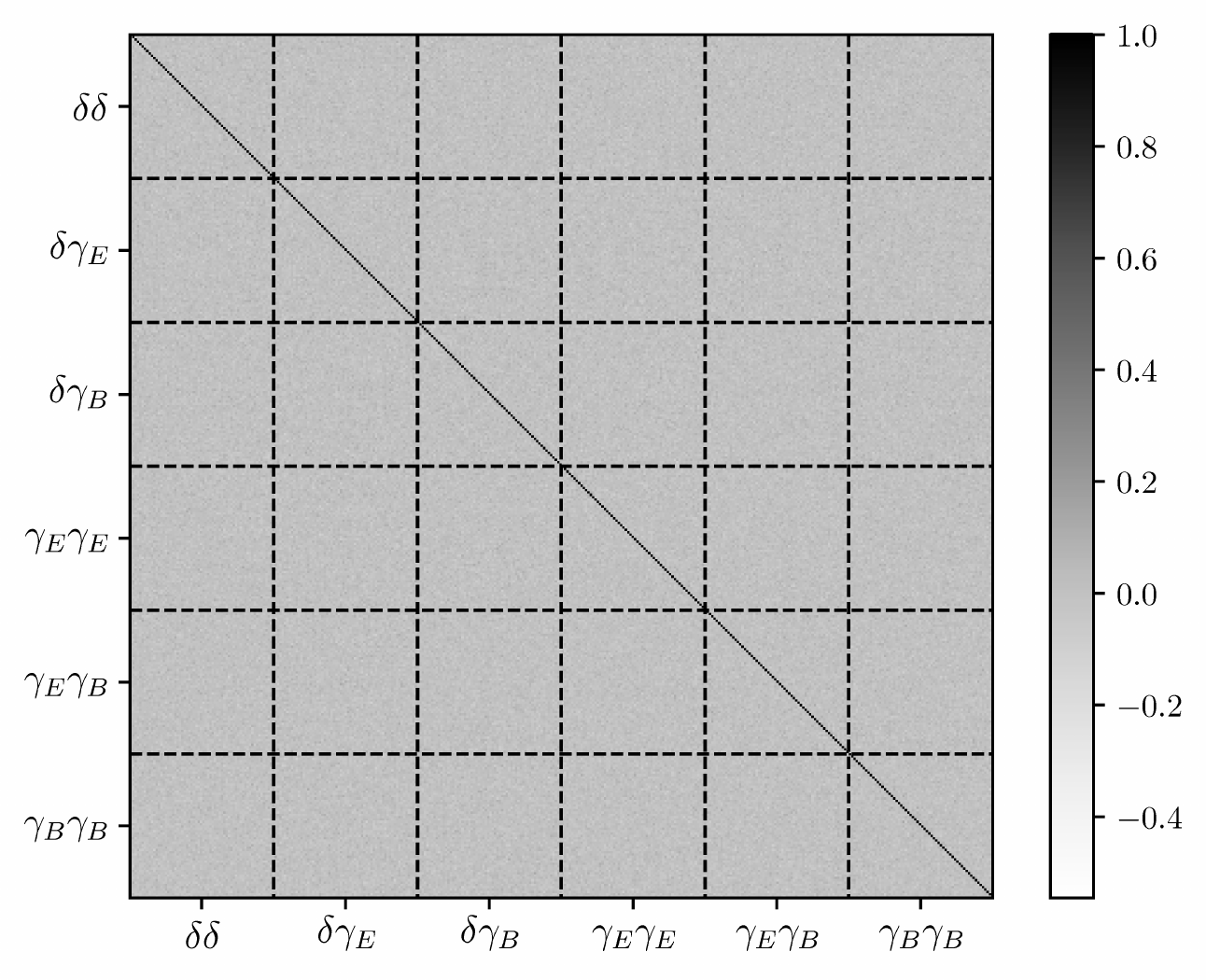}
        \caption{{\sl Left panel:} average over all simulations in our curved-sky LSS validation suite of the six power spectra between $\delta$, $\gamma_E$ and $\gamma_B$ (solid lines). The analytical prediction for the input signal power spectra are shown as darker dashed lines, and lie mostly superposed on the solid line (for the non-zero cases $\delta\times\delta$, $\delta\times\gamma_E$ and $\gamma_E\times\gamma_E$). The null power spectra ($\delta\times\gamma_B$, $\gamma_E\times\gamma_B$ and $\gamma_B\times\gamma_B$) are compatible with zero. The middle subplot shows the mean residuals with respect to the input power spectra normalized by the 1$\sigma$ error over the mean (i.e. the scatter over simulations divided by $\sqrt{N_{\rm sim}}$). The residuals are scattered around zero and their fluctuations are well within those expected under the assumption of Gaussian statistics given 1000 independent realizations (see Fig. \ref{fig:lss_val_chi2}). For comparison, the lower panel show the mean relative difference for the power spectra of contaminated maps without contaminant deprojection. {\sl Right panel:} correlation matrix ($C_{ij}/\sqrt{C_{ii}C_{jj}}$) of all measured power spectra estimated from 1000 realizations.} \label{fig:lss_val_cls}
      \end{figure*}
      \begin{figure*}
        \centering
        \includegraphics[width=0.80\textwidth]{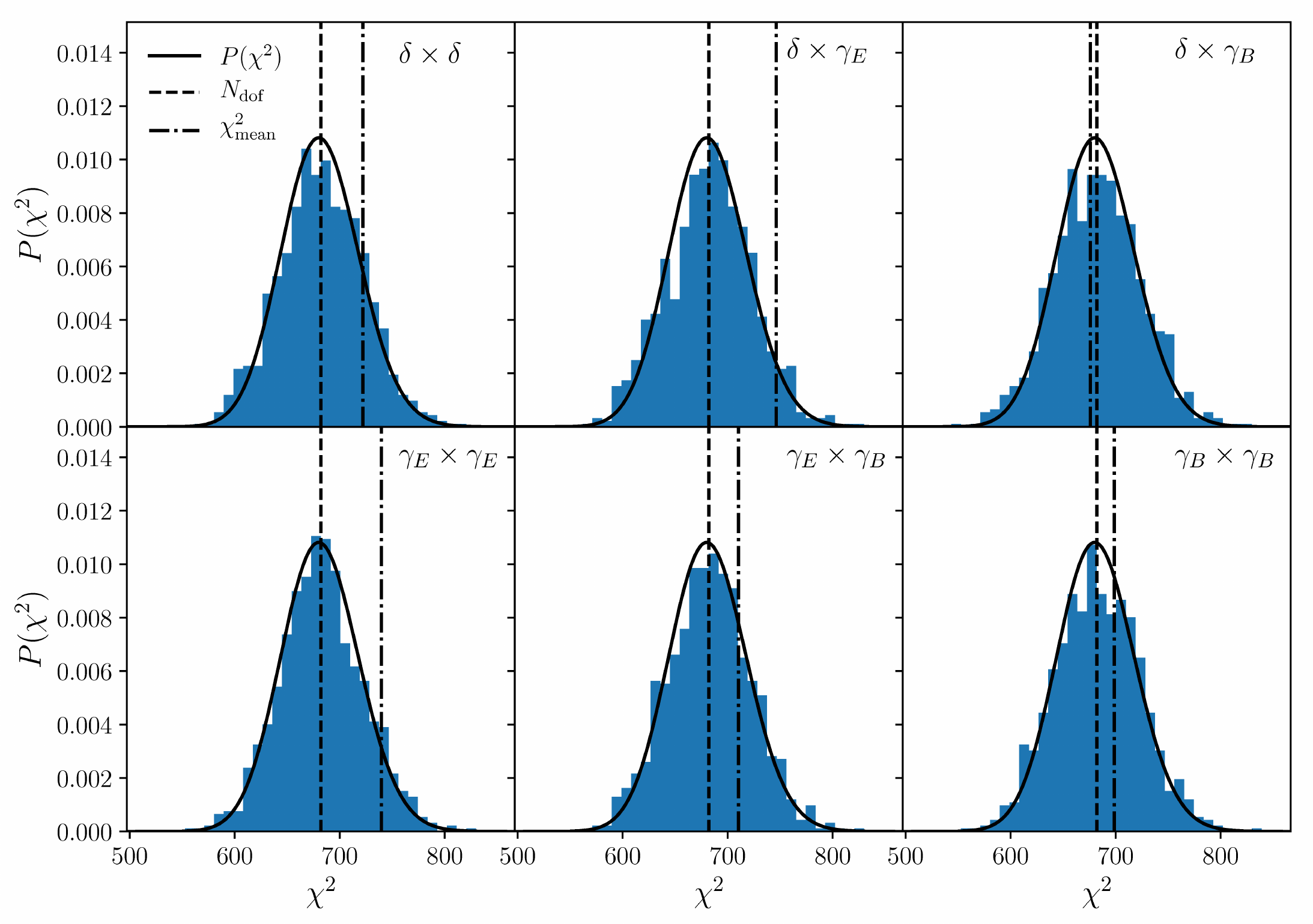}
        \caption{Distribution of the $\chi^2$ values found for each of the 1000 simulations in our curved-sky LSS validation suite (blue histograms). Results are shown for all possible cross-correlations, labeled in the top right corner of each subplot. The simulated distributions are compared with the expected $\chi^2$ distribution (solid lines) assuming Gaussian statistics for $N_{\rm dof}=681$ degrees of freedom (denoted by the dashed vertical lines). The dot-dashed vertical lines mark the $\chi^2$ value found for the mean residual with respect to the input power spectrum (see middle panel of the left panel in Fig. \ref{fig:lss_val_cls}). The associated PTE values are $\gtrsim0.1$, and thus no statistically significant bias is found by our validation suite.} \label{fig:lss_val_chi2}
      \end{figure*}
      We generate 1000 realizations of contaminated maps of the galaxy overdensity $\delta$ and cosmic shear $\gamma$ using the models described in Section \ref{sssec:validation.suite.lss} with resolution $N_{\rm side}=1024$. In our fiducial run, we included all contaminants, and used a non-binary mask that traces the shot-noise fluctuations due to the presence of stars and depth variations. For each simulation, we use \nmt~to compute all six possible cross-correlations between the $\delta$ and the $E$ and $B$ modes of $\gamma$ (which we will label $\gamma_{E,B}$), including map-level contaminant deprojection and removal of the noise and deprojection bias in the power spectrum. We use band-powers defined by averaging over groups of $\Delta\ell=3$ multipoles. The typical running time for one realization is $\sim 200$s using 32 \texttt{OpenMP} threads. Subsequent runs are typically faster since we cache the coupling matrices, which are the slowest step in the pipeline. More details about timing and scalability of \nmt~can be found in Appendix \ref{app:scalability}.
      
      The upper plot in the left panel of Fig. \ref{fig:lss_val_cls} shows the six deconvolved bandpowers averaged over simulations (solid lines) compared with the analytical prediction for the input signal power spectra (darker dashed lines). The middle panel of the same figure shows the average residual with respect to the analytical prediction normalized by the error on the mean (given by the standard deviation of all $N_{\rm sim}=1000$ simulations divided by $\sqrt{N_{\rm sim}}$). We see that the latter are accurately recovered by \nmt~in all cases, including all null correlations ($\delta\times\gamma_B$, $\gamma_E\times\gamma_B$ and $\gamma_B\times\gamma_B$), and that the normalized residuals fluctuate around zero with typical variations of $\Delta C_\ell/\sigma_\ell\lesssim3$, corresponding to expected $\lesssim3\sigma$ fluctuations. For comparison, the bottom panel of the figure shows the mean relative difference for power spectra of contaminated maps without deprojection, where we can appreciate that the residuals, $\Delta C'_{\ell}/\sigma'_{\ell}$, appear to not be centered around zero in some cases, i.e., the estimated power-spectra are biased.
      
      \begin{figure*}
        \centering
        \includegraphics[width=0.99\columnwidth]{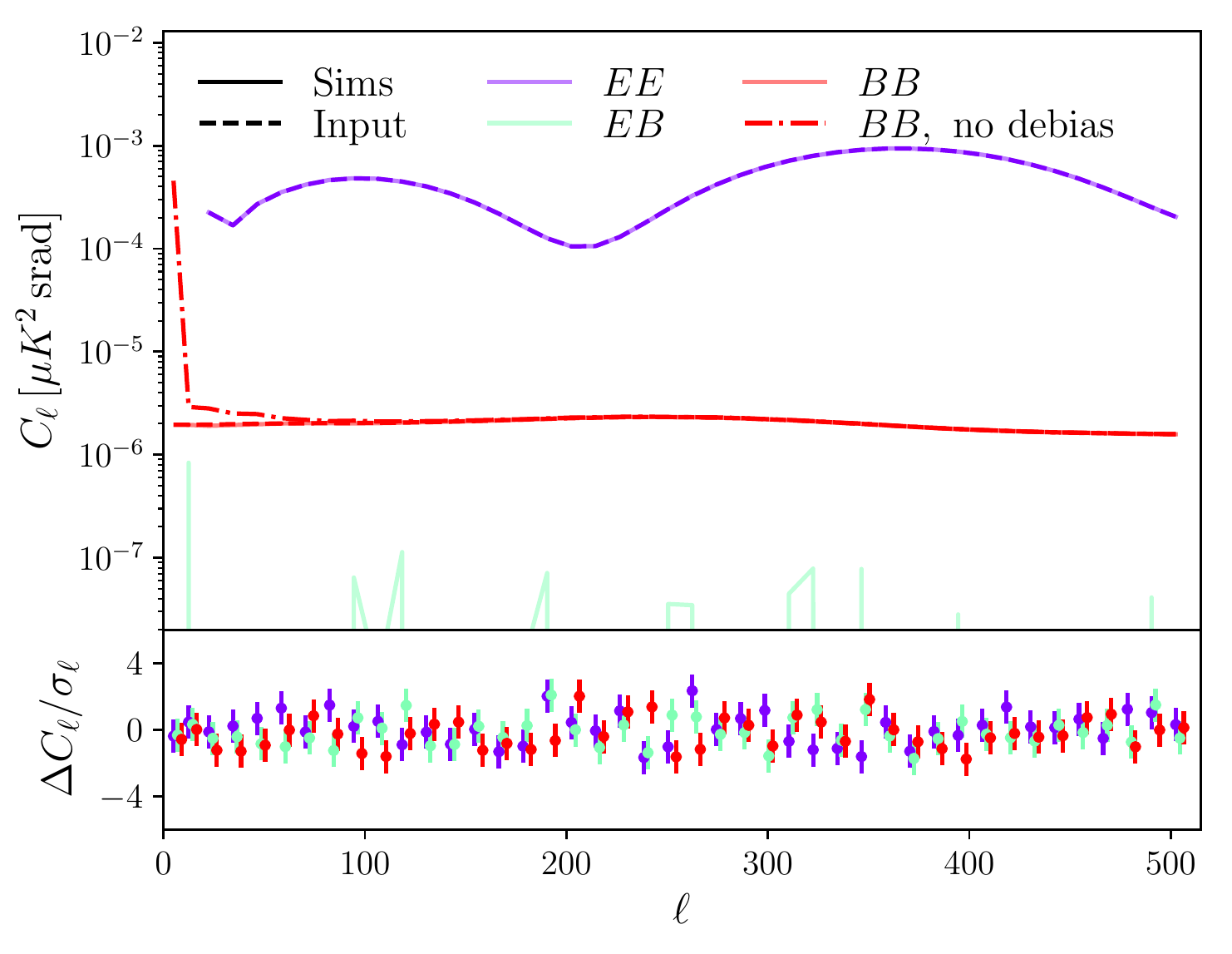}
        \includegraphics[width=0.99\columnwidth]{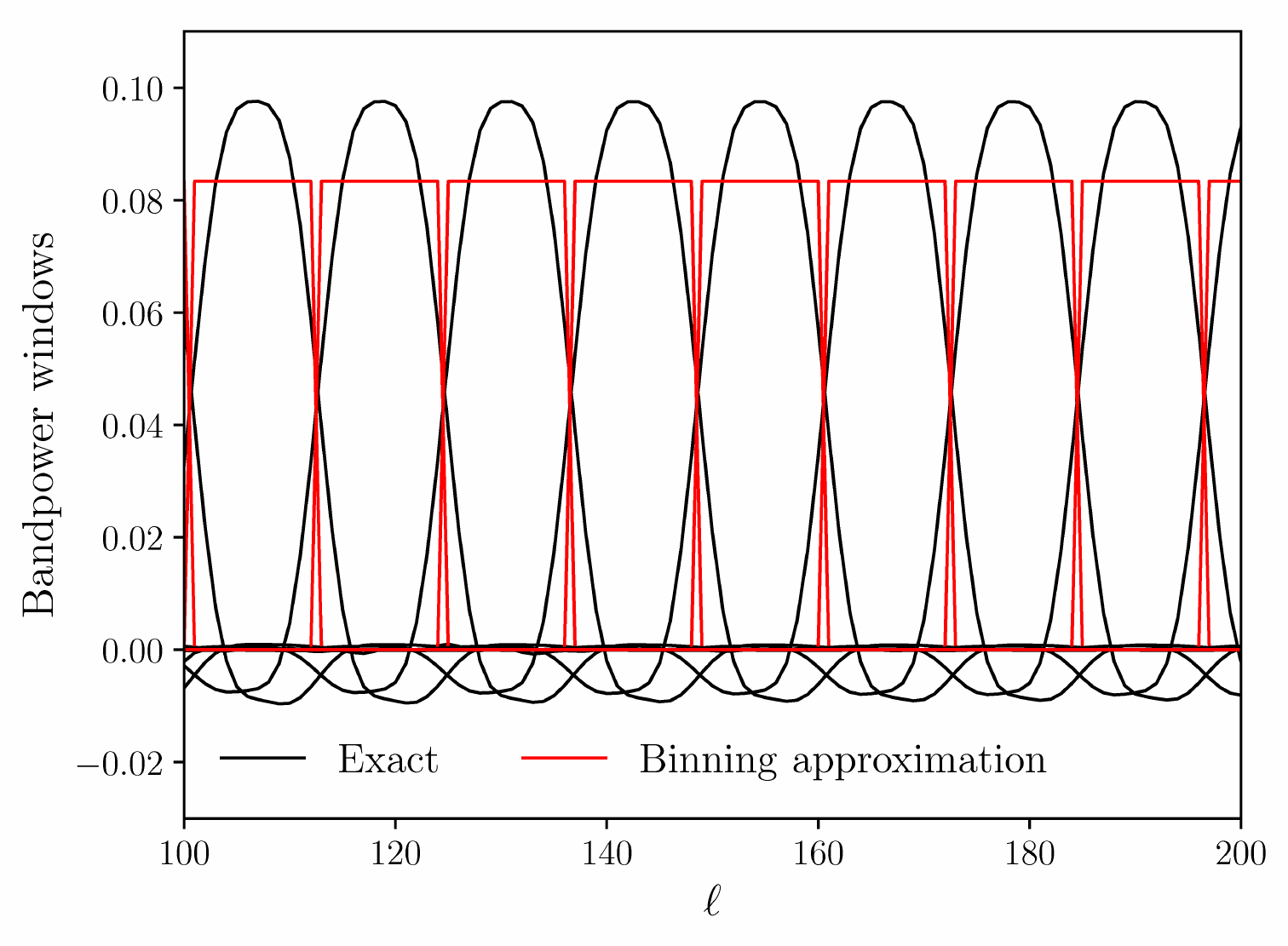}
        \caption{{\sl Left:} average over all simulations in our curved-sky CMB validation suite of the three spin-2 power spectra ($EE$, $EB$ and $BB$, solid lines) using mode deprojection and $B$-mode purification. The analytical prediction for the input signal power spectra are shown as dashed lines, and line mostly under the solid lines for the non-zero cases ($EE$ and $BB$). The null power spectrum $EB$ is compatible with zero. For comparison, the red dot-dashed line shows the result of not correcting for the mode-deprojection bias when estimating the $B$-mode power spectrum, which leads to a strong bias on small $\ell$. The bottom panel of this figure shows the mean residuals with respect to the input power spectra normalized by the $1\sigma$ error on the mean (i.e. the scatter over simulations divided by $\sqrt{N_{\rm sim}}$). The residuals are scattered around zero with random fluctuations of up to $\sim3\sigma$. {\sl Right:} exact bandpower window functions (solid black), which fully account for the effects of mode-coupling, compared with the input top-hat bandpower weights (red). Using the correct window functions is in general important in order to avoid parameter biases. Results are shown for the $BB$ power spectrum.} \label{fig:cmb_val_cls}
      \end{figure*}
      \begin{figure}
        \centering
        \includegraphics[width=0.99\columnwidth]{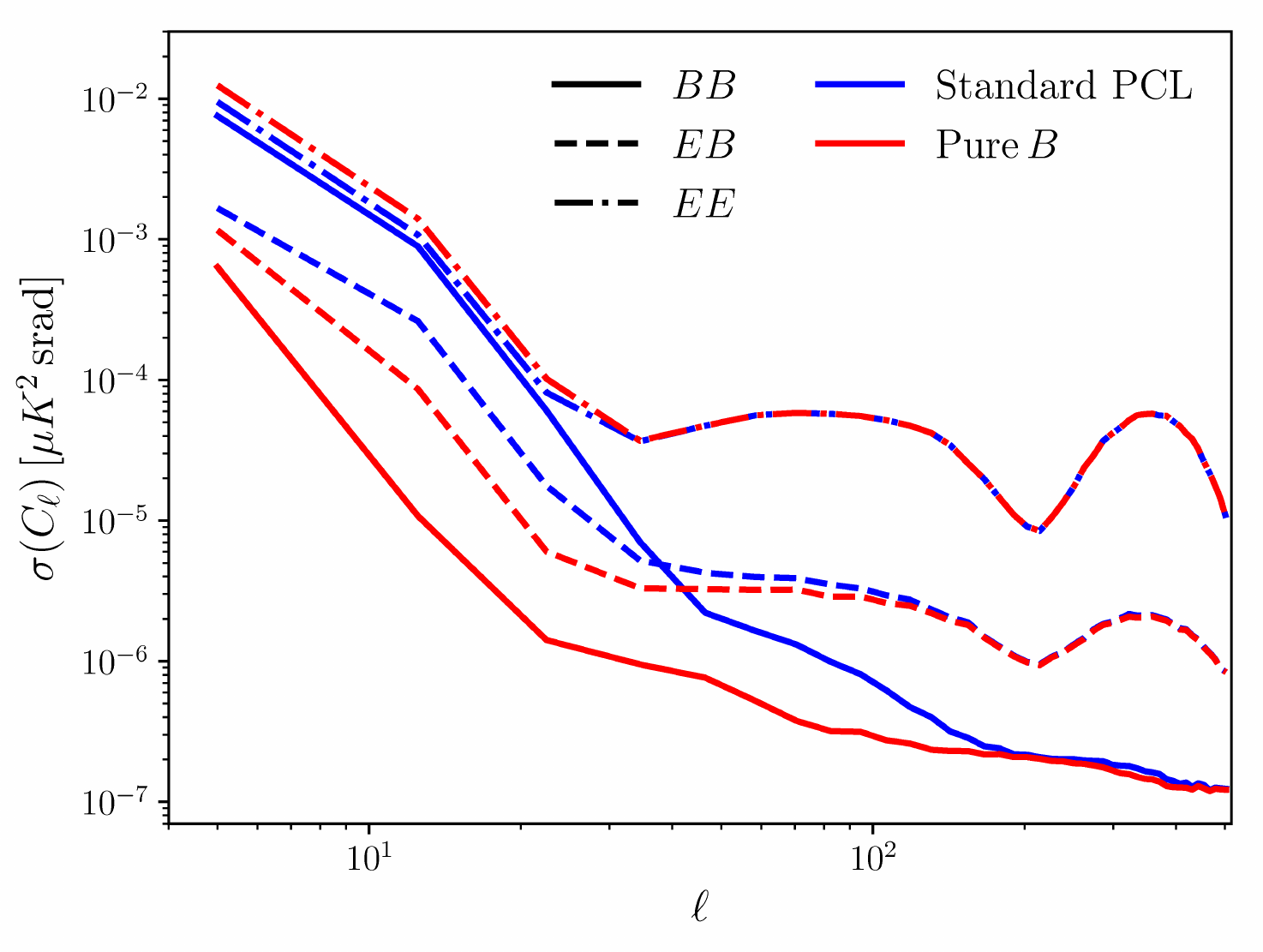}
        \caption{1$\sigma$ errors on the $BB$, $EB$ and $EE$ power spectra (solid, dashed and dot-dashed lines respectively) for the standard pseudo-$C_\ell$ estimator (blue) and for the estimator with purified $B$ modes (red). $B$-mode purification is able to produce significant gains in sensitivity for the $B$-mode power spectrum, particularly at low $\ell$.}\label{fig:cmb_val_sigma}
      \end{figure}
      To quantify the presence of any residual bias in the estimator we start by computing, for each simulation and cross-power spectrum in the validation suite, the quantity:
      \begin{equation}
        \chi^2_i=({\bf s}_i-{\bf t})^T\,{\sf Cov}^{-1}\,({\bf s}_i-{\bf t}),
      \end{equation}
      where ${\bf s}_i$ is the vector of $N_{\rm dof}$ power spectrum values estimated in the $i$-th simulation, ${\bf t}$ is the analytical prediction for the input power spectrum, and ${\sf Cov}$ is the covariance matrix of the estimator, which we compute by averaging over all simulations:
      \begin{equation}
        {\sf Cov}=\frac{1}{N_{\rm sim}}\sum_{i=1}^{N_{\rm sim}}({\bf s}_i-\bar{\bf s})({\bf s}_i-\bar{\bf s})^T,\hspace{6pt} \bar{\bf s}\equiv\frac{1}{N_{\rm sim}}\sum_{i=1}^{N_{\rm sim}}{\bf s}_i.
      \end{equation}
      The full covariance matrix is very close to diagonal, as shown in the right panel of Fig. \ref{fig:lss_val_cls}. Under the assumption that the estimator is unbiased ($\langle{\bf s}\rangle={\bf t}$), the values of $\chi^2_i$ should follow a $\chi^2$ distribution with a $N_{\rm dof}$ degrees of freedom. Figure \ref{fig:lss_val_chi2} shows the distribution of $\chi^2_i$ across simulations approximating the covariance matrix as diagonal (blue histograms), compared with the predicted $\chi^2$ distribution for $N_{\rm dof}=681$ (marked by the vertical dashed lines) for the six different power spectrum combinations. Both distributions
      are found to be compatible. Moreover, the dot-dashed lines show the values of $\chi^2$ found for the mean residual with respect to the input power spectra (middle panel of Fig. \ref{fig:lss_val_cls}). The probability to exceed (PTE) associated with these $\chi^2$ values are higher than $10\%$ in all cases, and therefore our validation suite does not show any statistically significant bias in the estimator. This bias, if any, must be at least $\sqrt{1000}\sim30$ times smaller than the statistical uncertainties associated with our validation suite, which are representative of the statistical power of next-generation galaxy surveys.
      
      Our validation suite also allows us to explore the impact of the sub-optimal inverse-variance weighting of the pseudo-$C_\ell$ estimator. The comparison with the uncertainties associated with optimal quadratic estimators has been made before in the literature \citep{2013MNRAS.435.1857L}, and we will not explore this here. Instead, we can study the impact on the final uncertainties of using a mask that implements exact inverse-noise weighting (i.e. the mask shown in the top left corner of Figure \ref{fig:lss_mocks_sph}, inversely proportional to the number density variations induced by stars and depth fluctuations) against a top-hat mask that weights all pixels equally. Repeating our analysis for the latter mask, we observe a negligible increase of $\sim0.5-1\%$ in the diagonal error bars. Thus, except in the case of large variations in noise properties (as will be the case for inhomogeneous scanning strategies in CMB experiments, for instance), the exact weighting scheme used with the pseudo-$C_\ell$ estimator is unlikely to produce significantly better uncertainties.

    \subsection{Curved-sky CMB \texorpdfstring{$B$}{B}-mode validation}\label{ssec:validation.cmb_full}
      \begin{figure*}
        \centering
        \includegraphics[width=0.99\columnwidth]{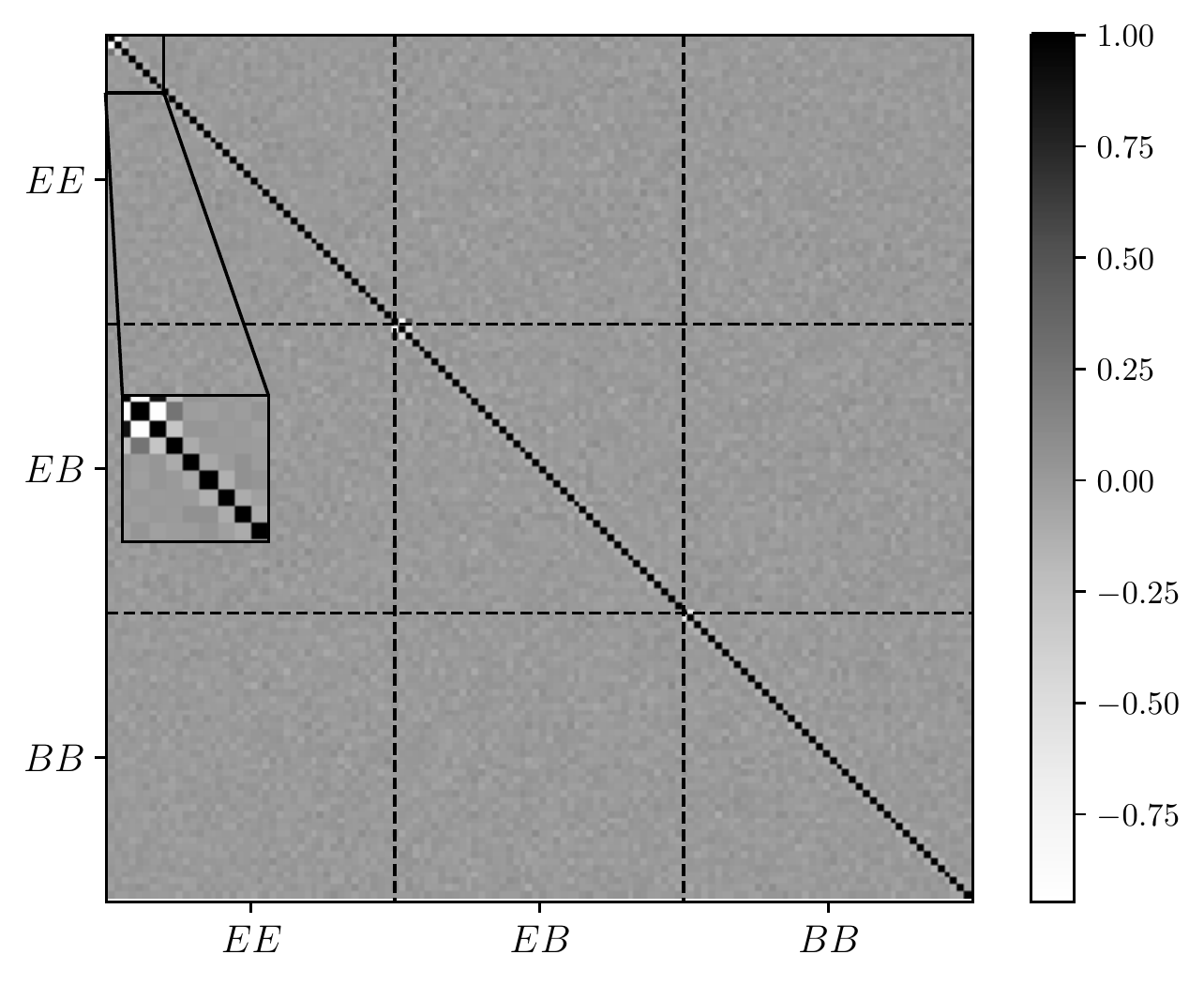}
        \includegraphics[width=0.99\columnwidth]{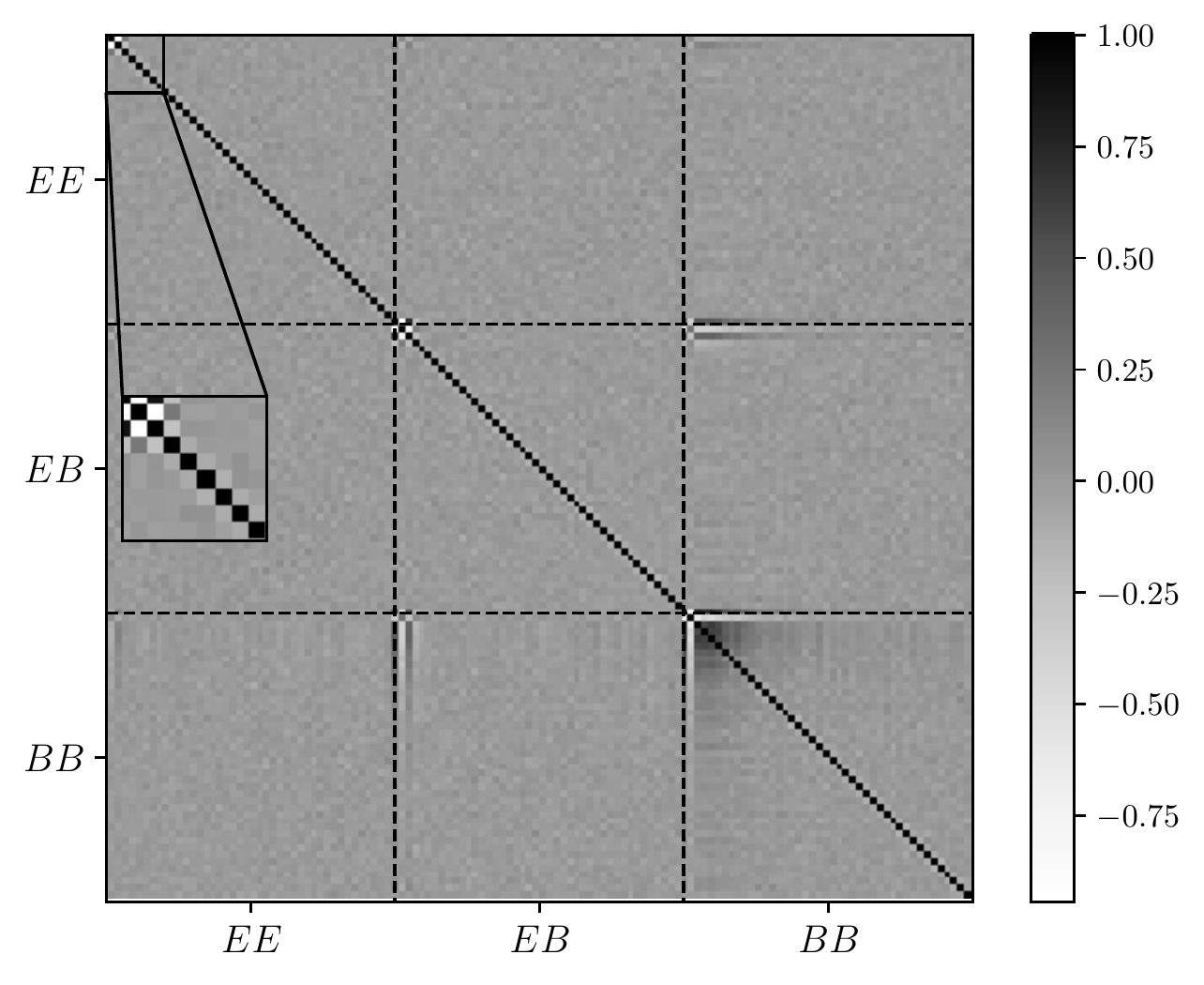}
        \caption{{\sl Left:} covariance matrix for power spectra computed in the absence of foreground contaminants computed from the 1000 simulations of our curved-sky CMB validation suite. The incomplete sky coverage causes a noticeable anti-correlation of neighboring bandpowers at low $\ell$ for the three power spectra ($EE$, $EB$ and $BB$, see inset for the $EE$ case), but beyond that, the covariance matrix is close to diagonal. {\sl Right:} the same as the left panel but this time for simulations containing foreground contaminants. The combination of $B$-mode purification and contaminant deprojection produces noticeable correlations at low $\ell$, especially in the $BB$ spectrum, that must be taken into account.} \label{fig:cmb_val_cov}
      \end{figure*}
      \begin{figure*}
        \centering
        \includegraphics[width=0.80\textwidth]{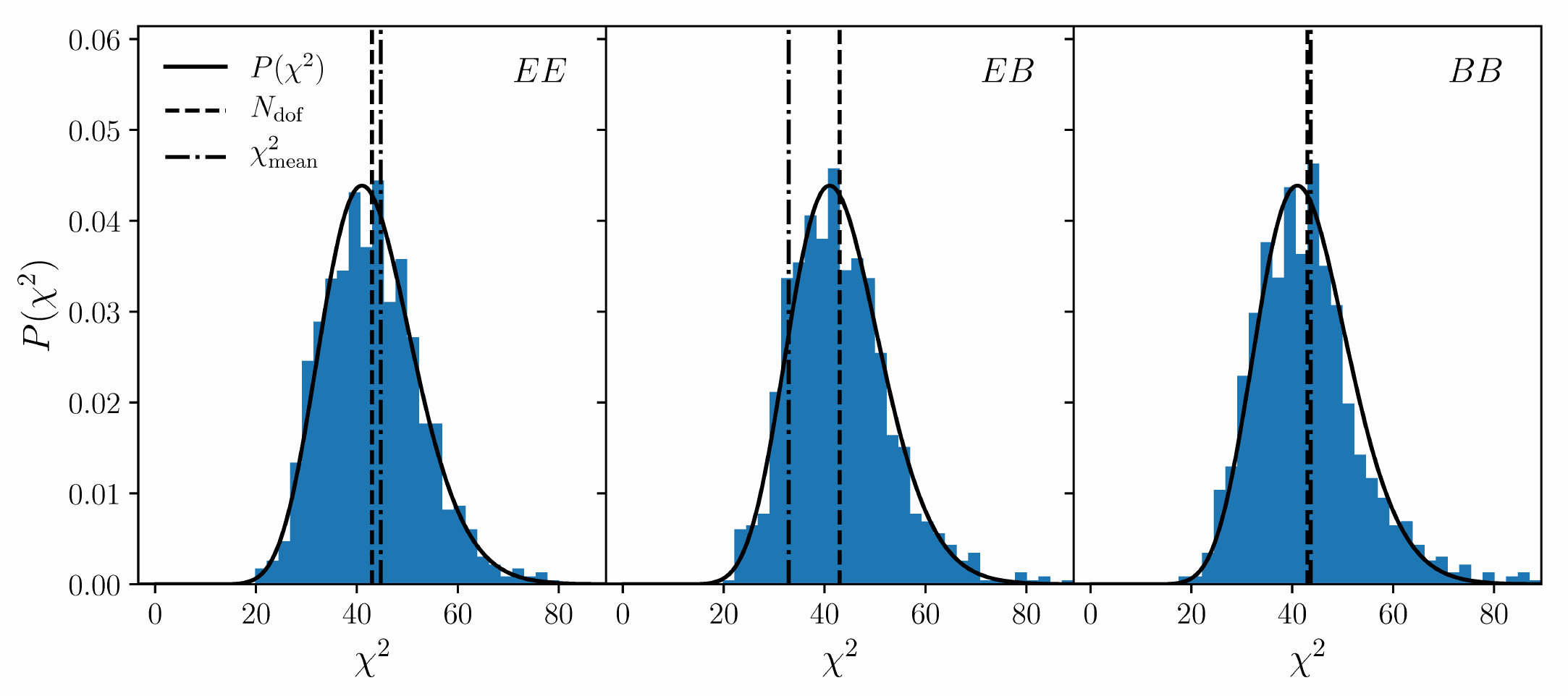}
        \caption{Distribution of $\chi^2$ values found for the 1000 simulations in our curved-sky CMB validation suite (blue histograms). Results are shown for the three polarized spectra ($EE$, $EB$ and $BB$), labeled in the top right corner of each panel. The simulated distributions agree well with the expected $\chi^2$ distribution (solid line) assuming Gaussian statistics for $N_{\rm dof}=43$ degrees of freedom (denoted by the dashed vertical lines). The dot-dashed vertical lines mark the $\chi^2$ value found for the mean residual with respect to the input power spectrum (see lower half of the left panel in Fig. \ref{fig:cmb_val_cls}). The associated PTE values are $\gtrsim0.4$, and thus no statistically significant bias is found by our validation suite.} \label{fig:cmb_val_chi2}
      \end{figure*}
      We have carried out a similar set of tests on an ensemble of 1000 simulations from the CMB validation suite described in Section \ref{sssec:validation.suite.cmb}. In our fiducial run we both deprojected the foreground contaminants and subsequently used $B$-mode purification to decouple $E$ and $B$-modes at the map level, since this is the main feature of the code we want to validate. The left panel of Figure \ref{fig:cmb_val_cls} shows the mean of the $EE$, $EB$ and $BB$ power spectra computed for all simulations (solid lines) compared with the expectation from the input spectra for $EE$ and $BB$ (dashed lines, mostly superposed on the solid ones). The average residuals normalized over the error on the mean, shown in the lower panel of the same figure, fluctuate around zero with an amplitude of up to $3\sigma$, as would be expected in the absence of a bias in the estimator. 

      The above figure was created using $B$-mode purification. Disabling this option has no impact on the bias of the estimates, but increases its variance as we show in the Figure \ref{fig:cmb_val_sigma}. We plot diagonal error bars with (red) and without (blue) $B$-mode purification for the $BB$ (solid), $EB$ (dashed) and $EE$ (dot-dashed) power spectra. Although there is no practical improvement in the error bars for the $EE$ spectrum (there is even a slight increase due to the loss of the ambiguous $E$ modes to purification), and the improvement is only mild for the $EB$ case, very significant gains can be obtained for the $BB$ power spectrum, reducing uncertainties by up to a factor of $\sim10$ on large scales.
      \begin{figure*}
        \centering
        \includegraphics[width=0.99\columnwidth]{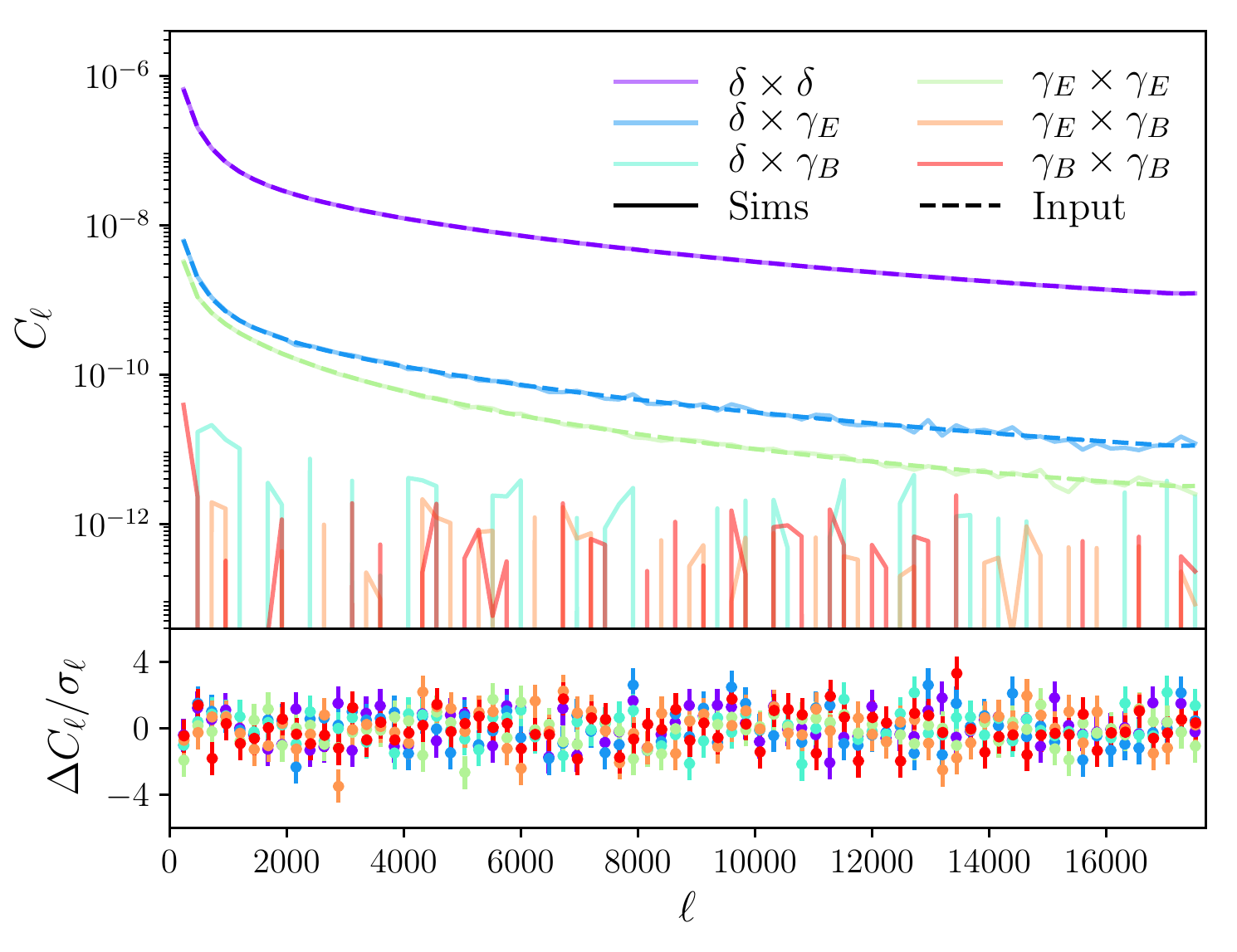}
        \includegraphics[width=0.99\columnwidth]{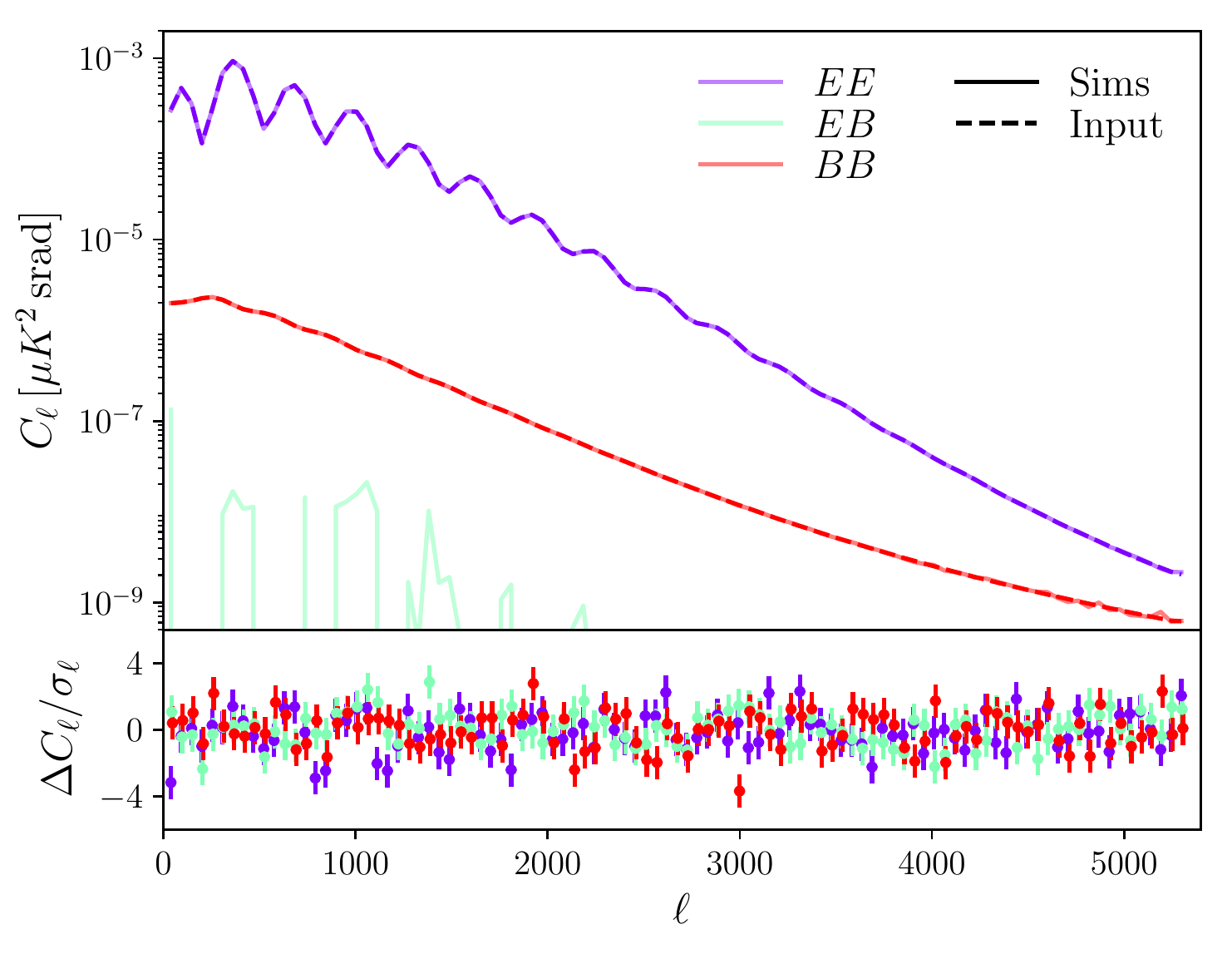}
        \caption{{\sl Left and Right:} the analogues of the left panels of Figures \ref{fig:lss_val_cls} and \ref{fig:cmb_val_cls} for the flat-sky LSS and CMB validation sets respectively.} \label{fig:val_cls_flat}
      \end{figure*}
      
      At this point we can use the validation suite to illustrate the relation between the input power spectrum and the theoretical prediction for the de-convolved bandpowers. As described by Equations \ref{eq:theory_correct} and \ref{eq:bpresponse}, the mode-coupling asociated with the sky mask and the subsequent de-coupling into bandpowers performed by inverting the binned mode-coupling matrix implies that the window function ${\cal F}_{q\ell}$ in Eq. \ref{eq:bpresponse} differs in general from a simple binning operator, even if the bandpower weights in Eq. \ref{eq:bpws} are chosen to be top-hat functions with a fixed width $\Delta\ell=12$. This is illustrated in the right panel of Figure \ref{fig:cmb_val_cls}, which shows, in black, the exact window function for different bandpowers (solid black), in comparison with the input top-hat bandpower weights (solid red). Results are shown for the $BB$ power spectrum. Depending on the specific problem, and on the range of scales under study using the simpler top-hat windows or even simply evaluating the theory power spectra at the bandpower centres may lead to e.g. significant biases in subsequent parameter inference stages. \nmt~provides simple and fast functions to apply the exact window functions to theory power spectra.
      
      As in the case of the LSS validation set, we also study the distribution of $\chi^2$ values from each simulation in order to quantify the presence of any possible bias. Unlike in the LSS suite, we find that in this case the covariance matrix is noticeably non-diagonal, and cannot be replaced by the diagonal errors when computing the $\chi^2$. The non-zero off-diagonal elements are caused by two effects. First, the limited footprint area, map apodization and choice of bandpowers leads to a noticeable level of anti-correlation between neighboring bandpowers at low $\ell$. This can be seen in the left panel of Figure \ref{fig:cmb_val_cov}, which shows the covariance matrix of simulations without foreground contaminants or deprojection. Second, introducing contaminant deprojection produces additional correlations that affect the lowest bandpowers and also involves a level of cross-talk between the three different cross-spectra.
      
      The distribution of $\chi^2$ values found from the simulations is shown in Fig. \ref{fig:cmb_val_chi2} for the three power spectra. As before, the vertical dashed lines show the number of degrees of freedom ($N_{\rm dof}=43$), while the  dot-dashed line marks the $\chi^2$ value of the mean residuals (bottom panel of Fig. \ref{fig:cmb_cls}). The associated PTE values are all above $40\%$ and therefore, as in the case of the LSS validation set, we do not detect any significant bias in the estimator when including both contaminant deprojection and $B$-mode purification.

    \subsection{Flat-sky validation}\label{ssec:validation.flat}
      \begin{figure*}
        \centering
        \includegraphics[width=0.80\textwidth]{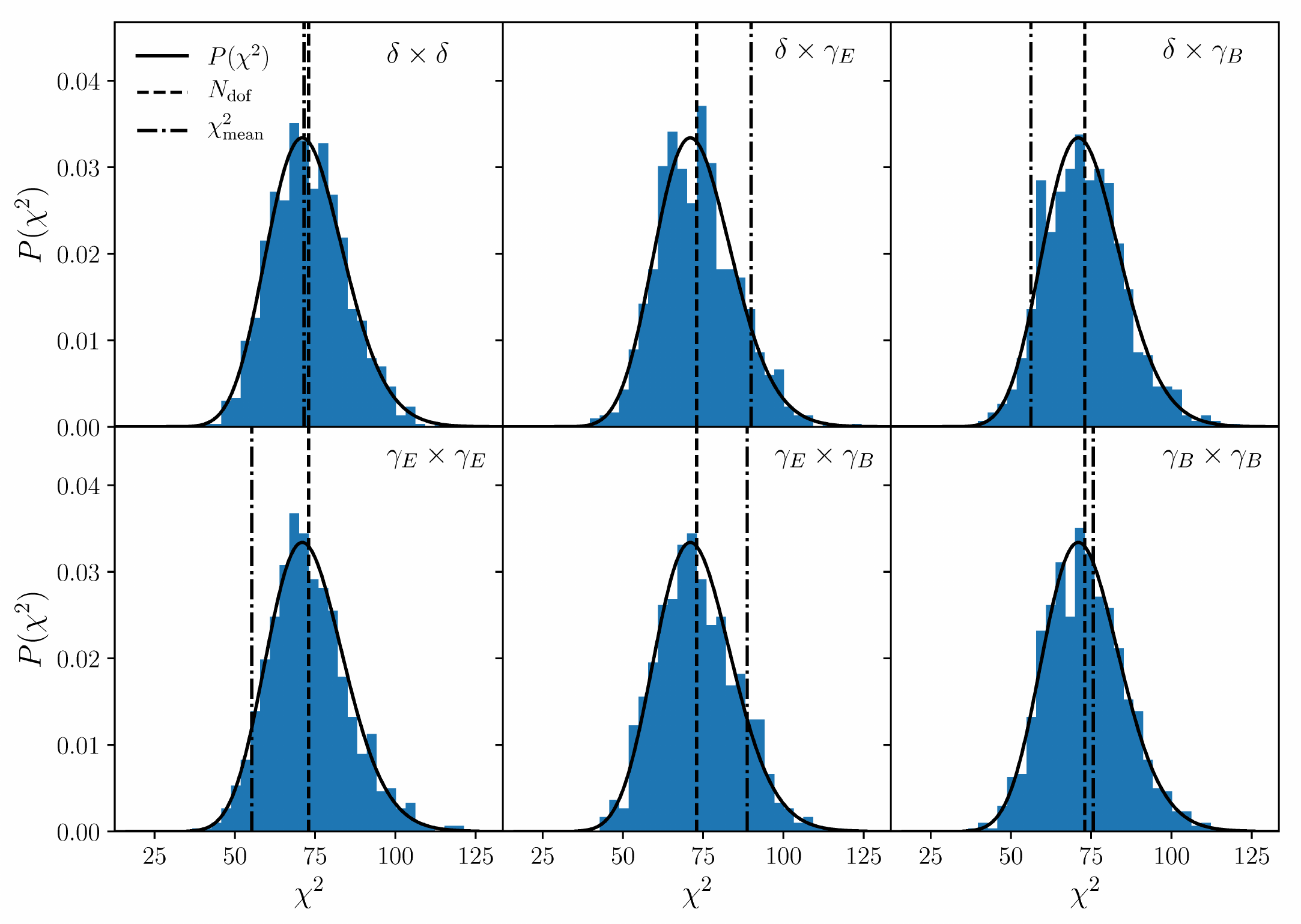}
        \includegraphics[width=0.80\textwidth]{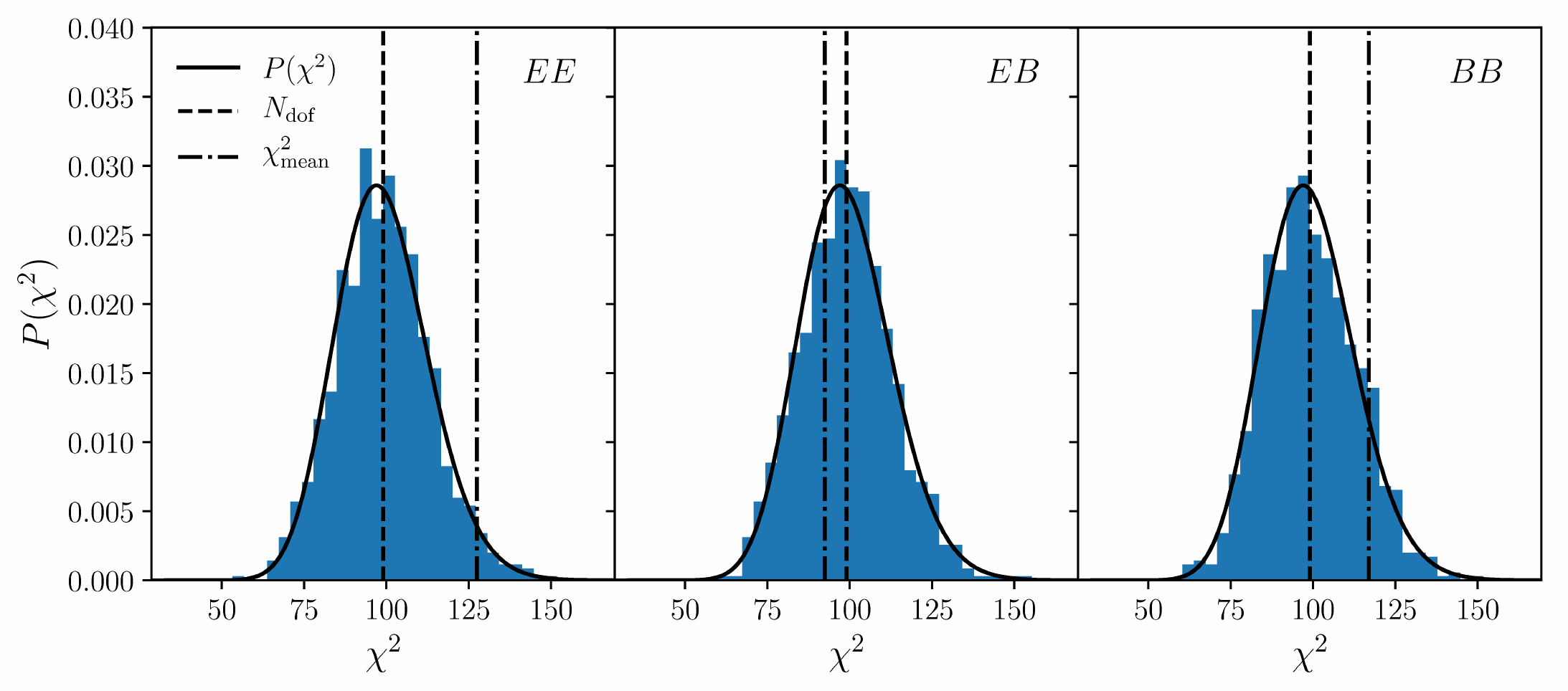}
        \caption{{\sl Top and bottom:} the analogues of the Figures \ref{fig:lss_val_chi2} and \ref{fig:cmb_val_chi2} for the flat-sky LSS and CMB validation sets respectively.} \label{fig:val_chi2_flat}
      \end{figure*}
      
      In order to validate the flat-sky features of \nmt, we follow the same procedure used in Sections \ref{ssec:validation.lss_full} and \ref{ssec:validation.cmb_full} for the LSS and CMB validation suites respectively.
      
      Before we discuss these results, a technical point, specific to the flat-sky case, must first be addressed. The theory prediction for the pseudo-$C_\ell$ method is given by Eq. \ref{eq:predflat} (before multiplying by the inverse binned mode-coupling matrix, Eq. \ref{eq:modcoup_flat}). This involves a convolution of the power spectra interpolated into the discretized Fourier plane with the un-binned mode-coupling matrix. This is an $\mathcal{O}(N_{\rm pix}^2)$ operation that would be too expensive to perform at every step of a Monte-Carlo chain, and therefore \nmt~uses a different approximate approach. This consists of first interpolating the theory power spectrum into the discrete Fourier plane, followed by an averaging of the interpolated power spectrum into a set of radial rings with a width given by the minimum wavenumber probed (given by ${\rm min}(2\pi/L_x,2\pi/L_y)$), and spanning all possible values of $|{\bf l}|$. This binned power spectrum, an object of size $\sim\sqrt{N_{\rm pix}}$, is then convolved with a high-resolution version of the mode-coupling matrix. This high-resolution matrix is given by a variation of Eq. \ref{eq:modcoup_flat} in which $S_{q'}$ represents all the values of the wavenumber ${\bf k}$ in one of the narrow rings described above (instead of the final wide bandpowers). The convolved power spectrum is then multiplied by the inverse binned coupling matrix (given by Eq. \ref{eq:modcoup_flat} exactly) to produce the final theoretical prediction. As we show below, this procedure, which is $\mathcal{O}(N_{\rm pix})$, is able to reproduce the exact theoretical prediction with a negligible error in all cases explored here.
      
      Figure \ref{fig:val_cls_flat} shows, with solid lines, the mean over all simulations of the 6 LSS power spectra and the 3 polarized CMB power spectra in the left and right panels respectively. The analytical predictions for the non-zero input power spectra are shown as dashed lines in both cases, and agree with the simulation mean almost perfectly. The mean residuals normalized by the 1$\sigma$ error on the mean are shown in the lower half of both figures, and feature fluctuations with amplitudes smaller than $\sim3$-$4\sigma$.
      
      Again, we quantify the presence of a bias in the estimator by studying the distribution of the $\chi^2$ values across simulations. In analogy with Figures \ref{fig:lss_val_chi2} and \ref{fig:cmb_val_chi2}, these are shown in the upper and lower panels of figure \ref{fig:val_chi2_flat} for the LSS and CMB suites respectively. In all cases the distributions (blue histograms) agree well with the expected $\chi^2$ distribution (solid lines) for the appropriate number of degrees of freedom (vertical dashed lines). The $\chi^2$ values of the mean residuals shown in the lower panels of Fig. \ref{fig:val_cls_flat} and marked by vertical dot-dashed lines in Fig. \ref{fig:val_chi2_flat} have PTE values above $10\%$. We therefore do not observe any significant bias in the estimator, proving \nmt's usefulness in the analysis of current and next-generation experiments in situations requiring the use of the flat-sky approximation.
 
  \section{Discussion}\label{sec:discussion}
    The pseudo-$C_\ell$ power spectrum estimator is one particular example of quadratic minimum-variance estimators as presented by \cite{1997PhRvD..55.5895T} in which the proposal covariance matrix of the data is assumed to be diagonal in real space. This allows us to significantly reduce the computational complexity of the problem, from ${\cal O}(N_{\rm pix}^3)$ to ${\cal O}(N_{\rm pix}^{3/2})$ through both analytical and numerical simplifications, as described in Section \ref{sec:maths}. A number of techniques that are naturally included in the standard quadratic estimator, such as mode deprojection and $B$-mode purification, can also be implemented in the standard pseudo-$C_\ell$ algorithm without altering the computational complexity of the method. Although this speed boost is achieved at the expense of optimality (in terms of the estimator's variance), the degradation in sensitivity will be small or negligible as long as the data does not deviate significantly from the diagonal assumption. This implies that the pseudo-$C_\ell$ approach is a good choice for non-steep power spectra for which the typical correlation length is smaller than the mask structure, and in general on small scales, where the impact of the power spectrum shape drops as the number of available modes increases \citep{2004MNRAS.349..603E,2006MNRAS.370..343E}. As we have argued, this will be the case for a wide range of relevant science cases pursued by next-generation cosmological observations.
    
    This paper presents \nmt, the first public, validated and easy-to-use power spectrum estimation software package that implements the pseudo-$C_\ell$ method to compute angular cross-power spectra for any pair of spin-0 or spin-2 fields, including all their generalizations (flat-sky, deprojection and purification). We have described the steps taken to validate the code, showing that it is able to provide unbiased results in a wide variety of scenarios. In particular, we have constructed a realistic validation suite that resembles the type of data, masks and contaminants that will affect future wide-area optical galaxy surveys and ground-based CMB polarization experiments. As shown in Section \ref{sec:validation}, we are able to recover the input power spectrum in all cases (which include varying degrees of contamination, purification, different spins etc.) with no detectable bias down to $\lesssim1/30$-th of the statistical uncertainties.
    
    Besides describing and validating the code, and in order to generalize the pseudo-$C_\ell$ estimator to all the cases supported by \nmt, this paper has also presented a number of results that are, to the best of our knowledge, new. These include:
    \begin{itemize}
      \item Extending the analytical treatment of contaminant deprojection presented in \cite{2016MNRAS.456.2095E} to fields of arbitrary spin.
      \item Combining contaminant deprojection and $E/B$ purification.
      \item Consistently deriving all of this functionality (standard pseudo-$C_\ell$, deprojection and purification for arbitrary spins) in the flat-sky approximation.
    \end{itemize}
    
    Although \nmt~is intended to make power spectrum estimation as easy as possible, some tasks are too dependent upon the properties of the data to be analyzed to automatize them in a general way, and users are therefore in charge of them. We list the most important of these here:
    \begin{itemize}
      \item \nmt~operates directly on {\sl maps}. Galaxy survey data, however, are normally provided in the form of object catalogs, and the conversion those into the desired maps (e.g. galaxy overdensity or shear) must be done externally. 
      \item $E/B$ purification can only be carried out in the context of pseudo-$C_\ell$ estimators when the mask is differentiable up to its second derivatives, which can usually be achieved with an appropriate apodization. Although \nmt~provides tools to carry out this type of operations, some of the properties of the input mask (e.g. sharp corners) will in general spoil differentiability even after apodization. Users must therefore take care to inspect the mask derivatives before making use of \nmt's purification feature.
      \item A crucial part of any quadratic power spectrum estimation is the removal of noise bias in the auto-correlation of a given dataset. Although we have provided analytic formulas to estimate this bias in certain simple scenarios (see Section \ref{ssec:maths.beam_noise}), these are often not applicable. Users must therefore make sure that a sufficiently accurate estimate of the noise bias can be achieved (e.g. through simulations) or to use only cross-correlation data.
      \item Contaminant deprojection will be able to provide an accurate estimate of the power spectrum as long as two conditions are met. First, users must be able to provide a comprehensive list of possible contaminants as would be observed in their data (e.g. beam-smoothed, mean-subtracted etc.). Second, a good estimate of the underlying signal power spectra must be provided in order to calculate and subtract the deprojection bias. We have outlined a possible iterative process to do so in Section \ref{ssec:maths.deproj}.
    \end{itemize}

    \nmt~does not currently produce a covariance matrix estimate. Although computationally efficent analytical estimates of the Gaussian covariance matrix can be derived under certain approximations \citep{2004MNRAS.349..603E,2005MNRAS.360.1262B}, this is not widely supported by the code for several reasons. First, for many problems, such as large-scale structure power spectrum estimation, the covariance matrix contains important contributions from the connected four-point function that depend on the precise nature of the problem and go beyond the estimation step that \nmt~strives to do well. Second, with all additional options included, especially if a large number of contaminant templates are deprojected, the exact analytical estimates are either computational unfeasible or overly approximate. Since a majority of the forthcoming experiments will rely on a large number of mock datasets available for testing systematic effects, these maps also offer a natural path towards estimating the data covariance.
    
    \nmt~is open-source software and publicly available at \url{https://github.com/LSSTDESC/NaMaster}. To maximize the usefulness of this software for the community, it is accompanied by an extensive documentation and example code\footnote{See \url{https://github.com/LSSTDESC/NaMaster}.}. \nmt~is written in {\tt C} and {\tt OpenMP}-parallelized to maximize its performance, although we encourage its use through the companion {\tt python} wrapper, the fastest-growing language of choice in the analysis of astronomical data. The code is in constant development, and we will strive to implement further functionality to e.g. estimate covariance matrices, extend the range of supported sky pixelizations and enhance its computational performance.

\subsection*{Acknowledgements}
  This paper has undergone internal review in the LSST Dark Energy Science Collaboration. The internal reviewers were Scott Dodelson, Francois Lanusse and Boris Leistedt. Besides them, we would like to thank Rupert Allison, Jo Dunkley, Franz Elsner, Daniel Lenz, Thibaut Louis, Mathew Madhavacheril, Sigurd N\ae ss, Hiranya Peiris, Naomi Robertson and Ben Thorne for useful comments and discussions. 

  Author contributions are listed below. \\
  David Alonso: Co-led project; wrote the code; contributed to validation. \\
  Javier Sanchez: Co-led project; wrote code; contributed to validation. \\
  An\v{z}e Slosar: Co-led project; contributed to validation. \\

  DA acknowledges support from the Beecroft trust and from the Science and Technology Facilities Council (STFC) through an Ernest Rutherford Fellowship, grant reference ST/P004474/1. JS acknowledges support from the United States Department of Energy. The DESC acknowledges ongoing support from the Institut National de Physique Nucl\'eaire et de Physique des Particules in France; the Science \& Technology Facilities Council in the United Kingdom; and the Department of Energy, the National Science Foundation, and the LSST Corporation in the United States.  DESC uses resources of the IN2P3 Computing Center (CC-IN2P3--Lyon/Villeurbanne - France) funded by the Centre National de la Recherche Scientifique; the National Energy Research Scientific Computing Center, a DOE Office of Science User Facility supported by the Office of Science of the U.S.\ Department of Energy under Contract No.\ DE-AC02-05CH11231; STFC DiRAC HPC Facilities, funded by UK BIS National E-infrastructure capital grants; and the UK particle physics grid, supported by the GridPP Collaboration.  This work was performed in part under DOE Contract DE-AC02-76SF00515.

\setlength{\bibhang}{2.0em}
\setlength\labelwidth{0.0em}
\bibliography{main}

\begin{thebibliography}{}
\makeatletter
\relax
\def\mn@urlcharsother{\let\do\@makeother \do\$\do\&\do\#\do\^\do\_\do\%\do\~}
\def\mn@doi{\begingroup\mn@urlcharsother \@ifnextchar [ {\mn@doi@}
  {\mn@doi@[]}}
\def\mn@doi@[#1]#2{\def\@tempa{#1}\ifx\@tempa\@empty \href
  {http://dx.doi.org/#2} {doi:#2}\else \href {http://dx.doi.org/#2} {#1}\fi
  \endgroup}
\def\mn@eprint#1#2{\mn@eprint@#1:#2::\@nil}
\def\mn@eprint@arXiv#1{\href {http://arxiv.org/abs/#1} {{\tt arXiv:#1}}}
\def\mn@eprint@dblp#1{\href {http://dblp.uni-trier.de/rec/bibtex/#1.xml}
  {dblp:#1}}
\def\mn@eprint@#1:#2:#3:#4\@nil{\def\@tempa {#1}\def\@tempb {#2}\def\@tempc
  {#3}\ifx \@tempc \@empty \let \@tempc \@tempb \let \@tempb \@tempa \fi \ifx
  \@tempb \@empty \def\@tempb {arXiv}\fi \@ifundefined
  {mn@eprint@\@tempb}{\@tempb:\@tempc}{\expandafter \expandafter \csname
  mn@eprint@\@tempb\endcsname \expandafter{\@tempc}}}

\bibitem[\protect\citeauthoryear{{Abazajian} et~al.,}{{Abazajian}
  et~al.}{2016}]{2016arXiv161002743A}
{Abazajian} K.~N.,  et~al., 2016, preprint, \href
  {http://adsabs.harvard.edu/abs/2016arXiv161002743A} {} (\mn@eprint {arXiv}
  {1610.02743})

\bibitem[\protect\citeauthoryear{{Aihara} et~al.,}{{Aihara}
  et~al.}{2018}]{2018PASJ...70S...4A}
{Aihara} H.,  et~al., 2018, \mn@doi [\pasj] {10.1093/pasj/psx066}, \href
  {http://adsabs.harvard.edu/abs/2018PASJ...70S...4A} {70, S4}

\bibitem[\protect\citeauthoryear{{Alsing}, {Heavens}, {Jaffe}, {Kiessling},
  {Wandelt}  \& {Hoffmann}}{{Alsing} et~al.}{2016}]{2016MNRAS.455.4452A}
{Alsing} J.,  {Heavens} A.,  {Jaffe} A.~H.,  {Kiessling} A.,  {Wandelt} B.,
  {Hoffmann} T.,  2016, \mn@doi [\mnras] {10.1093/mnras/stv2501}, \href
  {http://adsabs.harvard.edu/abs/2016MNRAS.455.4452A} {455, 4452}

\bibitem[\protect\citeauthoryear{{Asgari}, {Taylor}, {Joachimi}  \&
  {Kitching}}{{Asgari} et~al.}{2016}]{2016arXiv161204664A}
{Asgari} M.,  {Taylor} A.,  {Joachimi} B.,   {Kitching} T.~D.,  2016, preprint,
  \href {http://adsabs.harvard.edu/abs/2016arXiv161204664A} {} (\mn@eprint
  {arXiv} {1612.04664})

\bibitem[\protect\citeauthoryear{{Asorey}, {Crocce}, {Gazta{\~n}aga}  \&
  {Lewis}}{{Asorey} et~al.}{2012}]{2012MNRAS.427.1891A}
{Asorey} J.,  {Crocce} M.,  {Gazta{\~n}aga} E.,   {Lewis} A.,  2012, \mn@doi
  [\mnras] {10.1111/j.1365-2966.2012.21972.x}, \href
  {http://adsabs.harvard.edu/abs/2012MNRAS.427.1891A} {427, 1891}

\bibitem[\protect\citeauthoryear{{BICEP2 Collaboration} et~al.,}{{BICEP2
  Collaboration} et~al.}{2016}]{2016PhRvL.116c1302B}
{BICEP2 Collaboration} et~al., 2016, \mn@doi [Physical Review Letters]
  {10.1103/PhysRevLett.116.031302}, \href
  {http://adsabs.harvard.edu/abs/2016PhRvL.116c1302B} {116, 031302}

\bibitem[\protect\citeauthoryear{{Bond}}{{Bond}}{1995}]{1995PhRvL..74.4369B}
{Bond} J.~R.,  1995, \mn@doi [Physical Review Letters]
  {10.1103/PhysRevLett.74.4369}, \href
  {http://adsabs.harvard.edu/abs/1995PhRvL..74.4369B} {74, 4369}

\bibitem[\protect\citeauthoryear{{Bond}, {Jaffe}  \& {Knox}}{{Bond}
  et~al.}{1998}]{1998PhRvD..57.2117B}
{Bond} J.~R.,  {Jaffe} A.~H.,   {Knox} L.,  1998, \mn@doi [\prd]
  {10.1103/PhysRevD.57.2117}, \href
  {http://adsabs.harvard.edu/abs/1998PhRvD..57.2117B} {57, 2117}

\bibitem[\protect\citeauthoryear{{Bonvin} \& {Durrer}}{{Bonvin} \&
  {Durrer}}{2011}]{2011PhRvD..84f3505B}
{Bonvin} C.,  {Durrer} R.,  2011, \mn@doi [\prd] {10.1103/PhysRevD.84.063505},
  \href {http://adsabs.harvard.edu/abs/2011PhRvD..84f3505B} {84, 063505}

\bibitem[\protect\citeauthoryear{{Brown}, {Castro}  \& {Taylor}}{{Brown}
  et~al.}{2005}]{2005MNRAS.360.1262B}
{Brown} M.~L.,  {Castro} P.~G.,   {Taylor} A.~N.,  2005, \mn@doi [\mnras]
  {10.1111/j.1365-2966.2005.09111.x}, \href
  {http://adsabs.harvard.edu/abs/2005MNRAS.360.1262B} {360, 1262}

\bibitem[\protect\citeauthoryear{{Bunn}}{{Bunn}}{2011}]{2011PhRvD..83h3003B}
{Bunn} E.~F.,  2011, \mn@doi [\prd] {10.1103/PhysRevD.83.083003}, \href
  {http://adsabs.harvard.edu/abs/2011PhRvD..83h3003B} {83, 083003}

\bibitem[\protect\citeauthoryear{{Bunn}, {Zaldarriaga}, {Tegmark}  \& {de
  Oliveira-Costa}}{{Bunn} et~al.}{2003}]{2003PhRvD..67b3501B}
{Bunn} E.~F.,  {Zaldarriaga} M.,  {Tegmark} M.,   {de Oliveira-Costa} A.,
  2003, \mn@doi [\prd] {10.1103/PhysRevD.67.023501}, \href
  {http://adsabs.harvard.edu/abs/2003PhRvD..67b3501B} {67, 023501}

\bibitem[\protect\citeauthoryear{{Chon}, {Challinor}, {Prunet}, {Hivon}  \&
  {Szapudi}}{{Chon} et~al.}{2004}]{2004MNRAS.350..914C}
{Chon} G.,  {Challinor} A.,  {Prunet} S.,  {Hivon} E.,   {Szapudi} I.,  2004,
  \mn@doi [\mnras] {10.1111/j.1365-2966.2004.07737.x}, \href
  {http://adsabs.harvard.edu/abs/2004MNRAS.350..914C} {350, 914}

\bibitem[\protect\citeauthoryear{{Connolly} et~al.,}{{Connolly}
  et~al.}{2014}]{2014SPIE.9150E..14C}
{Connolly} A.~J.,  et~al., 2014, in Society of Photo-Optical Instrumentation
  Engineers (SPIE) Conference Series. p.~14, \mn@doi{10.1117/12.2054953}

\bibitem[\protect\citeauthoryear{{DES Collaboration} et~al.,}{{DES
  Collaboration} et~al.}{2017}]{2017arXiv170801530D}
{DES Collaboration} et~al., 2017, preprint, \href
  {http://adsabs.harvard.edu/abs/2017arXiv170801530D} {} (\mn@eprint {arXiv}
  {1708.01530})

\bibitem[\protect\citeauthoryear{{Dark Energy Survey Collaboration}
  et~al.}{{Dark Energy Survey Collaboration}
  et~al.}{2016}]{2016MNRAS.460.1270D}
{Dark Energy Survey Collaboration} et~al., 2016, \mn@doi [MNRAS]
  {10.1093/mnras/stw641}, \href
  {http://adsabs.harvard.edu/abs/2016MNRAS.460.1270D} {460, 1270}

\bibitem[\protect\citeauthoryear{{De Bernardis} et~al.,}{{De Bernardis}
  et~al.}{2016}]{2016SPIE.9910E..14D}
{De Bernardis} F.,  et~al., 2016, in Observatory Operations: Strategies,
  Processes, and Systems VI. p. 991014 (\mn@eprint {arXiv} {1607.02120}),
  \mn@doi{10.1117/12.2232824}

\bibitem[\protect\citeauthoryear{{Delgado}, {Saha}, {Chandrasekharan}, {Cook},
  {Petry}  \& {Ridgway}}{{Delgado} et~al.}{2014}]{2014SPIE.9150E..15D}
{Delgado} F.,  {Saha} A.,  {Chandrasekharan} S.,  {Cook} K.,  {Petry} C.,
  {Ridgway} S.,  2014, in Modeling, Systems Engineering, and Project Management
  for Astronomy VI. p. 915015, \mn@doi{10.1117/12.2056898}

\bibitem[\protect\citeauthoryear{{Doroshkevich} et~al.,}{{Doroshkevich}
  et~al.}{2011}]{2011IJMPD..20.1053D}
{Doroshkevich} A.~G.,  et~al., 2011, \mn@doi [International Journal of Modern
  Physics D] {10.1142/S0218271811019219}, \href
  {http://adsabs.harvard.edu/abs/2011IJMPD..20.1053D} {20, 1053}

\bibitem[\protect\citeauthoryear{Driscoll \& Healy}{Driscoll \&
  Healy}{1994}]{Driscoll:1994:CFT:184069.184073}
Driscoll J.~R.,  Healy D.~M.,  1994, \mn@doi [Adv. Appl. Math.]
  {10.1006/aama.1994.1008}, 15, 202

\bibitem[\protect\citeauthoryear{{Efstathiou}}{{Efstathiou}}{2004}]{2004MNRAS.349..603E}
{Efstathiou} G.,  2004, \mn@doi [\mnras] {10.1111/j.1365-2966.2004.07530.x},
  \href {http://adsabs.harvard.edu/abs/2004MNRAS.349..603E} {349, 603}

\bibitem[\protect\citeauthoryear{{Efstathiou}}{{Efstathiou}}{2006}]{2006MNRAS.370..343E}
{Efstathiou} G.,  2006, \mn@doi [\mnras] {10.1111/j.1365-2966.2006.10486.x},
  \href {http://adsabs.harvard.edu/abs/2006MNRAS.370..343E} {370, 343}

\bibitem[\protect\citeauthoryear{{Elsner}, {Leistedt}  \& {Peiris}}{{Elsner}
  et~al.}{2016}]{2016MNRAS.456.2095E}
{Elsner} F.,  {Leistedt} B.,   {Peiris} H.~V.,  2016, \mn@doi [\mnras]
  {10.1093/mnras/stv2777}, \href
  {http://adsabs.harvard.edu/abs/2016MNRAS.456.2095E} {456, 2095}

\bibitem[\protect\citeauthoryear{{Elsner}, {Leistedt}  \& {Peiris}}{{Elsner}
  et~al.}{2017}]{2017MNRAS.465.1847E}
{Elsner} F.,  {Leistedt} B.,   {Peiris} H.~V.,  2017, \mn@doi [\mnras]
  {10.1093/mnras/stw2752}, \href
  {http://adsabs.harvard.edu/abs/2017MNRAS.465.1847E} {465, 1847}

\bibitem[\protect\citeauthoryear{{Eriksen} et~al.,}{{Eriksen}
  et~al.}{2004}]{2004ApJS..155..227E}
{Eriksen} H.~K.,  et~al., 2004, \mn@doi [\apjs] {10.1086/425219}, \href
  {http://adsabs.harvard.edu/abs/2004ApJS..155..227E} {155, 227}

\bibitem[\protect\citeauthoryear{Frigo \& Johnson}{Frigo \&
  Johnson}{2005}]{FFTW05}
Frigo M.,  Johnson S.~G.,  2005, Proceedings of the IEEE, 93, 216

\bibitem[\protect\citeauthoryear{{Galassi} et~al.}{{Galassi}
  et~al.}{2009}]{gslbook}
{Galassi} M.,  et~al., 2009, GNU Scientific Library Reference Manual, 3 edn.
The name of the publisher

\bibitem[\protect\citeauthoryear{{Gorski}}{{Gorski}}{1994}]{1994ApJ...430L..85G}
{Gorski} K.~M.,  1994, \mn@doi [\apjl] {10.1086/187444}, \href
  {http://adsabs.harvard.edu/abs/1994ApJ...430L..85G} {430, L85}

\bibitem[\protect\citeauthoryear{{G{\'o}rski}, {Hivon}, {Banday}, {Wandelt},
  {Hansen}, {Reinecke}  \& {Bartelmann}}{{G{\'o}rski}
  et~al.}{2005}]{2005ApJ...622..759G}
{G{\'o}rski} K.~M.,  {Hivon} E.,  {Banday} A.~J.,  {Wandelt} B.~D.,  {Hansen}
  F.~K.,  {Reinecke} M.,   {Bartelmann} M.,  2005, \mn@doi [\apj]
  {10.1086/427976}, \href {http://adsabs.harvard.edu/abs/2005ApJ...622..759G}
  {622, 759}

\bibitem[\protect\citeauthoryear{{Grain}, {Tristram}  \& {Stompor}}{{Grain}
  et~al.}{2009}]{2009PhRvD..79l3515G}
{Grain} J.,  {Tristram} M.,   {Stompor} R.,  2009, \mn@doi [\prd]
  {10.1103/PhysRevD.79.123515}, \href
  {http://adsabs.harvard.edu/abs/2009PhRvD..79l3515G} {79, 123515}

\bibitem[\protect\citeauthoryear{{Hamilton}, {Tegmark}  \&
  {Padmanabhan}}{{Hamilton} et~al.}{2000}]{2000MNRAS.317L..23H}
{Hamilton} A.~J.~S.,  {Tegmark} M.,   {Padmanabhan} N.,  2000, \mn@doi [\mnras]
  {10.1046/j.1365-8711.2000.03888.x}, \href
  {http://adsabs.harvard.edu/abs/2000MNRAS.317L..23H} {317, L23}

\bibitem[\protect\citeauthoryear{{Hamimeche} \& {Lewis}}{{Hamimeche} \&
  {Lewis}}{2008}]{2008PhRvD..77j3013H}
{Hamimeche} S.,  {Lewis} A.,  2008, \mn@doi [\prd]
  {10.1103/PhysRevD.77.103013}, \href
  {http://adsabs.harvard.edu/abs/2008PhRvD..77j3013H} {77, 103013}

\bibitem[\protect\citeauthoryear{{Hansen}, {G{\'o}rski}  \& {Hivon}}{{Hansen}
  et~al.}{2002}]{2002MNRAS.336.1304H}
{Hansen} F.~K.,  {G{\'o}rski} K.~M.,   {Hivon} E.,  2002, \mn@doi [\mnras]
  {10.1046/j.1365-8711.2002.05878.x}, \href
  {http://adsabs.harvard.edu/abs/2002MNRAS.336.1304H} {336, 1304}

\bibitem[\protect\citeauthoryear{{Hivon}, {G{\'o}rski}, {Netterfield}, {Crill},
  {Prunet}  \& {Hansen}}{{Hivon} et~al.}{2002}]{2002ApJ...567....2H}
{Hivon} E.,  {G{\'o}rski} K.~M.,  {Netterfield} C.~B.,  {Crill} B.~P.,
  {Prunet} S.,   {Hansen} F.,  2002, \mn@doi [\apj] {10.1086/338126}, \href
  {http://adsabs.harvard.edu/abs/2002ApJ...567....2H} {567, 2}

\bibitem[\protect\citeauthoryear{{Ivezi{\'c}} et~al.,}{{Ivezi{\'c}}
  et~al.}{2008}]{2008arXiv0805.2366I}
{Ivezi{\'c}} {\v Z}.,  et~al., 2008, preprint, \href
  {http://adsabs.harvard.edu/abs/2008arXiv0805.2366I} {} (\mn@eprint {arXiv}
  {0805.2366})

\bibitem[\protect\citeauthoryear{{Joudaki} et~al.,}{{Joudaki}
  et~al.}{2018}]{2018MNRAS.474.4894J}
{Joudaki} S.,  et~al., 2018, \mn@doi [\mnras] {10.1093/mnras/stx2820}, \href
  {http://adsabs.harvard.edu/abs/2018MNRAS.474.4894J} {474, 4894}

\bibitem[\protect\citeauthoryear{{Kim} \& {Naselsky}}{{Kim} \&
  {Naselsky}}{2010}]{2010A&A...519A.104K}
{Kim} J.,  {Naselsky} P.,  2010, \mn@doi [\aap] {10.1051/0004-6361/201014739},
  \href {http://adsabs.harvard.edu/abs/2010A%26A...519A.104K} {519, A104}

\bibitem[\protect\citeauthoryear{{Kogut} et~al.,}{{Kogut}
  et~al.}{2003}]{2003ApJS..148..161K}
{Kogut} A.,  et~al., 2003, \mn@doi [\apjs] {10.1086/377219}, \href
  {http://adsabs.harvard.edu/abs/2003ApJS..148..161K} {148, 161}

\bibitem[\protect\citeauthoryear{{Krause} et~al.,}{{Krause}
  et~al.}{2017}]{2017arXiv170609359K}
{Krause} E.,  et~al., 2017, preprint, \href
  {http://adsabs.harvard.edu/abs/2017arXiv170609359K} {} (\mn@eprint {arXiv}
  {1706.09359})

\bibitem[\protect\citeauthoryear{{Landy} \& {Szalay}}{{Landy} \&
  {Szalay}}{1993}]{1993ApJ...412...64L}
{Landy} S.~D.,  {Szalay} A.~S.,  1993, \mn@doi [\apj] {10.1086/172900}, \href
  {http://adsabs.harvard.edu/abs/1993ApJ...412...64L} {412, 64}

\bibitem[\protect\citeauthoryear{{Laureijs} et~al.,}{{Laureijs}
  et~al.}{2011}]{2011arXiv1110.3193L}
{Laureijs} R.,  et~al., 2011, preprint, \href
  {http://adsabs.harvard.edu/abs/2011arXiv1110.3193L} {} (\mn@eprint {arXiv}
  {1110.3193})

\bibitem[\protect\citeauthoryear{{Leistedt}, {Peiris}, {Mortlock},
  {Benoit-L{\'e}vy}  \& {Pontzen}}{{Leistedt}
  et~al.}{2013}]{2013MNRAS.435.1857L}
{Leistedt} B.,  {Peiris} H.~V.,  {Mortlock} D.~J.,  {Benoit-L{\'e}vy} A.,
  {Pontzen} A.,  2013, \mn@doi [\mnras] {10.1093/mnras/stt1359}, \href
  {http://adsabs.harvard.edu/abs/2013MNRAS.435.1857L} {435, 1857}

\bibitem[\protect\citeauthoryear{{Leistedt} et~al.,}{{Leistedt}
  et~al.}{2016}]{2016ApJS..226...24L}
{Leistedt} B.,  et~al., 2016, \mn@doi [\apjs] {10.3847/0067-0049/226/2/24},
  \href {http://adsabs.harvard.edu/abs/2016ApJS..226...24L} {226, 24}

\bibitem[\protect\citeauthoryear{{Lewis}, {Challinor}  \& {Turok}}{{Lewis}
  et~al.}{2002}]{2002PhRvD..65b3505L}
{Lewis} A.,  {Challinor} A.,   {Turok} N.,  2002, \mn@doi [\prd]
  {10.1103/PhysRevD.65.023505}, \href
  {http://adsabs.harvard.edu/abs/2002PhRvD..65b3505L} {65, 023505}

\bibitem[\protect\citeauthoryear{{Louis}, {N{\ae}ss}, {Das}, {Dunkley}  \&
  {Sherwin}}{{Louis} et~al.}{2013}]{2013MNRAS.435.2040L}
{Louis} T.,  {N{\ae}ss} S.,  {Das} S.,  {Dunkley} J.,   {Sherwin} B.,  2013,
  \mn@doi [\mnras] {10.1093/mnras/stt1421}, \href
  {http://adsabs.harvard.edu/abs/2013MNRAS.435.2040L} {435, 2040}

\bibitem[\protect\citeauthoryear{{Louis} et~al.,}{{Louis}
  et~al.}{2017}]{2017JCAP...06..031L}
{Louis} T.,  et~al., 2017, \mn@doi [\jcap] {10.1088/1475-7516/2017/06/031},
  \href {http://adsabs.harvard.edu/abs/2017JCAP...06..031L} {6, 031}

\bibitem[\protect\citeauthoryear{McEwen, Puy, Thiran, Vandergheynst, Ville  \&
  Wiaux}{McEwen et~al.}{2011}]{DBLP:journals/corr/abs-1110-6297}
McEwen J.~D.,  Puy G.,  Thiran J.,  Vandergheynst P.,  Ville D. V.~D.,   Wiaux
  Y.,  2011, CoRR, abs/1110.6297

\bibitem[\protect\citeauthoryear{{POLARBEAR Collaboration} et~al.,}{{POLARBEAR
  Collaboration} et~al.}{2017}]{2017ApJ...848..121P}
{POLARBEAR Collaboration} et~al., 2017, \mn@doi [\apj]
  {10.3847/1538-4357/aa8e9f}, \href
  {http://adsabs.harvard.edu/abs/2017ApJ...848..121P} {848, 121}

\bibitem[\protect\citeauthoryear{{Peebles}}{{Peebles}}{1973}]{1973ApJ...185..413P}
{Peebles} P.~J.~E.,  1973, \mn@doi [\apj] {10.1086/152431}, \href
  {http://adsabs.harvard.edu/abs/1973ApJ...185..413P} {185, 413}

\bibitem[\protect\citeauthoryear{{Pence}}{{Pence}}{1999}]{1999ASPC..172..487P}
{Pence} W.,  1999, in {Mehringer} D.~M.,  {Plante} R.~L.,   {Roberts} D.~A.,
  eds,  Astronomical Society of the Pacific Conference Series Vol. 172,
  Astronomical Data Analysis Software and Systems VIII. p.~487

\bibitem[\protect\citeauthoryear{{Planck Collaboration} et~al.,}{{Planck
  Collaboration} et~al.}{2016}]{2016A&A...594A..13P}
{Planck Collaboration} et~al., 2016, \mn@doi [\aap]
  {10.1051/0004-6361/201525830}, \href
  {http://adsabs.harvard.edu/abs/2016A%26A...594A..13P} {594, A13}

\bibitem[\protect\citeauthoryear{{Reinecke} \& {Seljebotn}}{{Reinecke} \&
  {Seljebotn}}{2013}]{2013A&A...554A.112R}
{Reinecke} M.,  {Seljebotn} D.~S.,  2013, \mn@doi [\aap]
  {10.1051/0004-6361/201321494}, \href
  {http://adsabs.harvard.edu/abs/2013A%26A...554A.112R} {554, A112}

\bibitem[\protect\citeauthoryear{{Rybicki} \& {Press}}{{Rybicki} \&
  {Press}}{1992}]{1992ApJ...398..169R}
{Rybicki} G.~B.,  {Press} W.~H.,  1992, \mn@doi [\apj] {10.1086/171845}, \href
  {http://adsabs.harvard.edu/abs/1992ApJ...398..169R} {398, 169}

\bibitem[\protect\citeauthoryear{{Schlegel}, {Finkbeiner}  \&
  {Davis}}{{Schlegel} et~al.}{1998}]{1998wfsc.conf..297S}
{Schlegel} D.,  {Finkbeiner} D.,   {Davis} M.,  1998, in {Colombi} S.,
  {Mellier} Y.,   {Raban} B.,  eds, Wide Field Surveys in Cosmology. p.~297
  (\mn@eprint {} {astro-ph/9809230})

\bibitem[\protect\citeauthoryear{{Slosar}, {Seljak}  \& {Makarov}}{{Slosar}
  et~al.}{2004}]{2004PhRvD..69l3003S}
{Slosar} A.,  {Seljak} U.,   {Makarov} A.,  2004, \mn@doi [\prd]
  {10.1103/PhysRevD.69.123003}, \href
  {http://adsabs.harvard.edu/abs/2004PhRvD..69l3003S} {69, 123003}

\bibitem[\protect\citeauthoryear{{Smith}}{{Smith}}{2006}]{2006PhRvD..74h3002S}
{Smith} K.~M.,  2006, \mn@doi [\prd] {10.1103/PhysRevD.74.083002}, \href
  {http://adsabs.harvard.edu/abs/2006PhRvD..74h3002S} {74, 083002}

\bibitem[\protect\citeauthoryear{{Suzuki} et~al.,}{{Suzuki}
  et~al.}{2016}]{2016JLTP..184..805S}
{Suzuki} A.,  et~al., 2016, \mn@doi [Journal of Low Temperature Physics]
  {10.1007/s10909-015-1425-4}, \href
  {http://adsabs.harvard.edu/abs/2016JLTP..184..805S} {184, 805}

\bibitem[\protect\citeauthoryear{{Suzuki} et~al.,}{{Suzuki}
  et~al.}{2018}]{2018JLTP..tmp..124S}
{Suzuki} A.,  et~al., 2018, \mn@doi [Journal of Low Temperature Physics]
  {10.1007/s10909-018-1947-7}, \href
  {http://adsabs.harvard.edu/abs/2018JLTP..tmp..124S} {}

\bibitem[\protect\citeauthoryear{{Szapudi}, {Prunet}  \& {Colombi}}{{Szapudi}
  et~al.}{2001}]{2001ApJ...561L..11S}
{Szapudi} I.,  {Prunet} S.,   {Colombi} S.,  2001, \mn@doi [\apjl]
  {10.1086/324312}, \href {http://adsabs.harvard.edu/abs/2001ApJ...561L..11S}
  {561, L11}

\bibitem[\protect\citeauthoryear{{Taylor}, {Ashdown}  \& {Hobson}}{{Taylor}
  et~al.}{2008}]{2008MNRAS.389.1284T}
{Taylor} J.~F.,  {Ashdown} M.~A.~J.,   {Hobson} M.~P.,  2008, \mn@doi [\mnras]
  {10.1111/j.1365-2966.2008.13630.x}, \href
  {http://adsabs.harvard.edu/abs/2008MNRAS.389.1284T} {389, 1284}

\bibitem[\protect\citeauthoryear{{Tegmark}}{{Tegmark}}{1997}]{1997PhRvD..55.5895T}
{Tegmark} M.,  1997, \mn@doi [\prd] {10.1103/PhysRevD.55.5895}, \href
  {http://adsabs.harvard.edu/abs/1997PhRvD..55.5895T} {55, 5895}

\bibitem[\protect\citeauthoryear{{Tegmark} \& {de Oliveira-Costa}}{{Tegmark} \&
  {de Oliveira-Costa}}{2001}]{2001PhRvD..64f3001T}
{Tegmark} M.,  {de Oliveira-Costa} A.,  2001, \mn@doi [\prd]
  {10.1103/PhysRevD.64.063001}, \href
  {http://adsabs.harvard.edu/abs/2001PhRvD..64f3001T} {64, 063001}

\bibitem[\protect\citeauthoryear{{The Simons Observatory Collaboration}
  et~al.,}{{The Simons Observatory Collaboration}
  et~al.}{2018}]{2018arXiv180807445T}
{The Simons Observatory Collaboration} et~al., 2018, preprint, \href
  {http://adsabs.harvard.edu/abs/2018arXiv180807445T} {} (\mn@eprint {arXiv}
  {1808.07445})

\bibitem[\protect\citeauthoryear{{Thorne}, {Dunkley}, {Alonso}  \&
  {N{\ae}ss}}{{Thorne} et~al.}{2017}]{2017MNRAS.469.2821T}
{Thorne} B.,  {Dunkley} J.,  {Alonso} D.,   {N{\ae}ss} S.,  2017, \mn@doi
  [\mnras] {10.1093/mnras/stx949}, \href
  {http://adsabs.harvard.edu/abs/2017MNRAS.469.2821T} {469, 2821}

\bibitem[\protect\citeauthoryear{{Tristram}, {Mac{\'{\i}}as-P{\'e}rez},
  {Renault}  \& {Santos}}{{Tristram} et~al.}{2005}]{2005MNRAS.358..833T}
{Tristram} M.,  {Mac{\'{\i}}as-P{\'e}rez} J.~F.,  {Renault} C.,   {Santos} D.,
  2005, \mn@doi [\mnras] {10.1111/j.1365-2966.2005.08760.x}, \href
  {http://adsabs.harvard.edu/abs/2005MNRAS.358..833T} {358, 833}

\bibitem[\protect\citeauthoryear{{Vanneste}, {Henrot-Versill{\'e}}, {Louis}  \&
  {Tristram}}{{Vanneste} et~al.}{2018}]{2018arXiv180702484V}
{Vanneste} S.,  {Henrot-Versill{\'e}} S.,  {Louis} T.,   {Tristram} M.,  2018,
  preprint, \href {http://adsabs.harvard.edu/abs/2018arXiv180702484V} {}
  (\mn@eprint {arXiv} {1807.02484})

\bibitem[\protect\citeauthoryear{{Wandelt} \& {Hansen}}{{Wandelt} \&
  {Hansen}}{2003}]{2003PhRvD..67b3001W}
{Wandelt} B.~D.,  {Hansen} F.~K.,  2003, \mn@doi [\prd]
  {10.1103/PhysRevD.67.023001}, \href
  {http://adsabs.harvard.edu/abs/2003PhRvD..67b3001W} {67, 023001}

\bibitem[\protect\citeauthoryear{{Wandelt}, {Hivon}  \& {G{\'o}rski}}{{Wandelt}
  et~al.}{2001}]{2001PhRvD..64h3003W}
{Wandelt} B.~D.,  {Hivon} E.,   {G{\'o}rski} K.~M.,  2001, \mn@doi [\prd]
  {10.1103/PhysRevD.64.083003}, \href
  {http://adsabs.harvard.edu/abs/2001PhRvD..64h3003W} {64, 083003}

\bibitem[\protect\citeauthoryear{{Zaldarriaga} \& {Seljak}}{{Zaldarriaga} \&
  {Seljak}}{1997}]{1997PhRvD..55.1830Z}
{Zaldarriaga} M.,  {Seljak} U.,  1997, \mn@doi [\prd]
  {10.1103/PhysRevD.55.1830}, \href
  {http://adsabs.harvard.edu/abs/1997PhRvD..55.1830Z} {55, 1830}

\bibitem[\protect\citeauthoryear{{Zhao} \& {Baskaran}}{{Zhao} \&
  {Baskaran}}{2010}]{2010PhRvD..82b3001Z}
{Zhao} W.,  {Baskaran} D.,  2010, \mn@doi [\prd] {10.1103/PhysRevD.82.023001},
  \href {http://adsabs.harvard.edu/abs/2010PhRvD..82b3001Z} {82, 023001}

\bibitem[\protect\citeauthoryear{{de Jong} et~al.,}{{de Jong}
  et~al.}{2017}]{2017A&A...604A.134D}
{de Jong} J.~T.~A.,  et~al., 2017, \mn@doi [\aap]
  {10.1051/0004-6361/201730747}, \href
  {http://adsabs.harvard.edu/abs/2017A%26A...604A.134D} {604, A134}

\makeatother
\end{thebibliography}

\appendix

  \section{Generalities and Spherical Harmonic Transforms}\label{app:shts}
    Let $\eth$ and $\bar{\eth}$ be the following complex differential operators defined on the sphere when acting on a spin-$s$ quantity $f_s$:
    \begin{equation}
      \begin{split}
        \eth f_s\equiv-(\sin\theta)^s\left(\partial_\theta+i\frac{\partial_\varphi}{\sin\theta}\right)(\sin\theta)^{-s}\,f_s(\theta,\varphi), \\
        \bar{\eth} f_s\equiv-(\sin\theta)^{-s}\left(\partial_\theta-i\frac{\partial_\varphi}{\sin\theta}\right)(\sin\theta)^{s}\,f_s(\theta,\varphi).
      \end{split}
    \end{equation}
    The following properties can be easily derived for the action of these operators, and are useful to derive some of the formulas presented here:
    \begin{itemize}
      \item If $f_s$ is a spin-$s$ quantity, $(f_s)^*$ is a spin-$(-s)$ quantity.
      \item $\eth f_s$ is a spin-$(s+1)$ quantity, and $\bar{\eth} f_s$ is a spin-$(s-1)$ quantity.
      \item $(\eth^nf_s)^*=\bar{\eth}^n(f_s)^*$
      \item $\eth(f\,g)=f\eth g+g\eth f$
      \item $\eth^2(f\,g)=f\eth^2g+g\eth^2f+\eth f\eth g$
    \end{itemize}
    
    We start by defining the spin-weighed spherical harmonics with spin $s\geq0$:
    \begin{equation}
      _sY_{\ell m}\equiv \beta_{\ell,s} \eth^s Y_{\ell m},\hspace{6pt}
      _{-s}Y_{\ell m}\equiv \beta_{\ell,s} (-1)^s\bar{\eth}^s Y_{\ell m},
    \end{equation}
    where $\beta_{\ell,s}\equiv\sqrt{(\ell-s)!/(\ell+s)!}$ and $Y_{\ell m}$ are the standard spherical harmonics. These functions satisfy the property: $(_sY_{\ell m})^*=(-1)^{s+m}\,_{-s}Y_{\ell-m}$. 
            
    We can then define the $E$-mode and $B$-mode spherical harmonic vectors as:
    \begin{align}\nonumber
      _s{\bf Y}^E_{\ell m}&\equiv {\bf D}^E_sY_{\ell m}
      \equiv-\frac{\beta_{\ell,s}}{2}\left(\begin{array}{c}
                              \eth^s+\bar{\eth}^s\\
                              -i(\eth^s-\bar{\eth}^s)
                            \end{array}\right)Y_{\ell m}\\\label{eq:ederiv}
        &=-\frac{1}{2}\left(\begin{array}{c}
                               _sY_{\ell m}+(-1)^s\,_{-s}Y_{\ell m}\\
                            -i(_sY_{\ell m}-(-1)^s\,_{-s}Y_{\ell m})
                            \end{array}\right) \\\nonumber
          _s{\bf Y}^B_{\ell m}&\equiv {\bf D}^B_sY_{\ell m}
          \equiv-\frac{\beta_{\ell,s}}{2}\left(\begin{array}{c}
                              i(\eth^s-\bar{\eth}^s)\\
                              \eth^s+\bar{\eth}^s
                            \end{array}\right)Y_{\ell m}\\\label{eq:bderiv}
        &=-\frac{1}{2}\left(\begin{array}{c}
                             i(_sY_{\ell m}-(-1)^s\,_{-s}Y_{\ell m})\\
                               _sY_{\ell m}+(-1)^s\,_{-s}Y_{\ell m}
                            \end{array}\right),
      \end{align}
      which also defines the differential operators ${\bf D}^{E,B}_s$. For $s=0$, these functions are simply $_0{\bf Y}^{E}_{\ell m}=(Y_{\ell m},0)$ and $_0{\bf Y}^{B}_{\ell m}=(0,Y_{\ell m})$.
      
      Finally we define the matrix operator $\hat{\sf Y}^s_{\ell m}$ to have $_s{\bf Y}^{E,B}_{\ell m}$ as columns:
      \begin{equation}
        \hat{\sf Y}^s_{\ell m}\equiv\left(_s{\bf Y}^E_{\ell m},_s{\bf Y}^B_{\ell m},\right).
      \end{equation}
      These matrices satisfy the following properties:
      \begin{align}
        &\hat{\sf Y}^{s\dag}_{\ell m}=(-1)^{m+s}\hat{\sf Y}^{-s}_{\ell -m}\\
        &\int d\nv\,\hat{\sf Y}^{s\dag}_{\ell m}\hat{\sf Y}^s_{\ell'm'}=\hat\mI\delta_{\ell\ell'}\delta_{mm'},\\
        &\int d\nv \left(\hat{\sf Y}^{s\dag}_{\bf l}(\nv)\hat{\sf Y}^{s}_{{\bf l}_1}(\nv)\right)\,\hat{\sf Y}^0_{{\bf l}_2}(\nv)\equiv\hat{\sf D}^s_{{\bf l}{\bf l}_1{\bf l}_2},
      \end{align}
      where we have abbreviated the pair $(\ell,m)$ as ${\bf l}$, and
      \begin{align}\label{eq:dmat}
        \hat{\sf D}^s_{{\bf l}{\bf l}_1{\bf l}_2}&=(-1)^{s+m}\sqrt{\frac{(2\ell+1)(2\ell_1+1)(2\ell_2+1)}{4\pi}}\\\nonumber
        &\hspace{10pt}\wtj{\ell}{\ell_1}{\ell_2}{-m}{m_1}{m_2}
        \wtj{\ell}{\ell_1}{\ell_2}{s}{-s}{0}\,\hat{\sf d}_{\ell+\ell_1+\ell_2}.
      \end{align}
      Here
      \begin{equation}
        \hat{\sf d}_n= \frac{1}{2}\left(\begin{array}{cc}
                                            1+(-1)^n & -i[1-(-1)^n]\\
                                            i[1-(-1)^n] & 1+(-1)^n
                                          \end{array}\right),
      \end{equation}
      and the Wigner 3$j$ symbols satisfy the orthogonality relation
      \begin{equation}\label{eq:3jorth}
        \sum_{mm_1}\wtj{\ell}{\ell_1}{\ell_2}{m}{m_1}{m_2}\wtj{\ell}{\ell_1}{\ell_3}{m}{m_1}{m_3}=\frac{\delta_{\ell_2\ell_3}\delta_{m_2m_3}}{2\ell_2+1}
      \end{equation}

      \subsection{Pixelization and signal band-limits}\label{app:shts.blim}
        SHT transforms as defined in Eq.~\ref{eq:sht1} are, strictly speaking, defined for continuous fields on the sky and results in an infinite number of spherical harmonic coefficients. In practice, we have to always deal with discretized sky maps with a finite number of pixels at which the field is sampled, and consequently use a finite number of spherical coefficients. If the sampling of the underlying signal is too sparse, we will get aliasing of higher frequency modes. In flat spaces, Nyquist's theorem provides a clear prescription for uniform sampling on a grid. Since there is no such natural sampling on a sphere, the situation is more complex and depends on the pixelization scheme used \citep{Driscoll:1994:CFT:184069.184073,DBLP:journals/corr/abs-1110-6297}. Nevertheless, clear heuristic prescriptions exist. Most importantly, the typical pixel separation needs to be smaller than the wavelength of the highest frequency mode present in the underlying signal. As a natural consequence, the smoothing of the map must always precede any downgrade in resolution (or equivalently downsampling) if it is necessary to do so for numerical expediency. Similarly, any mask should be applied at the highest possible resolution and the resulting map smoothed before being down-graded.

        For the same reasons, even if the signal is correctly sampled in a band-limited sense, one must always estimate power over the entire power band, even if one is not interested in certain region of $\ell$ space. Otherwise, the power spectrum estimator will try to ``explain'' the variance associated with $\ell$ modes not considered by artificially inflating the power in the estimated modes. For the {\tt HEALPix} pixelization scheme \citep{2005ApJ...622..759G}, commonly used in cosmology and also in this paper, \cite{2013MNRAS.435.1857L} suggest the following prescription: if one wants to measure power up to $\ell_{\rm max}$, one should employ pixelisation of \texttt{Nside}=$\ell_{\rm max}/2$ and estimate band-powers up to $2 \ell_{\rm max}$, but then discard measurements above $\ell_{\rm max}$ after correcting for pixel-response suppression and accounting for their contribution to the covariance matrix of measurements.

        In this paper we do not consider these issues any further as \nmt~is a generic tool and it is up to the user to ensure band-limit and pixelization constraints are properly accounted for.
      
  \section{Linear contamination from non-linear contamination} \label{app:lincont}
    Consider a contaminant field $c$ (e.g. a point-spread function (PSF) size) that affects the observable field $o$ (e.g. the number count densities) through a non-linear, but \emph{local} on some scale local functional $F$:
    \begin{equation}
      \delta_o(\nv) \rightarrow \delta_o + F [ \delta_c ]
    \end{equation}
    where we can, without loss of generality, take out the mean effects (i.e. there is some mean PSF size and fluctuations around this mean size produce additional fluctuations in the number density of fluctuations).

    In general, for sufficiently non-linear functionals $F$, the Taylor expansion around $\delta_c(\nv)$ is not valid. However, one can look at the effect of smoothing by a kernel of size $S(R)$, which in Fourier space is equivalent by multiplication by $e^{-k^2R^2}$. We get
    \begin{equation}
      \delta^R_o(\nv) \rightarrow \delta^R_o  + S(R)[F [ \delta_c  ]] =  \delta^R_o + F'_R [ \delta^R_c ],
    \end{equation}
    where we used a shorthand $\delta_x^R=S(R)[\delta_x]$ and  $F'_R$ is now a \emph{different} functional that takes a smoothed $\delta_c$ and returns out a smoothed $\delta_o$.  The crucial point is that $F'_R$ must still be local, since one cannot turn a local functional into a non-local one by application of smoothing.

    This means, that for a well behaved $\delta_c$, such that variances of both $\delta^R_c$ and $\delta^R_o$ are small for a large enough $R$, the Taylor expansion is valid for $F'_R$ (even if it is not for $F$), giving
    \begin{equation}
      \delta^R_o(\nv) \rightarrow \delta^R_o  + b \delta^R_c + c \nabla^2 \delta^R_c
    \end{equation}
    or in Fourier space
    \begin{equation}
      \delta_o(\nv) \rightarrow \delta_o  + b \delta_c + c k^2 \delta_c
    \end{equation}
    for $k\ll R^{-1}$. In other words a ``local'' contaminant becomes ``linearly biased'' in the $k\rightarrow 0 $ regime, very much like very non-linear physics of galaxy formations results in linearly biased galaxy field on large scales.

  \section{Flat skies and Fourier transforms}\label{app:flat}
    \subsection{Spin-\texorpdfstring{$s$}{s} fields in flat sky}\label{app:flat.fields}
      In the flat sky we will label a position in the plane by two coordinates $(x,y)\equiv{\bf x}$. These coordinates can be directly related to increments in the spherical coordinates $(\theta,\varphi)$ by considering a sufficiently small map centered around the equator (i.e. $\sin\theta\sim 1$). In this case we identify the $x$ coordinate with a latitude shift ($\delta\theta=-\delta x$) and the $y$ coordinate with longitude shifts ($\delta\varphi=\delta y$). The differential operator $\eth$ now takes the form:
      \begin{equation}
        \eth= (\partial_x-i\partial_y),\hspace{12pt}\bar{\eth}= (\partial_x+i\partial_y).
      \end{equation}
      The analog of the standard spherical harmonics for flat skies are plane waves $e^{i{\bf k}{\bf x}}$, and the action of $\eth$ on these is:
      \begin{equation}
        \eth^s e^{i{\bf k}{\bf x}}=(ik)^se^{-i\,s\varphi_k}e^{i{\bf k}{\bf x}},\hspace{12pt}
        \bar{\eth}e^{i{\bf k}{\bf x}}=(ik)^se^{i\,s\varphi_k}e^{i{\bf k}{\bf x}}
      \end{equation}
  
      In analogy with the full sky case, we start by defining the basis functions:
      \begin{align}
        \,_s{\cal Y}_{\bf k}({\bf x})\equiv k^{-s}\eth^se^{i{\bf k}{\bf x}}=i^s\,e^{-is\varphi_k}e^{i{\bf k}{\bf x}},\\
        _{-s}{\cal Y}_{\bf k}({\bf x})\equiv (-k)^{-s}\bar{\eth}^se^{i{\bf k}{\bf x}}=(-i)^se^{is\varphi_k}e^{i{\bf k}{\bf x}},
      \end{align}
      where $\varphi_k$ is the polar angle of ${\bf k}$. We then define the Fourier coefficients of a spin-$s$ complex field $a({\bf x})$ as:
      \begin{align}\nonumber
        &\,_sa_{\bf l}\equiv\int\frac{d{\bf x}^2}{2\pi}\,_s{\cal Y}^*_{\bf l}({\bf x})a({\bf x}),\hspace{6pt}
        \,_{-s}a_{\bf l}\equiv\int\frac{d{\bf x}^2}{2\pi}\,_{-s}{\cal Y}^*_{\bf l}({\bf x})a^*({\bf x}).\\\label{eq:fft0}
        &a({\bf x})  =\int\frac{d{\bf l}^2}{2\pi}\,_s{\cal Y}_{\bf l}({\bf x})_sa_{\bf l},\hspace{6pt}
        a^*({\bf x})=\int\frac{d{\bf l}^2}{2\pi}\,_{-s}{\cal Y}_{\bf l}({\bf x})_{-s}a_{\bf l}.
      \end{align}
      These are then related to the $E$ and $B$-mode coefficients as:
      \begin{align}
        _sE_{\bf l}&\equiv-\frac{1}{2}\left[_sa_{\bf l}+(-1)^s\,_{-s}a_{\bf l}\right],\\
        i\,_sB_{\bf l}&\equiv-\frac{1}{2}\left[_sa_{\bf l}-(-1)^s\,_{-s}a_{\bf l}\right].
      \end{align}
      Note the preceding $(-)$ sign. For scalar fields ($s\equiv0$) the $E$ and $B$ modes are defined omitting that sign.
    
      In analogy with our curved-sky nomenclature, let us now write $a$ as a vector such that in real space ${\bf a}({\bf x})\equiv({\rm Re}(a),{\rm Im}(a))$, and in Fourier space ${\bf a}_{\bf l}\equiv(\,_sE_{\bf l},\,_sB_{\bf l})$. We can rewrite the Eq. \ref{eq:fft0} in vectorial form:
      \begin{equation}
        {\bf a}({\bf x})\equiv\int\frac{d{\bf l}^2}{2\pi}\,_s{\sf E}_{\bf l}({\bf x}){\bf a}_{\bf l},\hspace{12pt}{\bf a}_{\bf l}\equiv\int\frac{d{\bf x}^2}{2\pi}\,_s{\sf E}^\dag_{\bf l}({\bf x}){\bf a}({\bf x}),
      \end{equation}
      where, in analogy with Eqs. \ref{eq:ederiv} and \ref{eq:bderiv},  we have defined the matrix basis functions:
      \begin{align}\nonumber
        \,_s{\sf E}_{\bf l}
        &\equiv-\frac{1}{2}
        \left[\begin{array}{cc}
                 \,_s{\cal Y}_{\bf l}+(-1)^s\,_{-s}{\cal Y}_{\bf l}  & i(\,_s{\cal Y}_{\bf l}-(-1)^s\,_{-s}{\cal Y}_{\bf l}) \\
              -i(\,_s{\cal Y}_{\bf l}-(-1)^s\,_{-s}{\cal Y}_{\bf l}) &   \,_s{\cal Y}_{\bf l}+(-1)^s\,_{-s}{\cal Y}_{\bf l}
            \end{array}\right]\\\nonumber
        &=-\frac{1}{2l^s}
        \left(\begin{array}{cc}
                 \eth^s+\bar{\eth}^s & i(\eth^s-\bar{\eth}^s) \\
              -i(\eth^s-\bar{\eth}^s) &  \eth^s+\bar{\eth}^s
            \end{array}\right)e^{i{\bf l}{\bf x}}\\\nonumber
        &=-i^s
        \left(\begin{array}{cc}
               \cos(s\varphi_l) & \sin(s\varphi_l) \\
              -\sin(s\varphi_l) & \cos(s\varphi_l)
            \end{array}\right)e^{i{\bf l}{\bf x}}\\\label{eq:flat_sht}
        &=-i^s{\sf R}^\dag(s\varphi_l)e^{i{\bf l}{\bf x}}.
      \end{align}
      Here ${\sf R}(\varphi)$ is a rotation matrix.
    
    \subsection{Discrete description and DFTs}\label{app:flat.dft}
      As we discuss in \ref{sssec:maths.flat.pcl101}, it is more convenient to describe power spectrum estimation methods in a discretized flat sky. Let the patch of the sky under inspection be contained by a rectangle of sides $L_x$ and $L_y$ (in units of radians), and let us discretize this rectangle by dividing it into an $N_x\times N_y$ grid with pixels of area $\Delta{\bf x}^2\equiv\Delta x\Delta y=(L_x/N_x)(L_y/N_y)$. Each pixel in this grid is then labeled by a pair of integers ${\bf n}\equiv(n_x,n_y)$, and is assigned coordinates ${\bf x}_{\bf n}\equiv(n_x\Delta x,n_y\Delta y)$. Each field in the pixelized map ${\bf a}({\bf x})$ is therefore defined for $n_x\in[0,N_x-1]$, $n_y\in[0,N_y-1]$.
    
      In this case, the spin Fourier transform (Eq. \ref{eq:flat_sht}) of the pixelized field can be computed as its discrete Fourier transform (DFT):
      \begin{equation}\label{eq:dft}
        {\bf a}_{\bf k}\equiv {\cal D}\left({\bf a}\right)^{s_a}_{\bf k}\equiv\sum_{\bf x}\frac{\Delta{\bf x}^2}{2\pi}\,_{s_a}{\sf E}^\dag_{\bf k}({\bf x}){\bf a}_{\bf x},
      \end{equation}
      where the wavenumber ${\bf k}$ is now discretized as ${\bf k}=(j_x\Delta k_x,j_y\Delta k_y)$, the integers $j_{(x,y)}$ run from $-N_{(x,y)}/2$ to $N_{(x,y)}/2-1$\footnote{This is the valid domain when $N_{(x,y)}$ is even. For odd $N_{(x,y)}$ the interval becomes $[-(N_{(x,y)}-1)/2,(N_{(x,y)}-1)/2]$.} and the pixel size is $\Delta k_{(x,y)}\equiv2\pi/L_{(x,y)}$.
    
      The following properties of the DFT are worth recalling:
      \begin{itemize}
       \item Periodicity:
             \begin{align}\nonumber
               {\bf a}_{(k_x,k_y)}&={\bf a}_{(k_x+N_x\Delta k_x,k_y)}\\\nonumber&={\bf a}_{(k_x,k_y+N_y\Delta k_y)}\\&={\bf a}_{(k_x+N_x\Delta k_x,k_y+N_y\Delta k_y)}.
             \end{align}
       \item For a real-valued scalar $a$, its DFT satisfies $a^*_{{\bf k}_{\bf j}}=a^{}_{{\bf k}_{{\bf N}-{\bf j}}}$ (where ${\bf N}\equiv(N_x,N_y)$).
       \item The orthogonality relation of the basis functions now takes the form:
             \begin{equation}
               \sum_{\bf x}\,_sE^\dag_{\bf l}({\bf x})\,_sE_{\bf k}({\bf x})=N_xN_y\delta_{{\bf l},{\bf k}}\hat\mI.
             \end{equation}
       \item The power spectrum of a pixelized field is defined as:
             \begin{equation}
               \langle{\bf a}_{\bf l}{\bf b}^\dag_{\bf k}\rangle\equiv\frac{\delta_{{\bf l},{\bf k}}}{\Delta{\bf k}^2}{\sf C}^{ab}_{\bf l}.
             \end{equation}
      \end{itemize}

  \section{Flat-sky pseudo-\texorpdfstring{$C_\ell$}{Cl}s in the continuum limit}\label{app:flat.pcl}
    In the continuum limit, the Fourier coefficients of the masked field (see Eq. \ref{eq:pcl_flat_masked}) are:
    \begin{align}\nonumber
      {\bf a}^v_{\bf l}&=\int\int\frac{d{\bf k}^2d{\bf q}^2}{2\pi}\left[\int\frac{d{\bf x}^2}{(2\pi)^2}\,_{s_a}{\sf E}^\dag_{\bf l}({\bf x})\,_{s_a}{\sf E}_{\bf k}({\bf x})\,_0E_{\bf q}({\bf x})\right]{\bf a}_{\bf k}\,v_{\bf q}\\
                       &=\int\frac{d{\bf k}^2}{2\pi}{\sf R}\left(s_a\Delta\varphi\right)\,{\bf a}_{\bf k}v_{{\bf l}-{\bf k}},
    \end{align}
    The covariance of the Fourier coefficients of two masked fields is then given by:
    \begin{equation}
      \left\langle{\bf a}^v_{\bf l}\,{\bf b}^{w\dag}_{\bf l}\right\rangle=\int\frac{d{\bf k}^2}{(2\pi)^2}{\sf R}(s_a\Delta\varphi){\sf C}^{ab}_k{\sf R}^\dag(s_b\Delta\varphi)v_{{\bf l}-{\bf k}}w^*_{{\bf l}-{\bf k}}.
    \end{equation}
    Defining the pseudo-$C_\ell$ as the nomalized angular average of ${\bf a}_{\bf l}{\bf b}^\dag_{\bf l}$ we obtain:
    \begin{align}
      \left\langle{\rm PCL}_\ell({\bf a},{\bf b})\right\rangle
      &\equiv\frac{(2\pi)^2}{S}\int\frac{d\varphi_l}{2\pi}\left\langle{\bf a}^v_{\bf l}\,{\bf b}^{w\dag}_{\bf l}\right\rangle\\\nonumber
      &=\int\frac{kdk\,qdq}{(2\pi)^2}\left[\frac{(2\pi)^2}{S}\int\frac{d\varphi_qd\varphi_kd\varphi_l}{2\pi}v_{{\bf q}}w^*_{{\bf q}}\right.\\\nonumber
      &\hspace{12pt}\left.{\sf R}(s_a\Delta\varphi){\sf C}^{ab}_k{\sf R}^\dag(s_b\Delta\varphi)\delta({\bf q}-{\bf l}+{\bf k})\right],
    \end{align}
    where $S$ is the observed sky area (in steradian), and where we have eliminated the dependence on ${\bf l}-{\bf k}$ by introducing an additional integral over $d{\bf q}^2\delta({\bf q}-{\bf l}+{\bf k})$.
    
    As shown in \cite{2016arXiv161204664A}, these expressions can be simplified through the following steps:
    \begin{enumerate}
     \item Substitute
       \begin{equation}
         \delta({\bf q}-{\bf l}+{\bf k})\rightarrow\int\frac{d{\bf r}^2}{(2\pi)^2}e^{i({\bf q}-{\bf l}+{\bf k}){\bf r}}
       \end{equation}
     \item Integrate over the angular parts of ${\bf r}$, ${\bf l}$ and ${\bf k}$ using the following relation:
       \begin{equation}
        \int_0^{2\pi} d\varphi\,e^{i\,x\,\cos\varphi}e^{i\,n\varphi}=2\pi\,i^n\,J_n(x),
       \end{equation}
       where $J_n$ is the cylindrical Bessel function of order $n$.
     \item Integrate over the angular part of ${\bf q}$, defining the power spectrum of the masks as:
       \begin{equation}
         \tilde{C}^{vw}_q\equiv\frac{(2\pi)^2}{S}\int d\varphi_qv_{\bf q}w^*_{\bf q}
       \end{equation}
     \item Solve the last isolated integral over the radial part of ${\bf r}$ by using the following relation:
       \begin{equation}
         \int_0^\infty dr\,r\,J_0(qr)\,J_n(kr)\,J_n(\ell r)=\frac{\cos n\theta}{\pi k\ell\sin\theta},
       \end{equation}
       where $\theta$ is the angle between sides $\ell$ and $k$ of the triangle formed by three sides of length $q$, $\ell$ and $k$.
     \item Make the change of variables $q^2(\ell,k,\theta)\equiv\ell^2+k^2-2k\ell\cos\theta$ to simplify the integral over the radial part of ${\bf q}$.
    \end{enumerate}
    
    This yields the following relation analogous to Eq. \ref{eq:modecoup}:
    \begin{equation}
      {\rm vec}\left[{\sf C}^{ab}_l\right]=\int_0^{\infty} dk\, {\sf M}^{s_as_b}_{\ell k}\cdot\,{\rm vec}\left[{\sf C}^{ab}_k\right],
    \end{equation}
    where:
    \begin{align}
      &{\sf M}^{00}_{\ell k}=\frac{k}{2\pi}\int_0^\pi \frac{d\theta}{\pi}\,\tilde{C}^{vw}_{q(\ell,k,\theta)}\\
      &{\sf M}^{02}_{\ell\ell'}=M^{0+}_{\ell\ell'}\,\hat\mI,
      \hspace{12pt}
      M^{0+}_{\ell k}=\frac{k}{2\pi}\int_0^\pi \frac{d\theta}{\pi}\,\tilde{C}^{vw}_{q(\ell,k,\theta)}\cos2\theta\\
      &{\sf M}^{22}_{\ell\ell'}=\left(
      \begin{array}{cccc}
        M^{++}_{\ell\ell'} &0 &0&M^{--}_{\ell\ell'}\\
        0&M^{++}_{\ell\ell'} &-M^{--}_{\ell\ell'}&0\\
        0&-M^{--}_{\ell\ell'} &M^{++}_{\ell\ell'}&0\\
        M^{--}_{\ell\ell'} &0 &0&M^{++}_{\ell\ell'}
      \end{array}\right)\\
      &M^{\pm\pm}_{\ell\ell'}=\frac{k}{2\pi}\int_0^\pi \frac{d\theta}{\pi}\,\tilde{C}^{vw}_{q(\ell,k,\theta)}\frac{1\pm\cos4\theta}{2}.
    \end{align}
    These can also be expressed as integrals over $q$:
    \begin{equation}
      M^{XY}_{\ell k}=\frac{k}{2\pi}\int\frac{q\,dq}{2\pi}\tilde{C}^{vw}_qF(\ell,k,q)G^{XY}(\ell,k,q),
    \end{equation}
    where $F(\ell,k,q)=0$ if $\ell$, $k$ and $q$ do not form a triangle, and
    \begin{equation}\nonumber
      F(\ell,k,q)\equiv\frac{4}{\sqrt{2\ell^2k^2+2k^2q^2+2q^2\ell^2-\ell^4-k^4-q^4}}
    \end{equation}
    otherwise. The functions $G^{XY}$ are given by
    \begin{align}
      &G^{00}\equiv1,\\
      &G^{0+}(\ell,k,q)\equiv\frac{\ell^4+k^4+q^4-2k^2q^2-2\ell^2q^2}{2k^2\ell^2},\\
      &G^{++}(\ell,k,q)\equiv\left[\frac{\ell^4+k^4+q^4-2k^2q^2-2\ell^2q^2}{2k^2\ell^2}\right]^2,\\
      &G^{--}(\ell,k,q)\equiv1-G^{++}(\ell,k,q).
    \end{align}
    
    Since in the flat-sky limit $\ell$ is a continuous variable, bandpowers ${\sf B}^{ab}_q$ can be defined as averages over a given interval in $\ell$, $[\ell_q^{\rm min},\ell_q^{\rm max}]$:
    \begin{equation}
      {\sf B}^{ab}_q\equiv \int_{\ell_q^{\rm min}}^{\ell_q^{\rm max}} \frac{d\ell}{\ell_q^{\rm max}-\ell_q^{\rm min}}\,{\sf C}^{ab}_\ell.
    \end{equation}
    The binned coupling matrix is therefore given by:
    \begin{equation}
      \mathcal{M}^{s_as_b}_{qq'}\equiv\int^{\ell_q^{\rm max}}_{\ell_q^{\rm min}} \frac{d\ell}{\ell_q^{\rm max}-\ell_q^{\rm min}}\int^{\ell_{q'}^{\rm max}}_{\ell_{q'}^{\rm min}} d\ell'\,{\sf M}^{s_as_b}_{\ell\ell'}.
    \end{equation}
    \begin{figure}
      \centering
      \includegraphics[width=0.99\columnwidth]{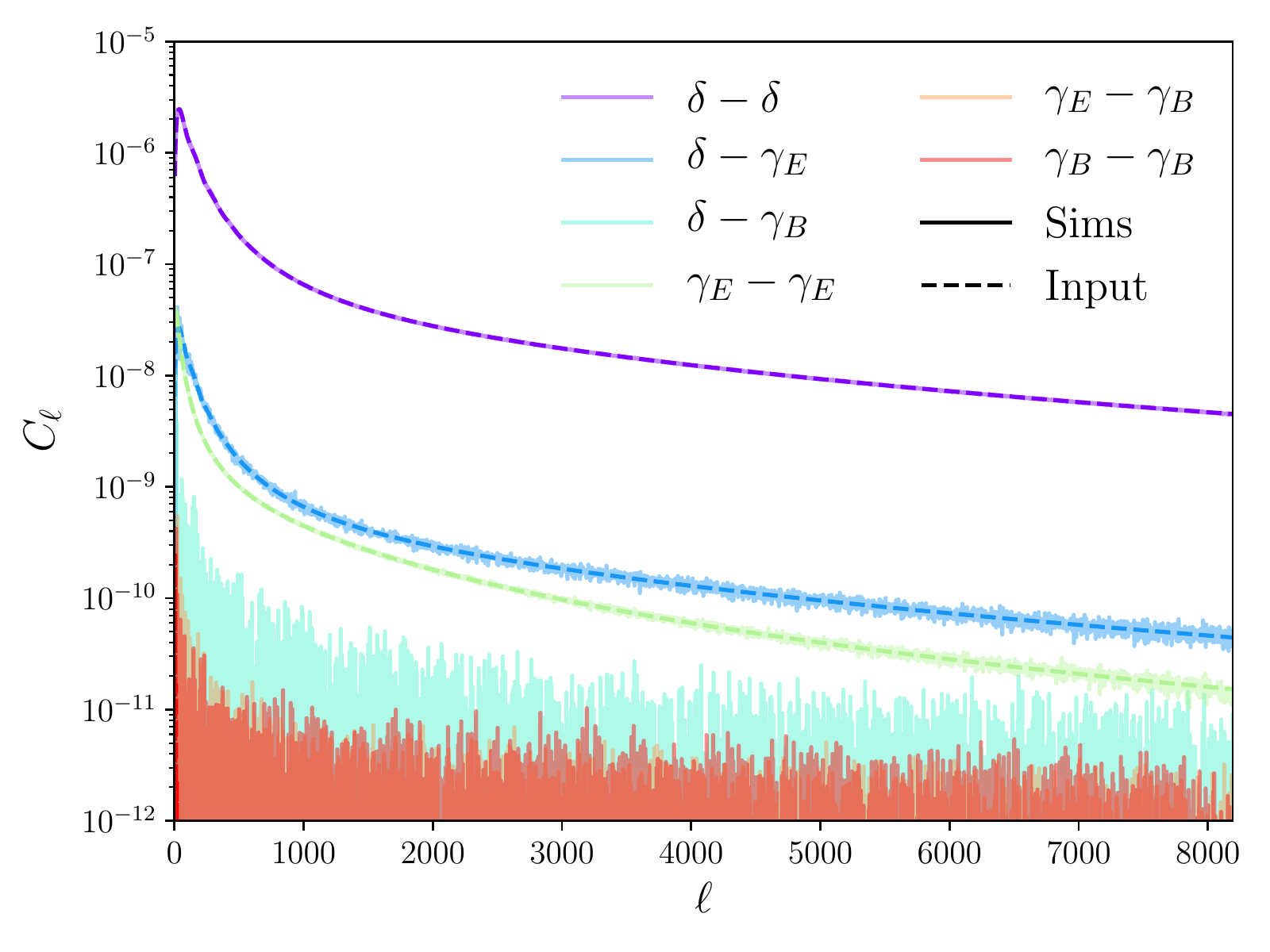}
      \caption{Input (dashed darker lines) and mean measured (solid semi-transparent lines) power spectra for simulations \texttt{HEALPix} $N_{\rm side}=4096$ using the curved-sky LSS validation suite (see Section \ref{sssec:validation.suite.lss}).}
      \label{fig:validation.fullsky.lss.4096}
    \end{figure}

    The main complication in using the pseudo-$C_\ell$ formalism in the continuum limit is the need to compute the angle-averaged mask pseudo power spectrum $C^{vw}_\ell$. Flat fields are most easily analyzed when pixelized in a Cartesian grid, and under this setup the mask power spectrum is not a well defined quantity for infinitesimally small intervals of $\ell$. This leads to non-negligible biases and poor performance \cite{2016arXiv161204664A} due to the need to use highly resolved finite intervals to compute the integrals presented in the previous section. This motivates the discrete formalism used throughout this paper, which connects directly with the storage format of flat-sky maps.

  \section{Curved-sky validation for different pixel resolutions}\label{app:full_sky_nside}
    We checked that \texttt{NaMaster} returns unbiased power-spectra for all possible values of the HEALPix resolution parameter $N_{\rm side}$. In this appendix we show that we recover the correct power spectra using the inputs from our curved-sky LSS validation suite (see Section \ref{sssec:validation.suite.lss}). In particular we tested $N_{\rm side}$ ranging from 8 to 8192. For the latter, the computational resources needed make it very challenging to generate and analyze a large enough number of simulations to have a good statistical power. However, we generate 100 simulations with $N_{\rm side}=4096$ and show the results in Figure~\ref{fig:validation.fullsky.lss.4096}. We can see that the power spectra are recovered without any noticeable bias in the expected range of validity $\ell \lesssim 2 N_{\rm side}$.
    
 \section{Performance and scalability}\label{app:scalability}
   \begin{figure}
     \centering
     \includegraphics[width=0.99\columnwidth]{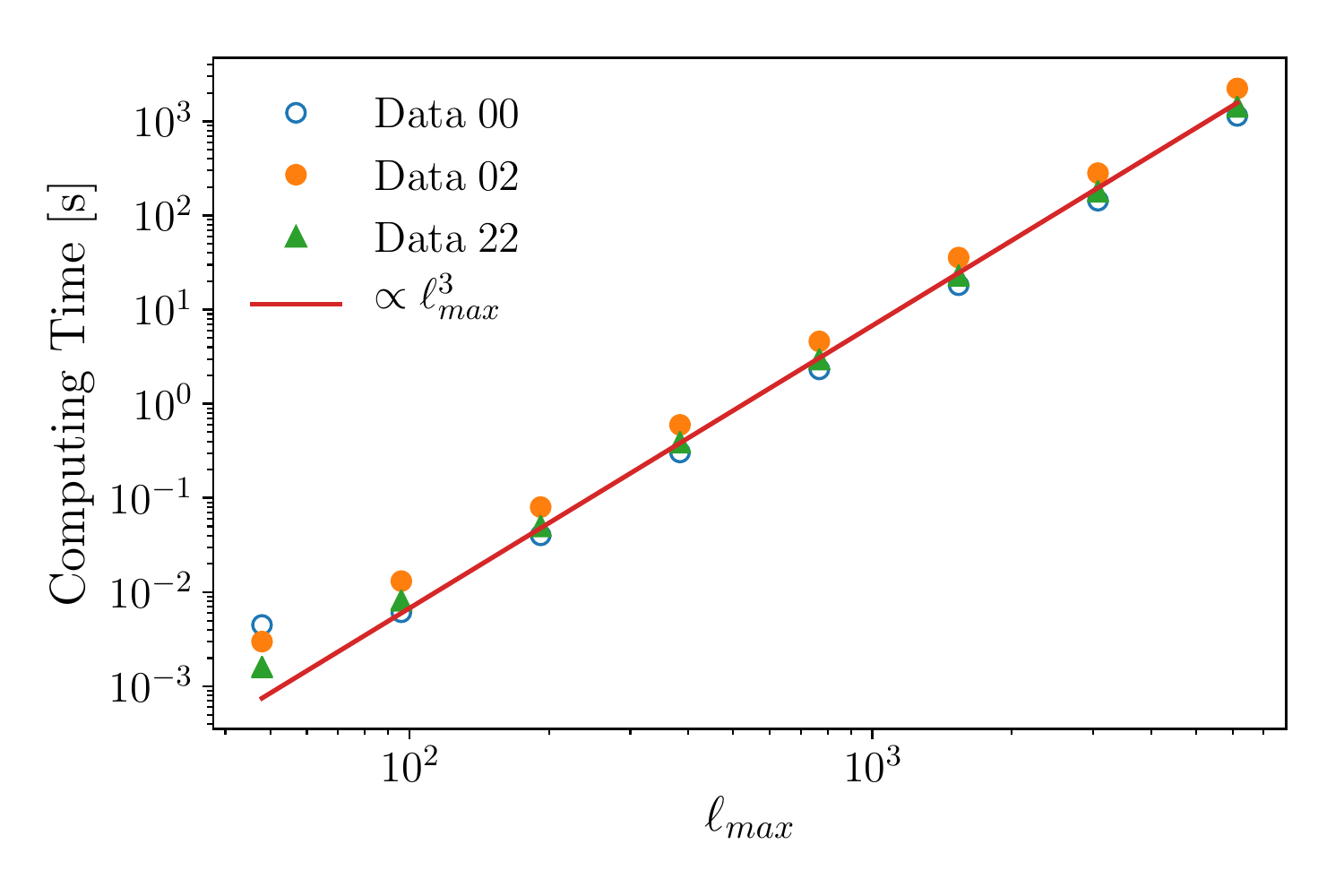}
     \includegraphics[width=0.99\columnwidth]{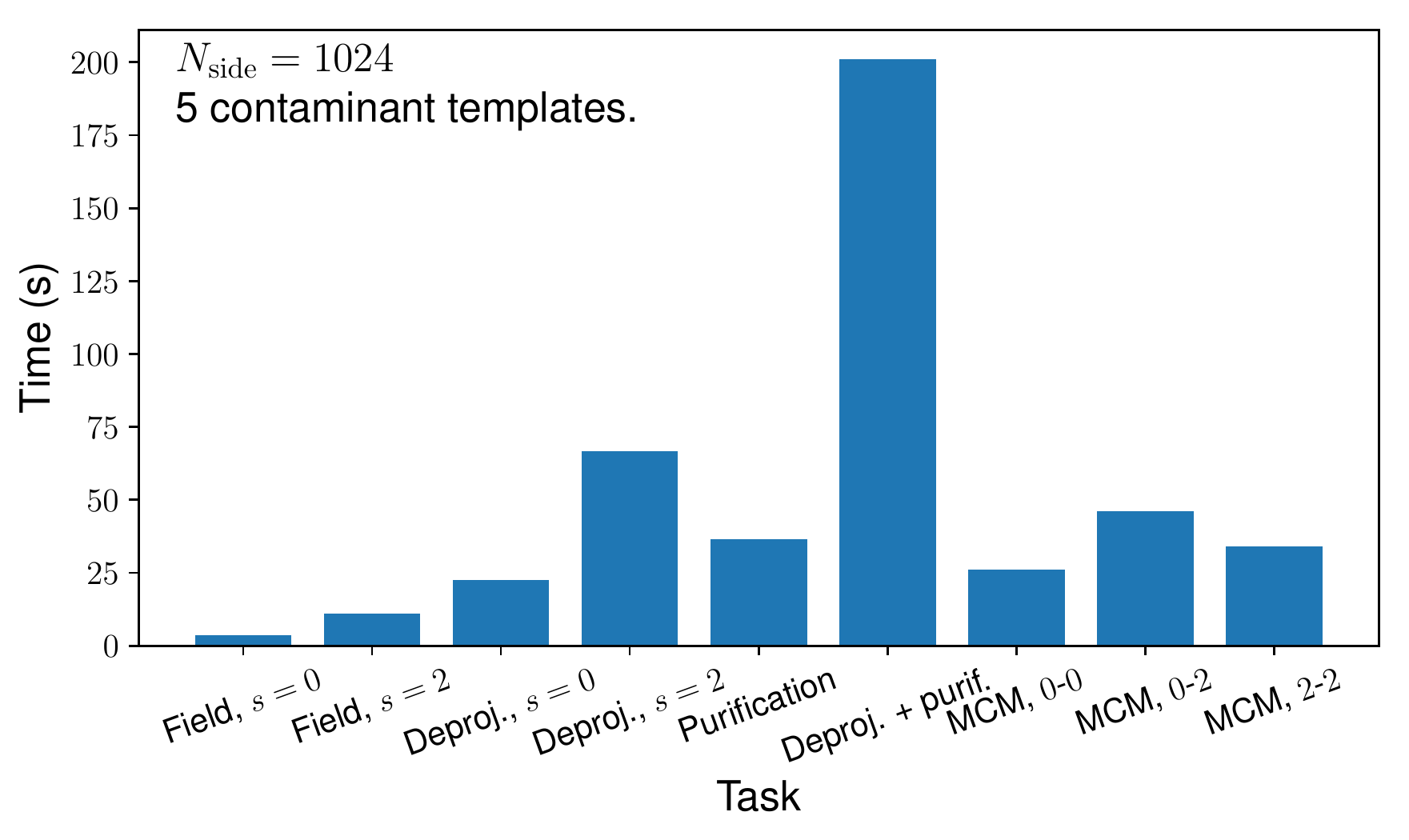}
     \caption{{\sl Top panel}: computing time for the auto-power spectra of curved-sky spin-0 (data 00, hollow circles), and spin-2 (data 22, solid triangles) contamination-free maps, as wells as their cross-power (data 02, solid circles) as a function of $\ell_{\rm max}$. We include the expected scaling $\propto\ell_{\rm max}$ for the pseudo-$C_{\ell}$ estimator (solid red line) for comparison. {\sl Bottom panel:} execution time for different tasks associated with the pseudo-$C_\ell$ estimator. We show times for fields with resolution $N_{\rm side}=1024$, which can be extrapolated to other resolutions using a $\propto N_{\rm side}^3$ scaling, as described above. The description of the different tasks shown is given in the main text.}\label{fig:scaling_lmax}
   \end{figure}
   One of the known advantages of the pseudo-$C_{\ell}$ estimator is that it is usually much faster that other known estimators~\citep{2017MNRAS.465.1847E} since the complexity scales as $\mathcal{O}(\ell_{\rm max}^{3})$, where $\ell_{\rm max}$ is the maximum multipole at which the power-spectra are calculated. In order to check how \nmt\, execution time changes with $\ell_{max}$, we compute the curved-sky angular auto and cross-power spectra of a spin-0 and a spin-2 field, with no contaminants. We change the resolution of the \texttt{HEALPix} maps by varying the $N_{\rm side}$ parameter in the range [16, 2048], and we calculate the power-spectra up to $\ell_{\rm max}=3N_{\rm side}$. We use one interactive Cori-Haswell node at the National Energy Research Scientific Computing Center (NERSC). This kind of nodes have two 16-core Intel Xeon (E5-2698 v3) processors at 2.3 GHz, and 128 GB DDR4 of RAM. We executed the code using 32 \texttt{OpenMP} threads. In the top panel of Figure~\ref{fig:scaling_lmax} we can see the results for this test. We show the results for the auto-power spectra for the spin-0 and spin-2 fields (Data 00, and Data 22, respectively), as well as the results for the cross-power spectra (Data 02). In all cases we find that the code scales as expected (the solid red line shows the expected $\propto\ell_{\rm max}^3$ behaviour). We can also see that, for very small $N_{side}$, data handling time starts to be comparable to the power-spectrum computation time, increasing the execution time with respect to the expectation.
     
   Not all parts of the power spectrum estimation take the same amount of time. E.g. even though the scaling with $\ell_{\rm max}$ is the same, the computation of the mode-coupling matrix typically takes longer than a performing single SHT. Computation times can be significantly increased if deprojection over a large number of maps is required or if $E/B$ purification is required. The bottom panel of Figure \ref{fig:scaling_lmax} shows the execution times associated to the most relevant types of operations. These are (using the labels provided in the figure):
   \begin{itemize}
     \item {\bf Field $s=0$ and $s=2$}: generating a {\tt NmtField} object. This entails taking the SHT of a given spin-0 or 2 map as well as that of its associated weights map.
     \item {\bf Deproj.}: generating a {\tt NmtField} object with contaminant deprojection. This is the same as the above in addition to computing the SHT of a set of contaminant templates. In this case we include 5 templates, and the additional computation time will increase linearly with the number of templates.
     \item {\bf Purification}: generating a {\tt NmtField} object with $B$-mode purification. This is the same as ``Field $s=2$'' in addition to computing the pure $B$-modes. This entails estimating a more accurate SHT of the apodized mask as well as a combination of harmonic-space and real-space operations described in Section \ref{ssec:maths.pureb}.
     \item {\bf Deproj. $+$ purif.}: generating a {\tt NmtField} object with both $B$-mode purification and contaminant deprojection. This entails all of the operations implied by ``Deproj.'' and ``Purification'' in addition to the computation of the $B$-mode-purified templates (see Eqs. \ref{eq:dp.g.aux} and \ref{eq:dp.f.aux}).
     \item {\bf MCM $s_1$-$s_2$}: generating the pseudo-$C_\ell$ mode-coupling matrix for a spin-$s_1$ field and a spin-$s_2$ field.
   \end{itemize}
   The times reported in this figure correspond to the time required to perform a single instance of the corresponding operation for fields with resolution $N_{\rm side}=1024$ using 16 \texttt{OpenMP} threads. Note that, even though the computation of the mode-coupling matrix dominates the total execution time in the simplest scenarios, in the most complex situation, involving both purification and deprojection, setting up a single field can take significantly longer.
     
\bsp
\end{document}